%% file: paper.tex
\documentclass[12pt,letterpaper]{article}

\usepackage{graphicx}
\usepackage[title]{appendix}
\usepackage{setspace,epstopdf,amsmath,amssymb,amsthm}
\usepackage{bbm,caption,float,longtable,pdflscape,rotating}
\usepackage[multiple]{footmisc}
\usepackage{natbib}
\usepackage[allcolors=blue,bookmarks=false,colorlinks=true]{hyperref}
\usepackage[margin=1cm]{geometry}
\usepackage[nolists,tablesfirst]{endfloat}

\begin{document}

\title{Do designated market makers provide liquidity \\ during downward extreme price movements?\thanks{We thank Giovanni Cespa, Jean-Eduard Colliard, Maria Flora, Thierry Foucault, John Hendershott, Fabrizio Lillo, Katya Malinova, Paolo Pasquariello, Athena Picarelli, Ryan Riordan, Andriy Shkilko, Xavier Vives, Ingrid Werner, and seminar participants at the Federal Reserve Board, Washington, Stevens Institute of Technology Business School, University of Trento, University of Bolzano, and participants at the 31st SNDE Annual Symposium (Padova, 2024); the XX Workshop on Quantitative Finance (Zurich, 2019); the 46th EFA Annual Meeting (Lisbon, 2019); and the 16th Paris December Finance Meeting (Paris, 2018) for valuable comments. We are solely responsible for any errors and omissions. Bellia: European Commission, Joint Research Centre,  and Department of Economics, Ca' Foscari University of Venice, Italy, \url{bellia@unive.it}. Christensen: Department of Economics and Business Economics, Aarhus University, Denmark, \url{kim@econ.au.dk}. Kolokolov: Alliance Manchester Business School, Manchester, UK, and New Economic School, Moscow, Russia, \url{alexeiuo@gmail.com}. Pelizzon: Leibniz Institute for Financial Research SAFE, Goethe University, Frankfurt, Germany, and Department of Economics, Ca' Foscari University of Venice, Italy. \url{pelizzon@safe.uni-frankfurt.de}. Ren\`{o}: ESSEC Business School, Paris, France, \url{reno@essec.edu}. Christensen was funded by a grant from the Independent Research Fund Denmark (DFF 1028--00030B). Kolokolov and Pelizzon acknowledge funding by the Deutsche Forschungsgemeinschaft (DFG--329107530). Pelizzon and Ren\`o acknowledge funding from the Italian Ministry for University and Research (MUR--2017RSMPZZ). Pelizzon also is grateful to the Leibniz Institute for Financial Research SAFE, funded by the State of Hessen initiative for research, for financial support. Part of this work was carried out when Bellia was employed at the SAFE research center and Ren\`o was employed at the Department of Economics of the University of Verona, Italy. Disclaimer: The views rendered in this paper are strictly those of the authors and may not in any circumstance be regarded as stating an official position of the European Commission.}}
\author{Mario Bellia \and Kim Christensen \and Aleksey Kolokolov \and Loriana Pelizzon \and Roberto Ren\`{o}\footnote{Corresponding author}}
\date{February, 2025}

\maketitle

\vspace*{-1.0cm}

\begin{abstract}
We study the trading activity of designated market makers (DMMs) in electronic markets using a unique dataset with audit-trail information on trader classification. DMMs may either adhere to their market-making agreements and offer immediacy during periods of heavy selling pressure, or they might lean-with-the-wind to profit from private information. We test these competing theories during extreme (downward) price movements, which we detect using a novel methodology. We show that DMMs provide liquidity when the selling pressure is concentrated on a single stock, but consume liquidity (leaving liquidity provision to slower traders) when several stocks are affected.


\end{abstract}

\thispagestyle{empty}

\pagebreak

\section{Introduction} \setcounter{page}{1} \setstretch{2.0}

Market makers facilitate the allocation of wealth and transfer of risk in financial markets, where participants may not otherwise be able to immediately locate a counterparty to take the opposite side of a deal. While any intermediary can supply liquidity on a voluntarily basis, today market making is overwhelmingly done by electronic traders, some of whom may commit to becoming a ``designated market maker'' (DMM) by entering into a written agreement with the exchange to offer immediacy to its customers.\footnote{The exact role of the DMM and the details of such agreements vary across exchanges.}

In this paper, we study the behavior of such electronic DMMs to examine whether they act according to their mandate during downward ``extreme price movements'' (EPMs), where stocks are selling off. Rather than maintaining a passive order flow during such EPMs, DMMs may decide to abstain from providing liquidity in extreme market conditions, or even switch to a ``lean-with-the-wind'' strategy with aggressive selling activity. For example, \citet{cespa-vives:25a} show that liquidity providers may lean-with-the-wind if the market lacks transparency. In these circumstances, the fear of trading against private information overshadows the foreseen revenue earned by providing liquidity, which increases market fragility. Further explanations for opportunistic trading include the backrunning theory of \citet{yang-zhu:20a} or predatory trading of \citet{brunnermeier-pedersen:05a}. Is the incentive economic presented to DMMs sufficient to overcome such effects? Moreover, DMMs often overlap with ``high-frequency traders'' (HFTs), as is also the case in our sample. The literature provides evidence of the lean-with-the-wind behavior of HFTs \citep[e.g.,][]{kervel-menkveld:19a, korajczyk-murphy:19a}. Which of these effects prevail for DMMs?

We provide compelling empirical evidence to suggest that the willingness of DMMs to supply liquidity in such a situation depends on the cross-sectional extent of the EPM. When it affects only a single stock, DMMs stick to their agreement and offer immediacy. However, when it affects several stocks, DMMs reverse course and start to consume liquidity. In their place, traditional ``slow'' traders start to purchase shares, presumably to take advantage of the discounted price. A policy implication of our study is thus that the compensation scheme offered by the exchange is not always a sufficient incentive for DMMs to maintain their presence during such adverse market conditions, and not even to discourage them from additional selling.

We analyze a unique dataset from the BEDOFIH database.\footnote{www.eurofidai.org/en/high-frequency-data-bedofih.} It consists of tick-by-tick order-level data on $37$ liquid French stocks from the CAC 40 Index and traded on NYSE Euronext Paris in 2013. The data are stamped with a flag indicating whether a participant is a HFT or not, and whether the participation falls under a DMM scheme or not---as determined by the French market authority, \citet{amf:17a}; AMF henceforth. Furthermore, it contains a trader category provided by NYSE Euronext (explained in Section \ref{section:data}). This granularity permits us to uncover the trading activity of various market participants, including DMMs.

A distinctive feature of our study is the definition of EPMs. In this paper, we rely on the ``drift burst'' approach of \citet{christensen-oomen-reno:22a}, which identifies periods where market prices are trending sharply. We investigate in detail the behavior of different trader groups during the downward EPMs detected by this statistic. We use only downward EPMs, because they are more likely to be associated with financial distress \citep[see][]{huang-wang:09a}. We show that drift burst EPMs are mostly originated by proprietary trading from investment banks, or by their clients trading via sponsored access.\footnote{We refer to ``client'' when a NYSE Euronext member is granting a counterparty access to the market via an Automated Routing System or a sponsored access.} This trading is, on average, informed, since it leads to a permanent change in the fundamental price.

In general, our results relate to a vast literature dealing with liquidity provision in electronic markets. This literature looks at the behavior of algorithmic traders, in particular HFTs, and broadly suggests that electronic traders enhance both market efficiency \citep[see, e.g.,][]{chaboud-chiquoine-hjalmarsson-vega:14a} and market liquidity \citep[see, e.g.,][]{hendershott-jones-menkveld:11a, jones:13a, clark-joseph-ye-zi:17a, bessembinder-hao-zheng:20a}.

However, the beneficial impact of algorithmic trading on market liquidity has recently been challenged. \citet{anand-venkataraman:16a} show that liquidity provision is highly correlated among HFTs, which increases the fragility of the market. They conclude that DMMs tend to mitigate illiquidity when HFTs withdraw synchronously from the market. In our framework, DMMs are also HFTs, and we show that DMMs do not always maintain their passive order flow and actually resort to aggressive selling when downward movements spread across stocks. Thus, we provide evidence that the HFT dimension prevails over designated liquidity provision.

An important closely related paper is \citet{brogaard-carrion-moyaert-riordan-shkilko-sokolov:18a}, who study HFT liquidity provision during EPMs. They label a 10-second return as an EPM if it belongs to the $99.9$th percentile of the absolute return distribution or is detected by the \citet{lee-mykland:08a} jump test.\footnote{For alternative definitions of an EPM, see \citet{goldstein-kavajecz:04a} and \citet{gao-mizrach:16a}.} Hence, they concentrate on HFTs during high volatility- and jump-based EPMs. We complement their paper in two directions. First, we look at drift burst-based EPMs. The drift burst EPMs is a largely different sample with limited overlap (see Appendix \ref{appendix:epm} for an in-depth comparison). This is because the testing procedure is based on the drift term, and thus identifies directional price moves, while it controls for volatility and excludes jumps. It allows us to extract a sequence of consecutive returns around which the market is directional. This has the advantage, important to our analysis, that we can exploit the time series dimension of our data, such as identifying the peak of an EPM, allowing us to disentangle various stages of its development. Moreover, our approach does not require us to fix the duration of the EPM in advance. Second, we concentrate on DMMs. \citet{brogaard-carrion-moyaert-riordan-shkilko-sokolov:18a} conclude that HFTs provide liquidity, on average, during EPMs, but they switch to the demand side if several stocks are affected. Our empirical analysis complements their findings by disentangling liquidity provision by trader group. We show that the incentive for liquidity provision is not sufficient to prevent DMMs from acting as non-designated market maker HFTs, thus casting the compensation scheme offered by the exchange into doubt.

The rest of the paper proceeds as follows. In Section \ref{section:description}, we describe the institutional structure of NYSE Euronext. In Section \ref{section:data}, we describe the data and presents the trader classification. In Section \ref{section:identification}, we describe the identification of drift burst EPMs. The econometric approach and our empirical results appear in Section \ref{section:empirical}. We conclude in Section \ref{section:conclusion}. In Appendix \ref{appendix:epm}, we compare the drift burst EPMs to those detected by \citet{brogaard-carrion-moyaert-riordan-shkilko-sokolov:18a}. Appendix \ref{appendix:supplemental} contains supplementary material. In Appendix \ref{appendix:kirilenko}, we implement an alternative model to analyze trading activity based on \citet{kirilenko-kyle-samadi-tuzun:17a}, which largely confirms the findings in the main text.

\section{Institutional structure} \label{section:description}

\subsection{The NYSE Euronext market}

NYSE Euronext operates as an order-driven stock market with a limit order book. The Paris division of the exchange includes all French instruments, including equities and derivatives. The daily trading schedule for the most liquid stocks is divided into different segments. The session starts at 7:15 a.m. with a pre-opening phase, followed by an auction at 9:00 a.m. The main trading phase starts at 9:00 a.m. and ends at 5:30 p.m. The daily schedule is followed by a closing auction and a further trading session called ``trading-at-last,'' where additional trades can be executed at the closing price. In this study, we only look at the main trading phase, where the vast majority of trading activity is concentrated, thus opening and closing activity are excluded from our analysis.

According to the Rule 4403/2 of Rule Book I \citep{euronext:25a}, during the continuous trading, Euronext has in place a set of trading safeguards that prevents price movements outside certain thresholds. Specifically, traded prices are constrained into a ``collar,'' defined by a reference price plus/minus a percentage price change. If the execution of an order causes the breach of the collar, two outcomes are possible: 1) the order is partially executed inside the collar, without halting the continuous trading; 2) the trading process is halted, and the market is put in ``reservation mode.'' Continuous trading resumes after a new auction. None of the EPMs included in our analysis trigger any of these measures. Thus, the collars are ineffective against preventing the occurrence of our detected events.

The Euronext group is the second-largest stock exchange in Europe (behind the LSE Group) and ranks fifth globally (the top four being NYSE, NASDAQ, the Japan Exchange Group, and the LSE Group) according to data from the World Federation of Exchanges and Statista. The French branch, Euronext Paris, accounts for approximately 70\% of the total market capitalization of the Euronext group, positioning the domestic market as a top 10 stock exchange worldwide on a standalone basis. Additionally, the markets exhibit levels of HFT activity comparable to those in the United States. In our sample, HFTs account for roughly 80\% of the euro trading volume. \citet{brogaard:10a} reports that HFTs contribute to 68.5\% of the dollar volume traded in the U.S. equities market, while \citet{benos-sagade:12a} find a much smaller participation in a small sample of U.K. stocks (27\% of all trading volume). \citet{menkveld:14a} posits, based on several sources, that HFT participation has grown from 30\% to 70\% in recent years. We conclude that the documented results for Euronext Paris ought to generalise to other limit order book markets, especially given the relative size of the market and the significance of the HFT activity therein.\footnote{As of 2013, NYSE Euronext Paris ranks fifth in the world in terms of market capitalization based on information from the World Federation of Exchanges. \citet{bellia-pelizzon-subrahmanyam-yuferova:25a} show that in our sample HFTs account for around 82\% of the gross turnover.}

\subsection{The contractual role of DMMs}

The market model of NYSE Euronext Paris relies on the provision of liquidity by electronic market makers. Since 2011, the NYSE Euronext operates a program called Supplemental Liquidity Provision (SLP), where electronic traders can agree to post two-sided quotes during the day and to provide a minimum passive execution volume. This program, described below, identifies our DMMs. The orders sent by DMMs must be electronic, use only their own funds, and exclude customer orders.\footnote{In contrast to the NYSE (US) market, the NYSE Euronext Paris does not have a single DMM for each stock, but relies on multiple DMMs that act as non-voluntarily liquidity providers under a market making agreement with the exchange.} We  emphasize that the agreement does not contain any special incentive for DMMs to provide liquidity during market distress. However, a recent review report by the European Securities and Markets Authority (ESMA) is now looking into this design \citep[see][]{mifid:21a}.

The SLP program was introduced in 2012 with the aim of protecting the market share of NYSE Euronext against other trading venues, such as Chi-X Europe, BATS Europe, and Equiduct. The Flash News of March 26, 2012 \citep{euronext:12a} covers the details of the implementation of the scheme, while the Flash News of May 9, 2013 \citep{euronext:13a} introduces new requirements and also extends the possibility of joining the program to other market participants starting June 3, 2013. The program initially required that each firm appointed as an SLP must commit to being present on one or more baskets of stocks and fulfill the following three rules: 1) be present at least 95\% of the time on both sides of the market during the continuous trading session; 2) display a minimum volume of at least 5,000 euros at the best limit; and 3) be present at the best limit prices for each assigned basket of securities, with a minimum time presence of 10\% for each security. The program was revised in June 2013 with significant changes to Rule 3. The exchange introduced a minimum passive execution level (liquidity provision) of 0.70\% in terms of the aggregate monthly volume traded on Chi-X, BATS, Turquoise, and NYSE Euronext, a minimum presence time of 25\% at the NYSE Euronext best limit for each assigned basket, weight-averaged over the entire basket and calendar month, and a minimum passive execution level of 0.1\%, as well as a minimum presence time of 10\%, at the NYSE Euronext best limit during the continuous trading session for each security, weight-averaged over the calendar month.

If DMMs meet the criteria, the minimum charge when they take liquidity is 0.30 bps and the maximum rebate for liquidity provision is 0.20 bps until May 2013, which increased to 0.22 bps from June 3, 2013. These amounts represent the bounds for fees and rebates. There are intermediate levels that can reduce the rebate or increase the fees up to 0.55 bps per trade, depending on the time presence and passive executions. Indeed, 0.55 bps per trade is the standard fee for both taking and providing liquidity. Additionally, the time priority of orders at the best limit price is not considered when determining SLP members' presence at the inside spread: as soon as there is an order at the top of the book flagged as SLP, the presence is counted.

Since August 1, 2012, the French government has imposed a financial transaction tax (FTT) of 20 bps on the purchase of French equities, along with an HFT tax. HFTs can avoid this taxation by either not holding inventories or signing an agreement with the exchange to perform market-making duties. If, after a certain period (2 months in the initial implementation), SLP members do not fulfil the requirements or their performance is below the commitment, the exchange terminates the contract. The member's activity will then be charged at 0.55 bps per order executed, regardless of whether it provides or takes liquidity. Other discounts for colocation and the dedicated data feed infrastructure will also be terminated if a member is excluded from the SLP program.

Overall, with the time presence (95\% of the time on both sides of the book, at any level) and top of the book presence (10\% of the time, weighted average across one month), the requirements provide substantial leeway for DMMs to move their quotes to a lower level or withdraw from the market in the case of extreme price movements without incurring substantial penalties.

\section{Data description} \label{section:data}

The database at our disposal is from the Base Europ\'{e}enne de Donn\'{e}es Financi\`{e}res \`{a} Haute Fr\'{e}quence (BEDOFIH). It is composed of tick-by-tick order-level data for 37 liquid stocks that are included in the CAC 40 Index in 2013.\footnote{Three stocks on the CAC 40 Index are excluded from our analysis: Arcelor Mittal, Gemalto, and Solvay, since their main trading venue is not the Paris branch of NYSE Euronext.} We can track the entire history of all orders, from the initial submission to the execution or cancellation, with a timestamp at the microsecond level. Each trade struck during the main trading phase has a flag indicating the initiator, allowing us to identify which trader-account is trading aggressively (demanding liquidity) and passively (supplying liquidity) for the matched order.

Euronext requires each trader to flag every order in compliance with the Rule Book I, according to the following list of possible categories \citep[see][]{euronext:12a}: orders submitted pursuant to the Supplemental Liquidity Provision agreement (MM); own account (OWN) for proprietary trading; account of an affiliate, or when operating from a parent company of the Euronext main member (PARENT);\footnote{From Rule 3.4 of Rule Book I, the ``Relevant Euronext Market Undertaking may consider an application from a Member who wishes to obtain direct access to an Euronext Market for its Affiliate(s). The Affiliates are a person who (i) owns 95 per cent or more of the Member; or (ii) is owned 95 per cent or more by the Member; or (iii) is owned 95 per cent or more by a third party who also owns 95 per cent or more of the Member.''} account of a third party (CLIENT), when the Euronext member grants access to the client via an Automated Routing System or through sponsored access;\footnote{Euronext members can grant access to the market and the facilities like co-location, via a sponsored access, according to Rules 3.2 and 3.3 of Rule Book I. However, ``All business undertaken by a Client via an Automated Order Routing System or via Sponsored Access on an Euronext Market will be done in the name of the Member and the Member retains full responsibility for the conduct of all such business'' (Rule 3201/2).} or orders submitted for retail investors only.\footnote{Orders for retail investors are flagged as RLP (Retail Liquidity Provider) or RMO (Retail Member Organisation) and belong to a specific program designated for them. Euronext member banks and brokers can execute their retail orders flow via the Retail Matching Facility (RMF) against new price-improving liquidity provided by RLPs. Details about the program are available at https://www.euronext.com/en/media/4053/download. However, the amount of liquidity provided via this channel is not visible in the central limit order book and is only accessible to retail clients under the RMO. The amount of trades and orders categorized with this flag is negligible compared to the overall market activity carried out by the remaining trading accounts.}

An intriguing feature of this database is that individual orders from the stock exchange are labelled with a high-frequency trading (HFT) identification flag from the French stock market regulator AMF. The classification is based on the lifetime of cancelled orders. A trader is classified with the HFT mark if one of two criteria is met:
\begin{enumerate}
\item The average lifetime of its cancelled orders is less than the average lifetime of all orders in the book and if it has cancelled at least 100,000 orders during the year.
\item The participant must have cancelled at least 500,000 orders with a lifetime of less than 0.1 second (i.e., the participant quickly updates the orders in the limit order book) and the top percentile of the lifetime of its cancelled orders must be less than 500 microseconds (i.e., the participant regularly uses fast access to the market).
\end{enumerate}
The HFTs are further divided into two sub-categories: pure HFT companies, such as Citadel or Virtu, and investment banks with HFT activity (named ``mixed'' by AMF), such as Goldman Sachs. The classification is revised yearly, and the groups are mutually exclusive.

\begin{figure}[t!]
\begin{center}
\caption{Trading activity by trader category}
\label{figure:trading_activity}
\begin{tabular}{c}
\includegraphics[height=10cm,width=0.9\textwidth]{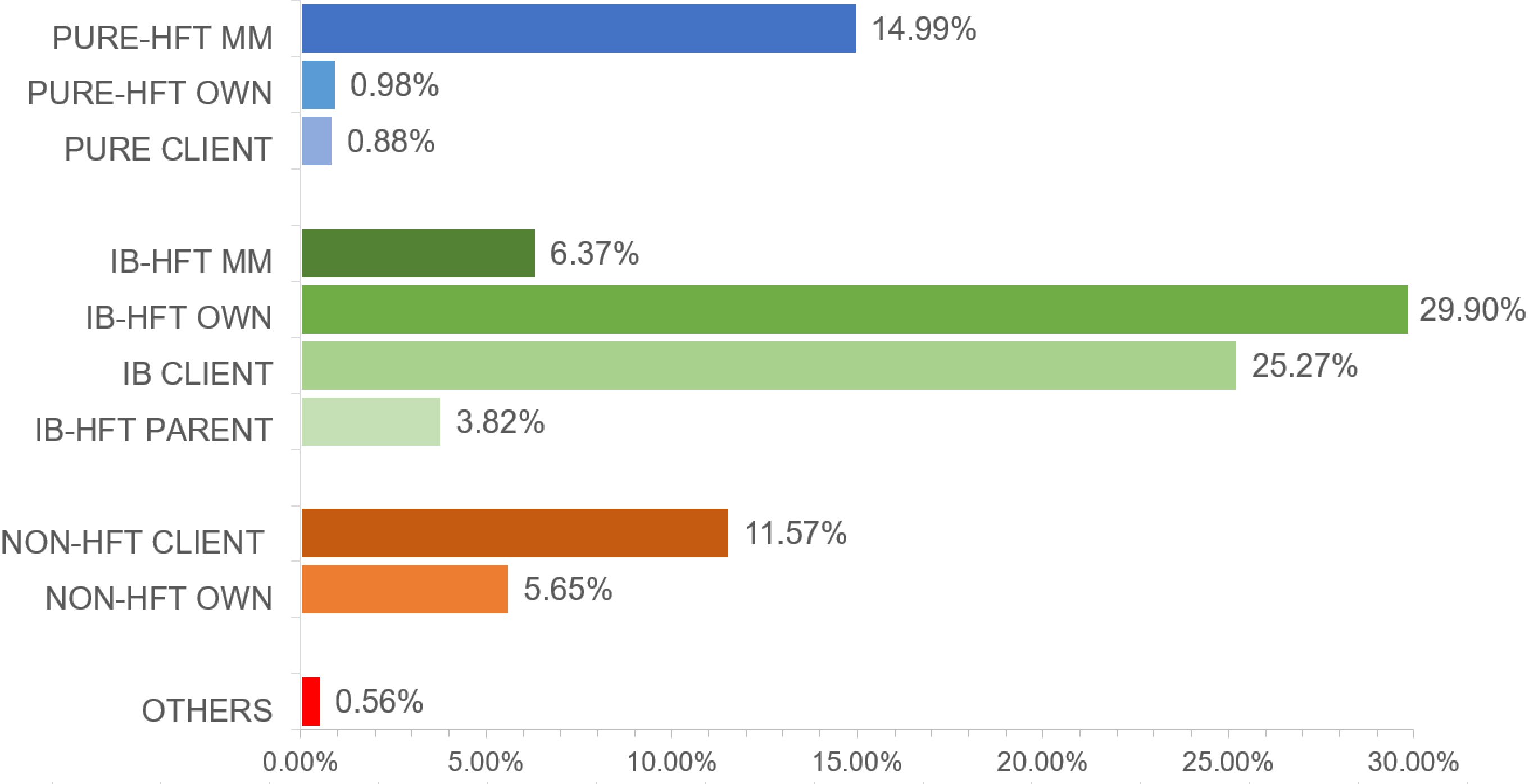}
\end{tabular}
\smallskip
\parbox{\textwidth}{\emph{Note.} This figure displays the percentage of total double-counted volume (buy and sell) for each trader group, averaged across stocks in our sample. The category is determined via the account flag from the exchange and the AMF flag for HFT.}
\end{center}
\end{figure}

We merge the AMF definition of HFT with the \citet{sec:14a} memorandum, which adds proprietary trading as a typical feature of HFTs, by grouping traders into the following categories (also displayed in Figure \ref{figure:trading_activity}):
\begin{itemize}
\item Pure HFT companies---according to AMF---are labelled by their trading classification, as PURE-HFT MM, PURE-CLIENT, and PURE-HFT OWN. We remove the flag ``HFT'' from clients given that the AMF classification does not guarantee that these trades are proprietary. The DMMs in this group are PURE-HFT MM.
\item Mixed HFT companies---according to AMF---are labelled IB-HFT MM, IB-CLIENT, IB-HFT OWN, and IB-HFT PARENT. Again, we remove the flag ``HFT'' from the clients. We prefer the label ``IB'' to ``MIXED,'' because we believe it adds less confusion in the naming convention of this trader category. The DMMs in this group are IB-HFT MM.
\item Non-HFTs---according to AMF---are labelled NON-HFT CLIENT and NON-HFT OWN. There are no DMMs in this last category.
\end{itemize}
We leave out a few trader categories with negligible trading activity (Others), accounting for only 0.56\% of total volume. Figure \ref{figure:trading_activity} shows that most of the trading activity in our sample is due to PURE-HFT MM, IB-HFT MM, IB-CLIENT, and IB-HFT OWN. Together, they account for roughly 76\% of the total double-counted volume (buy and sell). Non-HFTs account for 18\%. The rest is divided among PURE-HFT OWN, PURE-CLIENT, IB-HFT PARENT, and Others.

\section{EPM detection} \label{section:identification}

The detection of EPMs is based on a recently developed drift burst approach proposed by \citet{christensen-oomen-reno:22a}, which supports the notion that an EPM is (relative to the level of volatility) a large movement in the price over a short time horizon.

\begin{figure}[t!]
\begin{center}
\caption{The price of Technip on June 25, 2013}
\label{figure:technip}
\begin{tabular}{c}
\includegraphics[height=10cm,width=0.9\textwidth]{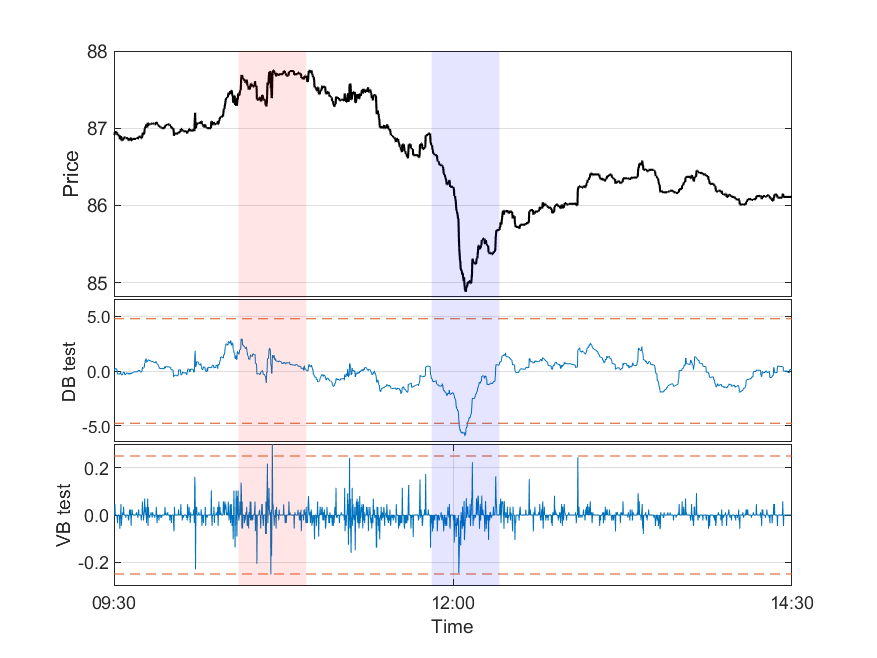}
\end{tabular}
\smallskip
\parbox{\textwidth}{\emph{Note.} In the upper panel, the figure reports the transaction price of Technip on June 25, 2013. The price is sampled at a 10-second grid. The middle panel reports the test statistic of \citet{christensen-oomen-reno:22a}, which detects a drift burst (DB) EPM around noon (as indicated by the shaded area around that time). The lower panel shows the approach of \citet{brogaard-carrion-moyaert-riordan-shkilko-sokolov:18a}. The authors examine the absolute value of 10-second returns and detects a volatility-based (VB) EPM around 10:30 (as indicated by the shaded area around that time). The dashed lines represent 99.9\% confidence bands.}
\end{center}
\end{figure}

To illustrate our approach, we look at an example of a detected event. In the top panel of Figure \ref{figure:technip}, we report the evolution of the price of Technip on June 25, 2013. Around 11:50 am the price starts to decline rapidly until it reaches a minimum a few seconds after 12:05 pm. The return over the 15-minute window is -2.35\%. After the crash, the price retraces somewhat. This is an example of an EPM that takes the form of a so-called ``flash crash'': a large price drop in a short time followed by a partial recovery. The intermediate panel shows the drift burst statistic of \citet{christensen-oomen-reno:22a}. The trough of the test statistic coincides with the trough of the price.

Figure \ref{figure:technip} highlights an important difference between our notion of an EPM compared to \citet{brogaard-carrion-moyaert-riordan-shkilko-sokolov:18a}. Their detection algorithm consists of extracting---often disconnected---10-second absolute returns exceeding the 99.9th percentile of the distribution of such returns for a stock. However, even though the cumulative price drop is large during the drift burst EPM displayed in Figure \ref{figure:technip}, the individual high-frequency returns, charted in the bottom panel, are typically small and in line with the overall volatility of that day. The largest negative 10-second return (-0.25\%) occurs inside the volatility cluster before 11:00 am, where the price level did not change much in the surrounding 30-minute window, as indicated by the shaded area. Hence, in this instance, \citet{brogaard-carrion-moyaert-riordan-shkilko-sokolov:18a} identify the interval with high volatility (or jumps) before the crash.\footnote{Appendix \ref{appendix:epm} provides a broader in-sample comparison of EPMs detected by the drift burst test statistic with those extracted by \citet{brogaard-carrion-moyaert-riordan-shkilko-sokolov:18a} showing that they largely constitute disjoint sets with limited overlap, in line with the example.}

To explain the mechanics of our EPM identification procedure in more detail, we need a minimal amount of notation. We assume that $(p_{t})_{t \geq 0}$ denotes the log-price process of an asset, which evolves according to the dynamic:
\begin{equation}\label{model1}
\mathrm{d}p_{t} = \mu_{t} \mathrm{d}t + \sigma_{t} \mathrm{d}W_{t} + \mathrm{d}J_{t},
\end{equation}
where $\mu_{t}$ is the drift, $\sigma_{t}$ is the volatility, $W_{t}$ is a standard Brownian motion, and $J_{t}$ is a jump process. Such a description of the instantaneous price change is standard in the continuous-time literature. The volatility and jump component are central to \citet{brogaard-carrion-moyaert-riordan-shkilko-sokolov:18a}, while we concentrate on the drift.

Suppose the log-price is observed on $[0,T]$ at irregular time points $0 = t_{0} < t_{1} < \ldots < t_{n} = T$, $(p_{t_{i}})_{i=0}^{n}$. An intuitive estimator of the drift is given by
\begin{equation} \label{equation:spot-drift}
\hat{ \mu}_{t} = \frac{1}{h} \sum_{i=1}^{n} K \Bigg( \frac{t_{i-1} - t}{h} \Bigg) r_{t_{i}},
\end{equation}
where $r_{t_{i}} = p_{t_{i}} - p_{t_{i-1}}$, for $i = 1, \ldots, n$ is the discretely sampled log-return on $[t_{i-1},t_{i}]$, $K$ is a kernel function, and $h$ is a bandwidth.

\citet{christensen-oomen-reno:22a} propose the drift burst test statistic:
\begin{equation} \label{equation:db-statistic}
T_{t} = \sqrt{ \frac{h}{ \mathcal{K}}} \frac{ \hat{ \mu}_{t}}{ \hat{ \sigma}_{t}},
\end{equation}
where
\begin{equation} \label{equation:spot-volatility}
\hat{ \sigma}_{t} = \sqrt{ \frac{1}{h} \sum_{i=1}^{n} K \Bigg( \frac{t_{i-1} - t}{h} \Bigg) r_{t_{i}}^{2}}
\end{equation}
is an estimator of the volatility and $\mathcal{K} = \int_{- \infty}^{ \infty} K^2(x)\text{d}x$.

The ratio $\hat{ \mu}_{t} / \hat{ \sigma}_{t}$ measures the current velocity of the price. In normal markets the ratio is small, as evident from Figure \ref{figure:technip} in the morning and afternoon. However, if the price starts to trend too fast relative to the volatility (i.e., price changes are directional), the ratio becomes large, as also evident from Figure \ref{figure:technip} during the crash at noon. \citet{christensen-oomen-reno:22a} formalize this idea by assuming that there exists a ``drift burst'' time point $\tau_{\text{db}}$, where $\mu_{t} \rightarrow \pm \infty$ as $t \rightarrow \tau_{\text{db}}$. They show that at such points the absolute value of the test statistic diverges to infinity. Otherwise, it follows a standard normal distribution. It can therefore be used to identify drift burst EPMs when the test statistic is larger than an appropriate quantile of the standard normal distribution.\footnote{To implement the drift burst test statistic, we set $K(x) = \exp(-|x|) \mathbbm{1}(x \leq 0)$, such that $\hat{ \mu}_{t}$ and $\hat{ \sigma}_{t}$ are computed with a left-sided exponential kernel based on backward-looking data to avoid a look-ahead bias. We employ a $h = 5$-minute bandwidth for the mean and a $h = 25$-minute bandwidth for the volatility. The transaction price record is pre-averaged following \citet{jacod-li-mykland-podolskij-vetter:09a} to soften the impact of microstructure noise. The volatility estimator is robustified with a HAC-type correction. This is the default choice in the MATLAB code made available by \citet{christensen-oomen-reno:22a}.}

\citet{christensen-oomen-reno:22a} show that the drift burst test statistic is robust to jumps. As evident from equation \eqref{equation:db-statistic}, it also controls for volatility. Hence, our EPM measure does not identify either jumps or large volatility, but only directional price moves. Furthermore, the procedure is robust to market microstructure frictions, such as bid-ask bounce. Moreover, extensive Monte Carlo simulations in their paper show that the test is correctly sized, so that for sufficiently high values of the test statistic, the probability of contamination by false positives is essentially zero.

We compute the drift burst statistic every second during the trading session, searching for downward EPMs after 9:30 am. We exploit the simulation-based algorithm from \citet{christensen-oomen-reno:22a} to set a critical value. We follow \citet{brogaard-carrion-moyaert-riordan-shkilko-sokolov:18a} and use a 99.9\% confidence level, which implies an average critical value of -4.9. A test statistic below this barrier is labelled a drift burst EPM.

To calculate the duration of an EPM, we denote by $t_{ \mathrm{trough}}$ the time point associated with the lowest (most negative) value of the drift burst test statistic during a significant event. The beginning of the EPM, which we call $t_{ \mathrm{start}}$, is identified as the latest time prior to $t_{ \mathrm{trough}}$ that the test statistic crosses the barrier -1. The difference $\tau = t_{ \mathrm{trough}} - t_{ \mathrm{start}}$ is the duration of the event. The drift burst EPM is assumed to stop at $t_{ \mathrm{end}} = t_{ \mathrm{trough}} + 3 \tau$. The interval from $t_{ \mathrm{trough}}$ to $t_{ \mathrm{end}}$ is called the recovery. We further include a pre-event window starting at time $t_{ \mathrm{pre-event}} = t_{ \mathrm{start}} - 2 \tau$. Our analysis of each EPM is based on price and order information from $t_{ \mathrm{pre-event}}$ to $t_{ \mathrm{end}}$.\footnote{Below we normalize the time unit of each drift burst EPM with the average duration across events. That is, each trade time (after subtracting $t_{ \mathrm{start}}$) is multiplied by the average duration and divided by the actual duration of the event. The resulting time unit, still in minutes, is indicated as ``average elapsed time.''}

\section{Empirical investigation} \label{section:empirical}

In this section, we first present a brief description of the drift burst EPM sample. Then, we test competing theories about DMM liquidity provision during EPMs using a VAR model for trading imbalances. Finally, we investigate whether DMM behavior depends on the EPMs or rather high selling pressure.

\subsection{Anatomy of a drift burst EPM}

Table \ref{table:descriptive} reports the descriptive statistics of the 148 identified downward EPMs extracted with the drift burst procedure. Moreover, an overview of each individual event is available in Table \ref{table:drift-burst-epm-individual} in Appendix \ref{appendix:supplemental}, while in Table \ref{table:drift-burst-epm-aggregated}, we aggregate them by stock, together with some firm-level information. This reveals that drift burst EPMs occur in 34 of the 37 CAC 40 Index stocks included in our sample. The largest number of EPMs in a single stock is 10 for Alstom, while the smallest number is 1 for Legrand. There is no obvious relationship among the number of EPMs and the stocks' market capitalization, volatility, or average return.

The duration of a drift burst EPM ranges anywhere from a few minutes to three-quarters of an hour, with a typical duration of $9.5$ minutes. The price concedes by -1.35\%, on average, which is an enormous move over such a short time interval. In the largest drop, which occurred in ST Microelectronics on March 12, 2013, the price changed by -5.18\% over 6.5 minutes, while during the smallest one, which occurred in Pernod Ricard on October 2, 2013, the price declined by -0.37\% over 3.1 minutes. Table \ref{table:descriptive} also shows that while the duration of a drift burst EPM represents only 1.87\% of the trading day, it accounts for 5.43\% of the daily trading volume, nearly 6\% of the transaction count, and roughly 21\% of the sell-side activity during that day.

\begin{table}[ht!]
\setlength{\tabcolsep}{0.25cm}
\begin{center}
\caption{Descriptive statistics of drift burst EPM}
\label{table:descriptive}
\begin{tabular}{lrrrrrrr}
\hline \hline
& & & & & \multicolumn{3}{c}{Quantile} \\
\cline{6-8}
& Mean & Std. & Min & Max & $q_{0.10}$ & Median & $q_{0.90}$ \\
Return & -1.35 & 0.80 & -5.18 & -0.37 & -2.42 & -1.11 & -0.62 \\
\\
Duration & 9.53 & 8.45 & 0.07 & 41.60 & 2.12 & 7.32 & 19.34 \\
in \%  & 1.87 & 1.66 & 0.01 & 8.16 & 0.42 & 1.44 & 3.81 \\
\\
Number of trades & 582.10 & 442.32 & 89.00 & 2612.00 & 182.90 & 452.50 & 1191.70 \\
in \% & 5.99 & 4.29 & 1.07 & 35.98 & 2.46 & 5.01 & 9.66 \\
\\
Trading volume & 4598.60 & 3789.11 & 537.53 & 18268.03 & 1210.75 & 3521.45 & 9530.02 \\
in \% & 5.43 & 4.06 & 0.71 & 34.71 & 2.02 & 4.50 & 9.03 \\
\\
Signed volume & -1276.37 & 1347.86 & -9291.14 & 1722.17 & -2942.18 & -1049.13 & -19.78 \\
in \% & 21.07 & 182.75 & -667.12 & 1592.26 & -58.15 & 9.19 & 64.44 \\
\hline \hline
\end{tabular}
\smallskip
\parbox{\textwidth}{\emph{Note.} This table shows descriptive statistics of a drift burst EPM. We calculate the return (in percent), the duration, the number of trades, the euro trading volume (in thousands), and the signed volume for each event. We report for each statistic its sample average, standard deviation, minimum value, maximum value, and the 10\%, 50\% (median), and 90\% quantile across events. The rows starting with ``in \%'' measure the statistic in percent relative to the associated number computed over the whole trading day.}
\end{center}
\end{table}

\begin{figure}[t!]
\begin{center}
\caption{Temporal distribution of drift burst EPMs}
\label{figure:epm-placement}
\begin{tabular}{c}
\includegraphics[height=10cm,width=0.9\textwidth]{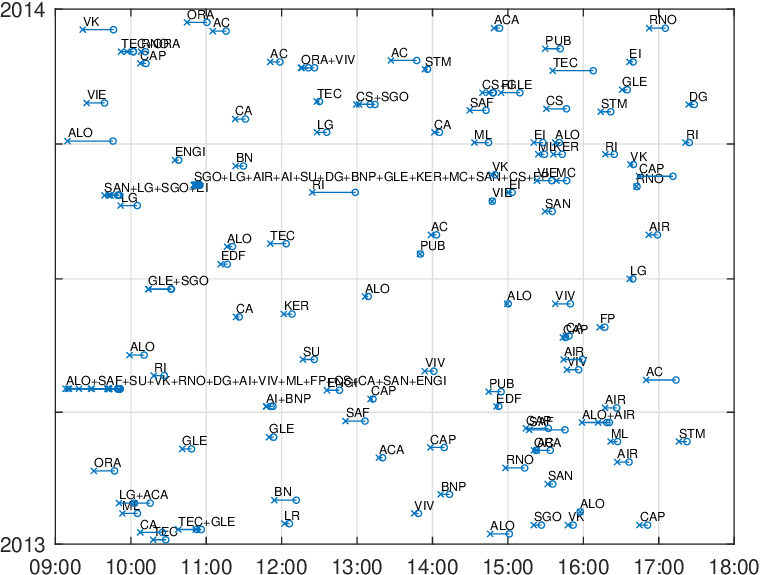}
\end{tabular}
\\
\medskip
\parbox{\textwidth}{\emph{Note.} In this figure, we present the timing of the 148 drift burst EPMs analyzed in the empirical application. On the x-axis, we plot the start of the event (or $t_{ \text{start}}$) with a cross and the time of the trough (or $t_{ \text{trough}}$) with a circle connected by a line segment to mark the duration. On the y-axis, we report the time of the year it occurred. The ticker symbols of the affected stocks are shown above the line (see Table \ref{table:drift-burst-epm-aggregated} for the company name). A ``+'' indicates drift burst EPMs that overlap.}
\end{center}
\end{figure}

Figure \ref{figure:epm-placement} presents a graphical representation of the temporal placement of the drift burst EPMs over the trading day and the year. This shows that the events are more or less randomly scattered over time, both intraday and interday, without any tendency to cluster around specific times. In a couple of cases---visible in the figure---drift burst EPMs affect a large number of stocks in the cross-section, namely on April 17, 2013 (with 14 stocks involved) and on September 3, 2013 (with 13 stocks involved).\footnote{Figure \ref{figure:systematic} in Appendix \ref{appendix:supplemental} shows the development of the equity prices for the 37 companies included in our sample during these events.}$^{,}$\footnote{On April 17, 2013, the French government announced new austerity budget measures that morning due to a pessimistic revision of GDP growth; see, e.g., https://lexpansion.lexpress.fr/actualite-economique/ce-qu-il-faut-retenir-du-nouveau-plan-budgetaire-de-la-france\_1404490.html. Meanwhile, to our knowledge the event on September 3, 2013 cannot be attributed to the release of any significant news. However, crashes in several stocks without news can be explained by liquidity spillover \citep{cespa-foucault:14a}.} In the remainder of the text, we label these $27$ events as ``systematic,'' while the rest ($121$ events) are labelled ``unsystematic.''\footnote{As evident from Figure \ref{figure:epm-placement}, there are additional drift burst EPMs affecting more than one stock. However, we do not label these as systematic, because the number of involved stocks is rather small (2--4).}

\begin{figure}[t!]
\begin{center}
\caption{Average cumulative return during a drift burst EPM}
\label{figure:drift-burst-EPM}
\begin{tabular}{c}
\includegraphics[height=10cm,width=0.9\textwidth]{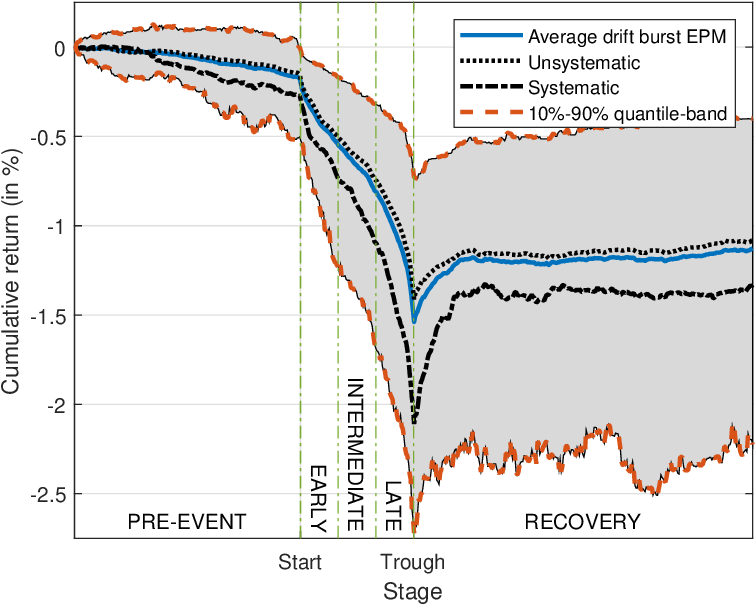}
\end{tabular}
\smallskip
\parbox{\textwidth}{\emph{Note.} This figure shows the evolution of the cumulative return over an average drift burst EPM, and also an unsystematic and systematic one, together with a 10\%-90\% quantile-band calculated across the combined 148 events. The vertical lines separate the duration of the drift burst EPM from start to through into an early, intermediate, and late stage, as detailed in the main text.}
\end{center}
\end{figure}

In Figure \ref{figure:drift-burst-EPM}, we plot the average cumulative return for unsystematic and systematic drift burst EPMs, together with a 10\%--90\% quantile-band computed across events. This figure provides a qualitative description of the price process during a drift burst EPM. It is consistent with a drift burst EPM being partly information-based, as it leads to a permanent price change. However, for systematic events there is also evidence of an inefficient market response due to overshooting.

We define, for each event, the permanent price impact (PPI) as:
\begin{equation} \label{equation:PPI}
\textrm{PPI} = p_{t_{ \mathrm{end}}} - p_{t_{ \mathrm{pre-event}}}.
\end{equation}
That is, the log-return from start to finish of each event.\footnote{Taking the price at a later time point to define the post-event efficient price follows the convention in the market microstructure literature, e.g., when estimating effective spreads or measuring price discovery, see \citet{glosten:87a} or \citet{barclay-warner:93a}. As a robustness check, we also employed the closing price at the end of the day or the next day, as in \citet{kervel-menkveld:19a}, but this had no discernible effect on the analysis.} Furthermore, the downward price impact (DPI) of the drift burst EPM is given by
\begin{equation} \label{equation:DPI}
\textrm{DPI} = p_{t_{ \mathrm{trough}}} - p_{t_{ \mathrm{pre-event}}},
\end{equation}
while the transient price impact (TPI) is:
\begin{equation} \label{equation:TPI}
\textrm{TPI} = \textrm{DPI} - \textrm{PPI} = p_{t_{ \mathrm{trough}}} - p_{t_{ \mathrm{end}}}.
\end{equation}
To understand who contributes to the PPI and TPI, we further subdivide the drift burst EPM (from $t_{ \mathrm{start}}$ to $t_{ \mathrm{trough}}$) into three stages with equal duration: early, intermediate, and late. This is also indicated in Figure \ref{figure:drift-burst-EPM}.

\begin{figure}[t!]
\begin{center}
\caption{Price impact of a drift burst EPM}
\label{figure:price-impact}
\begin{tabular}{cc}
Panel A: PPI vs. TPI. & Panel B: DPI vs. duration. \\
\includegraphics[height=8cm,width=0.48\textwidth]{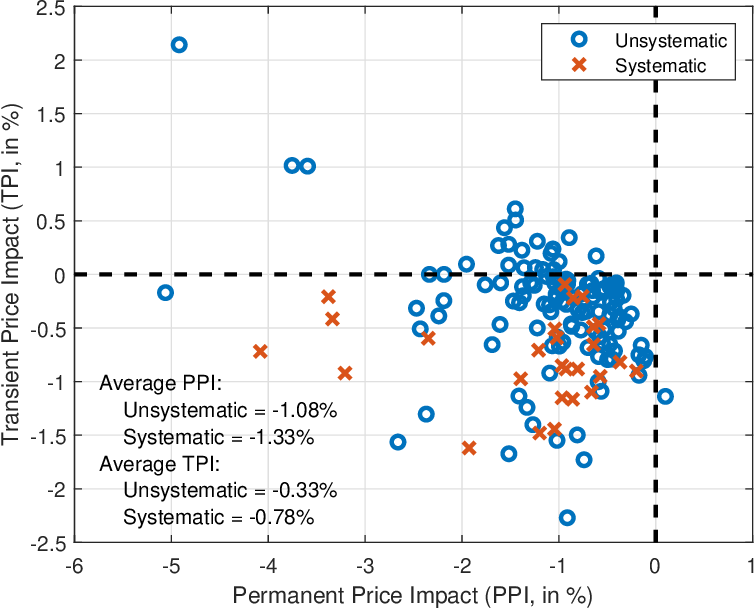} &
\includegraphics[height=8cm,width=0.48\textwidth]{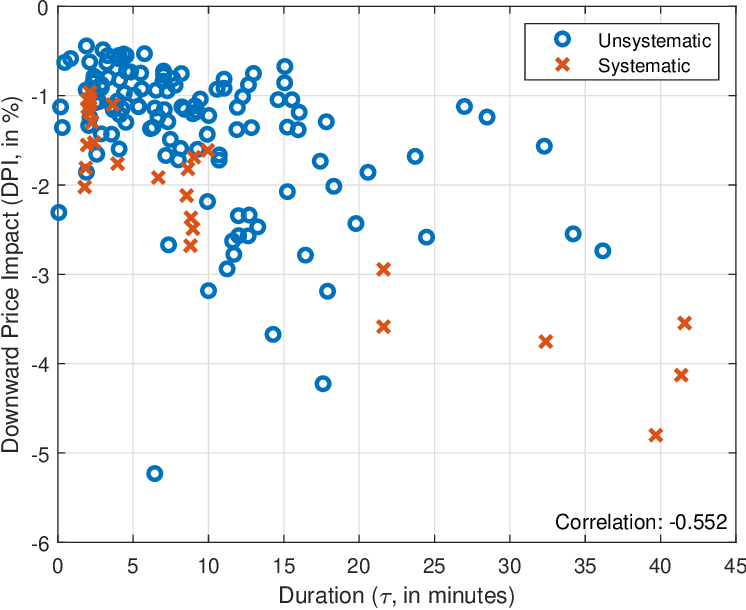}
\end{tabular}
\smallskip
\parbox{\textwidth}{\emph{Note.} In Panel A, we report a scatter plot of the permanent price impact (PPI, defined in equation \eqref{equation:PPI}) versus the transient price impact (TPI, defined in equation \eqref{equation:TPI}). We also compute the associated sample average and standard deviation (in parentheses) of the log-return, which is further divided into unsystematic and systematic drift burst EPMs. In Panel B, we show a scatter plot of the downward price impact (DPI, defined in equation \eqref{equation:DPI}) versus the duration of the drift burst EPM. The negative correlation implies that the longer a drift burst EPM lasts the deeper is the associated price drop.}
\end{center}
\end{figure}

In Panel A of Figure \ref{figure:price-impact}, we look at the relation between the PPI and TPI. PPI is mostly negative, which is again indicative of informed trading, while TPI is always negative (with a single exception), which is again indicative of overshooting. The average PPI and TPI are both highly significantly negative, which holds for unsystematic and systematic drift burst EPMs separately. Indeed, on average, TPI is -0.78\% for a systematic event, or about half of the PPI, and it is -0.33\% for an unsystematic event, or roughly one third of the PPI. Panel B of Figure \ref{figure:price-impact} includes a scatter of DPI and the duration of the drift burst EPM, which shows a strong negative association: the longer a drift burst EPM lasts, the more negative is the DPI.

\subsection{Trading imbalance}  \label{subsection:trade}

In this subsection, we investigate the trading behavior of DMMs during drift burst EPMs. We provide an empirical answer to the main research question of the paper: Do DMMs provide liquidity, in keeping with their role, or do they follow their ``HFT instinct''? Their behavior is synthesised by their trading imbalances (i.e., the net euro amount of their buying and selling volume), which we analyze with a VAR model.

We break the event window of identified drift burst EPMs into non-overlapping 10-second intervals, adopting the time frame of \citet{brogaard-carrion-moyaert-riordan-shkilko-sokolov:18a}. We then analyze the evolution of the trading imbalance of each trader group. We then calculate the euro volume of a single trade $t$, which is denoted by $V_{t} = Q_{t} \cdot P_{t}$, where $Q_{t}$ is the number of shares traded and $P_{t}$ is the transaction price. The trading imbalance for trader category $i$ over a given period is then:
\begin{equation} \label{equation:imbalance}
\mathcal{TI}^{i}_{ \textrm{period}} = \sum_{t \in { \textrm{period}}} s_{t} \cdot V_{t},
\end{equation}
where $s_{t} = +1$ for a buy order and $s_{t} = -1$ for a sell order.

The interpretation is as follows. A negative trading imbalance before the trough means that a trader group contributes to the price drop with selling activity (either aggressively or passively), while a positive trading imbalance after the trough indicates it is helping the price to recover (either aggressively or passively).\footnote{The measure is equivalent to the one employed by \citet{brogaard-carrion-moyaert-riordan-shkilko-sokolov:18a}, except they use the number of traded shares, $Q_{t}$. We further standardize inventories at the stock level to make them comparable across EPMs.}

We estimate a $G$-dimensional vector autoregression (VAR) for the trading imbalance across trader categories, which equation-by-equation reads as follows:
\begin{equation} \label{equation:var}
\mathcal{TI}^{i}_{j,t} =  \alpha_{0}^{i} + \text{FE}_{ \mathrm{stock}} + \sum_{s=1}^{5} \beta^{i}_{s} D^{s}_{j,t} + \sum_{g=1}^{G} \sum_{l=1}^{5} \gamma_{g,l} \mathcal{TI}^{g}_{j,t-l} + \alpha^{i} \textrm{Controls}_{j,t} + \epsilon^{i}_{j,t},
\end{equation}
where $\mathcal{TI}^{i}_{j,t} $ denotes the trading imbalance of trader category $i$, computed for the $t$-th 10-second interval of the $j$-th event, $\text{FE}_{ \mathrm{stock}}$ are stock fixed effects, $G$ is the number of trader categories, $D_{j,t}^{s}$ are dummy variables indicating the stage of the drift burst EPM with $s \in \{$pre-event, early drop, intermediate drop, late drop, recovery\}. For example, $D_{j,t}^{ \text{late drop}} = 1$ if the $t$-th interval of the $j$-th event belong to the period of the late drop, and is 0 otherwise. We add five lagged values of the trading imbalances of each trader category $g$ to account for serial correlation. $\textrm{Controls}_{j,t}$ include the contemporaneous 10-second return and log-euro volume, plus the percentage bid-ask spread.\footnote{As for the inventories, we standardize non-dummy variables at the stock level to make them more homogeneous.} $\epsilon_{j,t}^{i}$ is an error term.

We estimate the model after pooling observations into unsystematic and systematic drift burst EPMs. The reported standard errors are heteroskedasticity-robust and clustered at the stock and date level. In view of the 10-second time interval required to aggregate the trading imbalances, we exclude drift burst EPMs with a duration smaller than 100 seconds from the regression.

Table \ref{table:var} reports selected coefficient estimates of the VAR system. We restrict attention to the most active trader categories and group NON-HFT into a single category.\footnote{The full version of the table with all trader categories is presented in Table \ref{table:var-full} in Appendix \ref{appendix:supplemental}.} In Panel A, we report estimates for the unsystematic drift burst EPMs. Here, IB-HFT OWN displays significant selling activity at the beginning of the event, followed in a later stage by IB-CLIENT. IB-HFT MM supports the market and supplies liquidity in all stages of the drift burst EPM, in agreement with their role as DMMs. PURE-HFT MM are also liquidity providers, but to a lesser degree in terms of statistical significance. After the trough of the event, NON-HFT starts significantly buying the market during the partial recovery, arguably to take advantage of the discounted price.

In Panel B, we show the results for the systematic drift burst EPMs. The results are striking. In line with the above, the coefficients on the early drop dummy and the intermediate drop dummy are negative and statistically significant for IB-HFT OWN. However, as opposed to the above, the estimated coefficient for the late drop dummy is negative and statistically significant for IB-HFT MM, while the intermediate drop dummy is negative and statistically significant for PURE-HFT MM. By contrast, for NON-HFT, the coefficients are positive and statistically significant at all stages of the drop. We interpret these results as follows: The inception of the drift burst EPM can be associated with selling pressure from IB-HFT OWN (and also IB-PARENT as shown in Appendix \ref{appendix:supplemental}). In a systematic event, this is---correctly---interpreted by DMMs as informed trading, so both the PURE-HFT MM and, especially, IB-HFT MM start to lean-with-the-wind by following suit with opportunistic selling activity. This is, however, in contradiction to their role as DMMs. On the other side, NON-HFT absorbs some of the selling pressure in all stages of the event.

\begin{table}[p]
\setlength{ \tabcolsep}{0.30cm}
\begin{center}
\caption{Coefficient estimates from the VAR for trading imbalance}
\label{table:var}
\begin{footnotesize}
\begin{tabular}{lccccc}
\hline
\multicolumn{6}{l}{ \textit{Panel A: Unsystematic drift burst EPMs}} \\
 & PURE-HFT MM & IB-HFT MM & IB-HFT OWN & IB-CLIENT & NON-HFT \\
\cline{2-6}
Pre-event & -0.007 & 0.031{*}{*} & -0.012 & -0.019 & 0.014 \\
 & (-0.79) & (2.47) & (-1.08) & (-1.46) & (1.22) \\
Early drop & 0.078{*} & 0.290{*}{*}{*} & -0.145{*}{*} & -0.020 & -0.045 \\
 & (1.69) & (4.09) & (-2.28) & (-0.50) & (-0.98) \\
Intermediate drop & -0.025 & 0.176{*}{*}{*} & -0.056 & -0.084{*}{*} & -0.029 \\
 & (-0.74) & (3.83) & (-1.50) & (-2.02) & (-0.79) \\
Late drop & 0.017 & 0.228{*}{*}{*} & -0.007 & -0.266{*}{*}{*} & -0.012 \\
 & (0.30) & (3.38) & (-0.10) & (-3.47) & (-0.20) \\
Recovery & -0.024{*}{*} & -0.009 & -0.018 & -0.016 & 0.041{*}{*} \\
 & (-2.46) & (-0.62) & (-0.89) & (-0.90) & (2.39) \\
Bid-ask spread & 0.002 & 0.006 & -0.008 & 0.001 & -0.002 \\
 & (0.50) & (1.20) & (-1.45) & (0.15) & (-0.33) \\
Log(euro volume) & -0.018{*}{*} & 0.025{*}{*} & 0.008 & -0.026{*} & 0.011 \\
 & (-2.22) & (2.33) & (0.63) & (-1.83) & (0.73) \\
Return & 0.021{*} & -0.189{*}{*}{*} & 0.012 & 0.031{*}{*}{*} & 0.016{*} \\
 & (1.78) & (-12.06) & (1.23) & (3.68) & (1.83) \\
\\
$\bar{R}^{2}$ & 0.003 & 0.057 & 0.025 & 0.039 & 0.020 \\
\\
\multicolumn{6}{l}{ \textit{Panel B: Systematic drift burst EPMs}} \\
 & PURE-HFT MM & IB-HFT MM & IB-HFT OWN & IB-CLIENT & NON-HFT \\
\cline{2-6}
Pre-event & 0.015 & -0.010 & -0.024 & 0.018 & 0.017 \\
 & (0.80) & (-0.39) & (-1.27) & (1.02) & (0.93) \\
Early drop & -0.053 & 0.013 & -0.155{*}{*} & 0.048 & 0.177{*}{*} \\
 & (-0.97) & (0.15) & (-2.14) & (0.55) & (2.71) \\
Intermediate drop & -0.132{*}{*} & -0.162 & -0.350{*}{*}{*} & 0.135 & 0.472{*}{*}{*} \\
 & (-2.26) & (-1.27) & (-3.51) & (1.07) & (2.89) \\
Late drop & -0.089 & -0.373{*}{*} & -0.086 & -0.000 & 0.543{*}{*}{*} \\
 & (-1.27) & (-2.29) & (-0.95) & (-0.00) & (4.09) \\
Recovery & 0.011 & -0.042 & 0.028 & -0.015 & 0.013 \\
 & (0.47) & (-1.40) & (1.44) & (-0.59) & (0.85) \\
Bid-ask spread & 0.006 & -0.007 & -0.004 & 0.003 & 0.003 \\
 & (0.65) & (-0.71) & (-0.57) & (0.27) & (0.48) \\
Log(euro volume) & -0.023 & -0.143{*}{*}{*} & -0.022 & 0.047{*}{*} & 0.102{*}{*}{*} \\
 & (-1.29) & (-7.61) & (-1.14) & (2.33) & (4.54) \\
Return & 0.031 & -0.076{*}{*}{*} & 0.006 & 0.031{*}{*} & -0.034{*}{*}{*} \\
 & (1.09) & (-3.64) & (0.61) & (2.15) & (-3.26) \\
\\
$\bar{R}^{2}$ & 0.008 & 0.070 & 0.022 & 0.031 & 0.039 \\
\hline
\end{tabular}
\end{footnotesize}
\\
\medskip
\parbox{\textwidth}{\emph{Note.} We present the results of the VAR in equation \eqref{equation:var}. The model is estimated separately for unsystematic drift burst EPMs in Panel A and systematic ones in Panel B. The dependent variable is the 10-second trading imbalance of the trader category indicated by the column label. The explanatory variables ``Pre-event'', ``Early drop'', ``Intermediate drop'', ``Late drop'', and ``Recovery'' refer to phase dummies of a drift burst EPM. ``Bid-ask spread'' is the quoted spread (in percent), ``Log(euro volume)'' is the log-euro volume, and ``Return'' is the contemporaneous stock return. We also include stock fixed effects. The non-dummy variables are standardized at the stock level. The coefficient estimates of lagged trading imbalances are not reported for brevity. The $t$-statistics (in parentheses) are based on heteroscedasticity-robust standard errors clustered at the stock and date level. *$P$-value$<$0.1; **$P$-value$<$0.05; ***$P$-value$<$0.01. nObs = 342,032 in Panel A and nObs = 79,404 in Panel B.}
\end{center}
\end{table}

\begin{figure}[t!]
\begin{center}
\caption{Dynamic evolution of trading imbalance}
\label{figure:trading-imbalance}
\begin{tabular}{cc}
\small{Panel A: Unsystematic} & \small{Panel B: Systematic} \\
\includegraphics[height=8cm,width=0.48\textwidth]{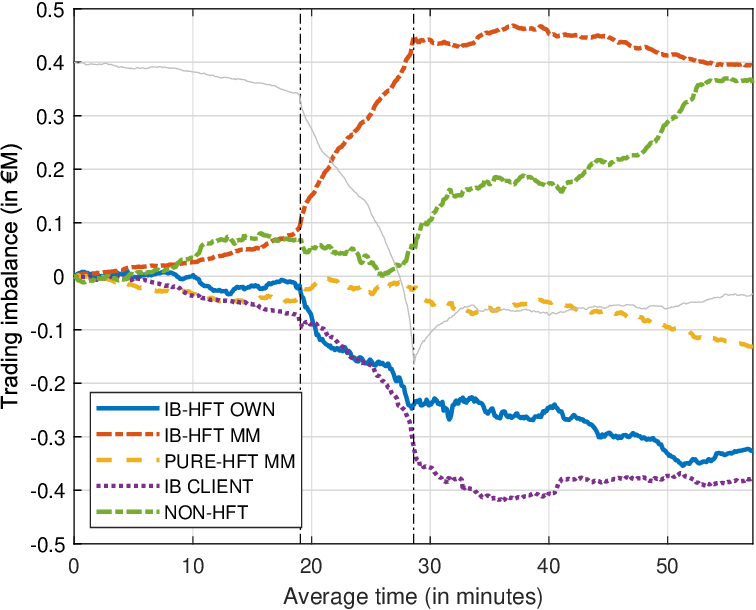} &
\includegraphics[height=8cm,width=0.48\textwidth]{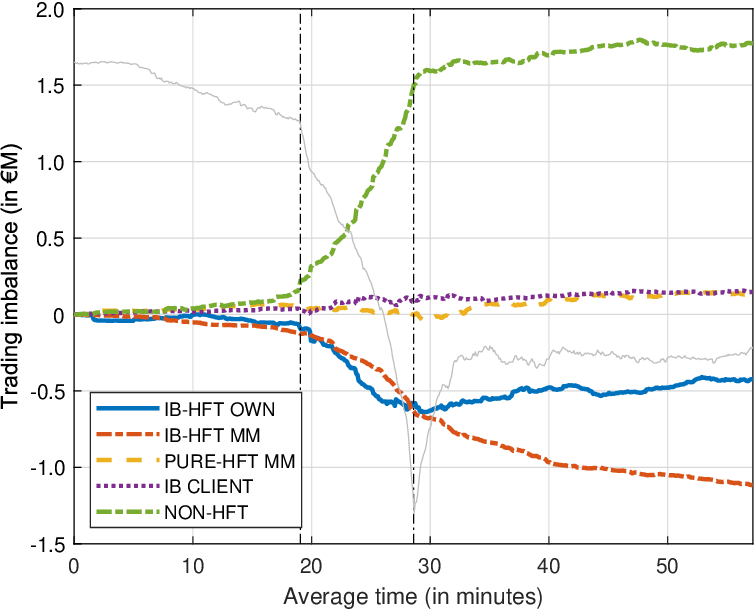}
\end{tabular}
\smallskip
\parbox{\textwidth}{\emph{Note.} The figure reports the trading imbalance of IB-HFT OWN, IB-HFT MM, PURE-HFT MM, IB-CLIENT, and NON-HFT (the sum of NON-HFT CLIENT and NON-HFT OWN) during a drift burst EPM. We average over the 121 unsystematic events in Panel A, while Panel B holds the results for the 27 systematic events. In the background of each figure, we showcase the average cumulative return dynamic from Figure \ref{figure:drift-burst-EPM} as a visual landmark.}
\end{center}
\end{figure}

Figure \ref{figure:trading-imbalance} illustrates the dynamic evolution of the average trading imbalance, computed as in equation \eqref{equation:imbalance}, during a drift burst EPM. Panel A (Panel B) reports the results for unsystematic (systematic) events. As a visual benchmark, we add the average cumulative return from Figure \ref{figure:drift-burst-EPM} as a grey-colored line. The figure shows how the selling activity of IB-HFT OWN (together with IB-CLIENT in the unsystematic setting) can be associated with the beginning of the downturn. This trading is informed, since it shifts the price toward a lower fundamental value. The graph further illuminates the difference in the behavior of PURE-HFT MM and IB-HFT MM.

\begin{figure}[p]
\begin{center}
\caption{Changes in trading imbalance for a systematic drift burst EPM}
\label{figure:delta-imbalance-s}
\begin{tabular}{ccc}
\small{Panel A: IB-HFT OWN} & \small{Panel B: IB-HFT MM} & \small{Panel C: PURE-HFT MM} \\
\includegraphics[height=4.5cm,width=0.31\textwidth]{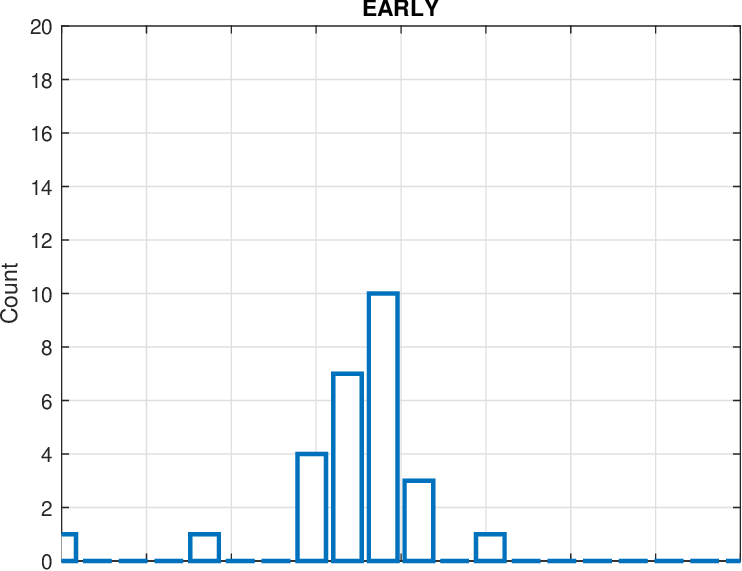} &
\includegraphics[height=4.5cm,width=0.31\textwidth]{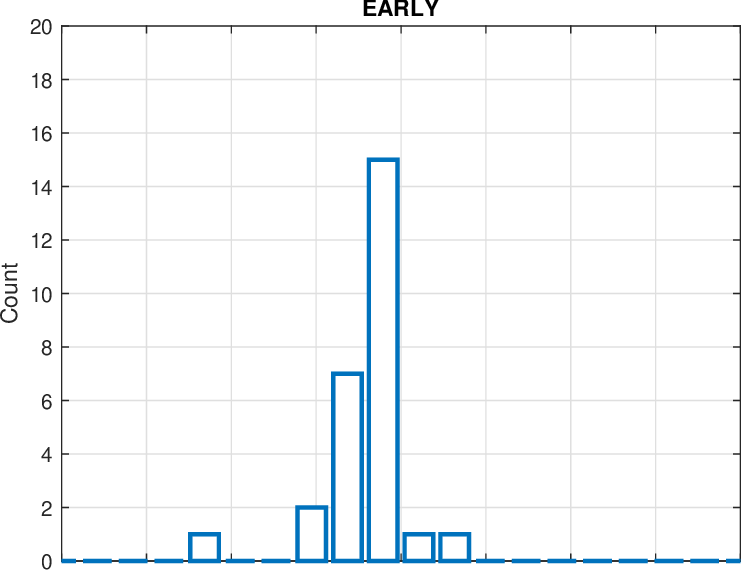} &
\includegraphics[height=4.5cm,width=0.31\textwidth]{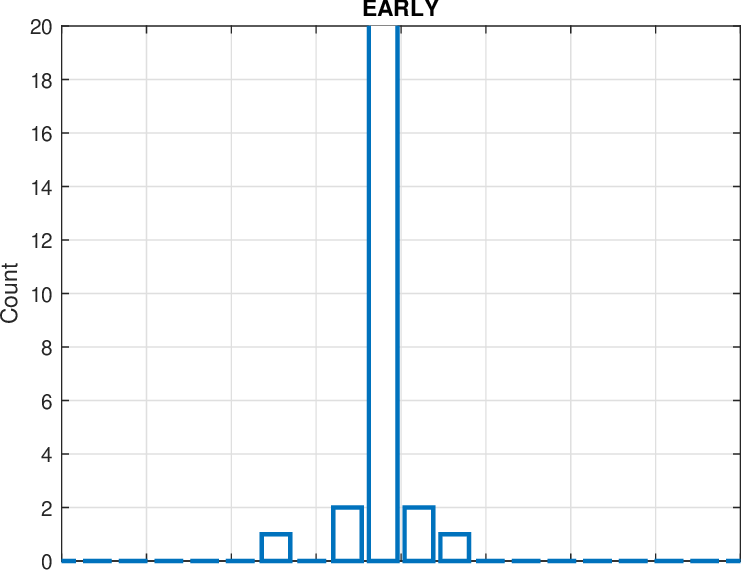} \\
\includegraphics[height=4.5cm,width=0.31\textwidth]{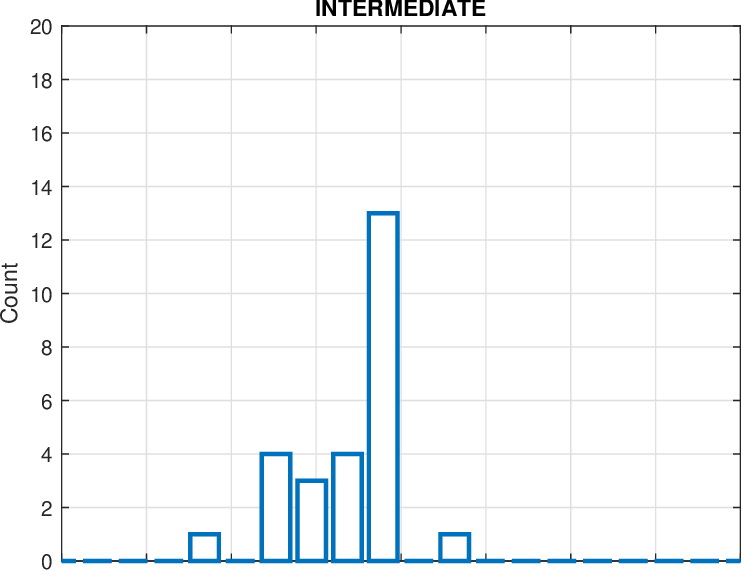} &
\includegraphics[height=4.5cm,width=0.31\textwidth]{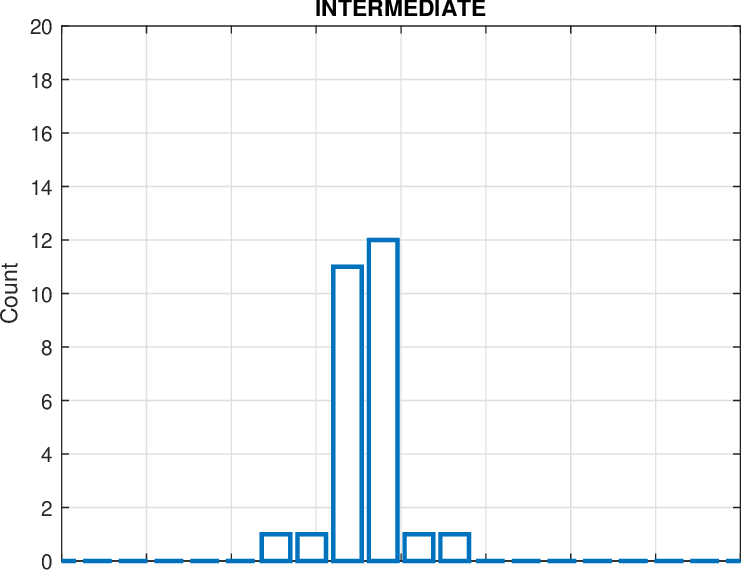} &
\includegraphics[height=4.5cm,width=0.31\textwidth]{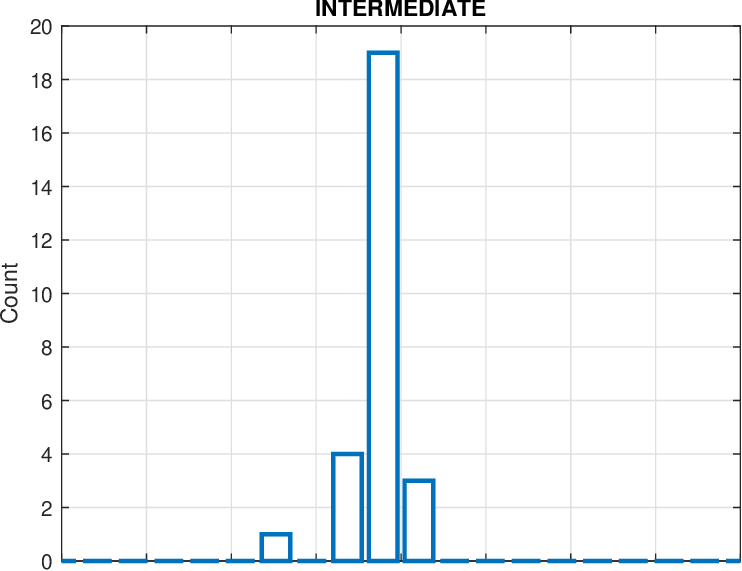} \\
\includegraphics[height=4.5cm,width=0.31\textwidth]{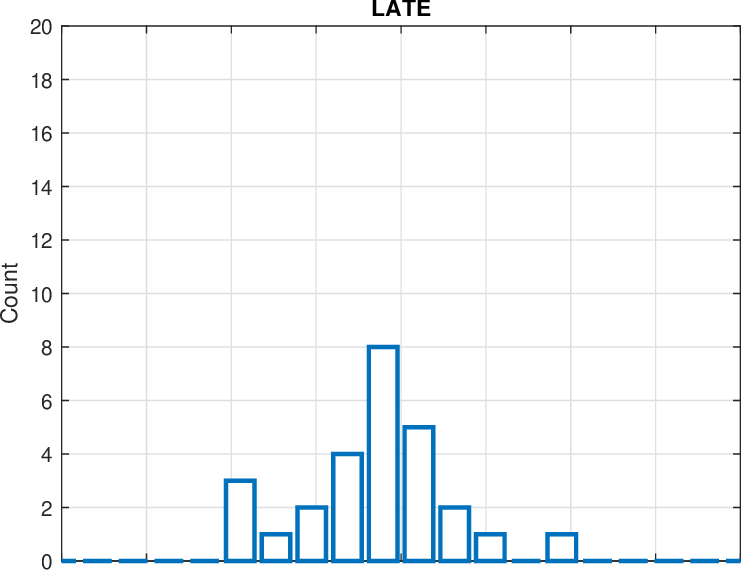} &
\includegraphics[height=4.5cm,width=0.31\textwidth]{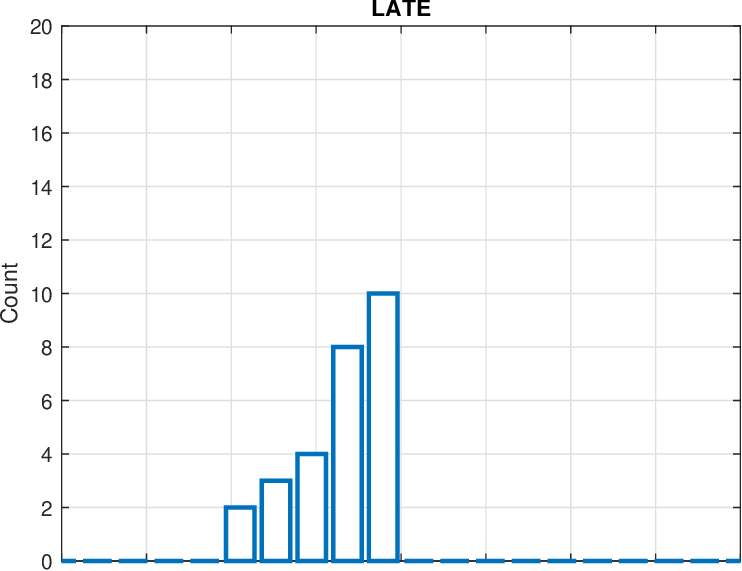} &
\includegraphics[height=4.5cm,width=0.31\textwidth]{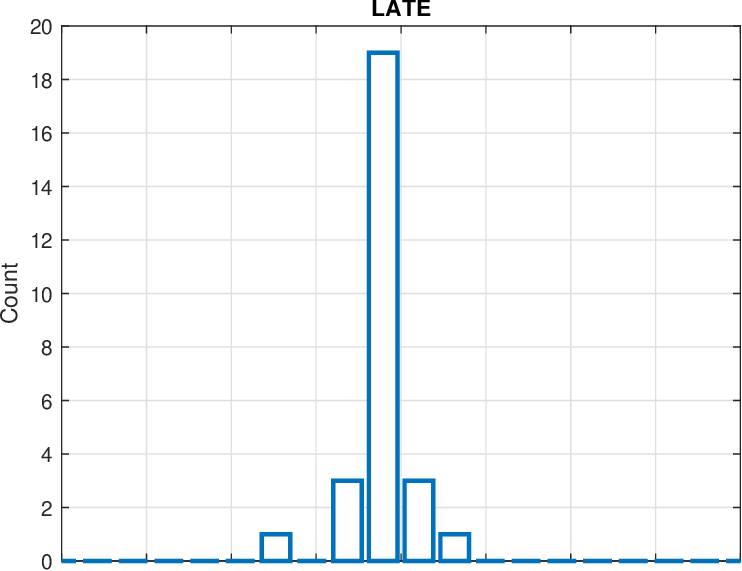} \\
\includegraphics[height=4.5cm,width=0.31\textwidth]{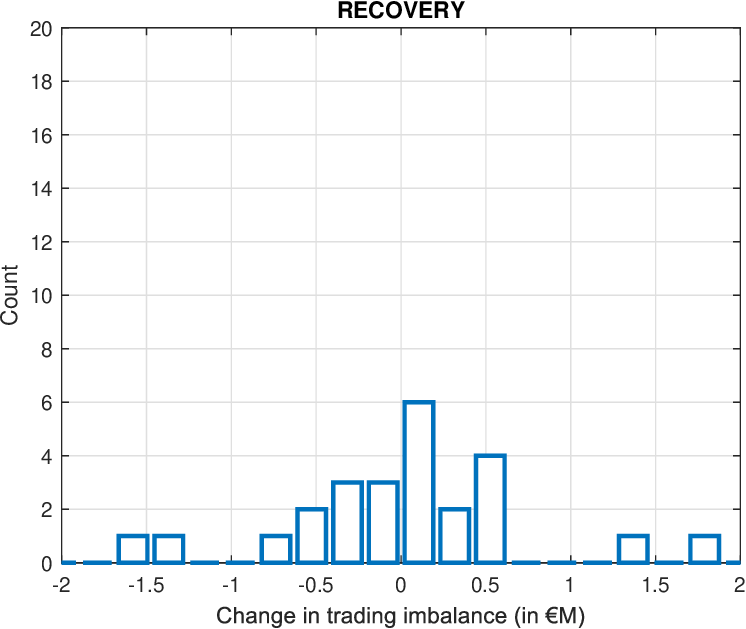} &
\includegraphics[height=4.5cm,width=0.31\textwidth]{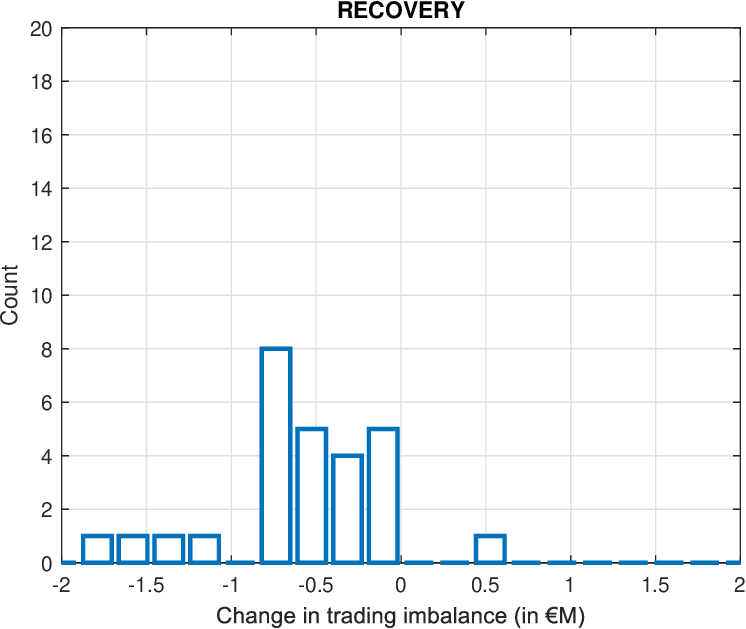} &
\includegraphics[height=4.5cm,width=0.31\textwidth]{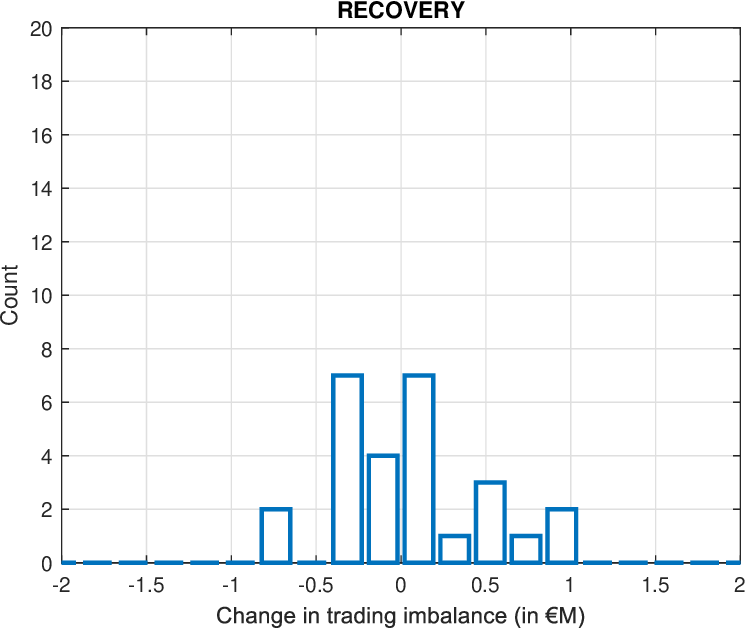} \\
\end{tabular}
\smallskip
\parbox{\textwidth}{\emph{Note.} The figure depicts a histogram of the empirical distribution of one-minute changes in trading imbalance for IB-HFT OWN (Panel A), IB-HFT MM (Panel B), and PURE-HFT MM (Panel C) at different stages of a systematic drift burst EPM. A negative number means that the trader category is collectively selling either via aggressive or passive trades, and vice versa.}
\end{center}
\end{figure}

In Figure \ref{figure:delta-imbalance-s}, we show a histogram of the one-minute changes in the trading imbalances of IB-HFT MM, PURE-HFT MM, and IB-HFT OWN during different stages of the drop and recovery for systematic drift burst EPMs.\footnote{Figure \ref{figure:delta-imbalance-u} in Appendix \ref{appendix:supplemental} reports the associated results for unsystematic drift burst EPMs.} The figure shows that IB-HFT OWN mostly sell at the beginning of the drop, whereas IB-HFT MM start selling immediately thereafter. In particular, IB-HFT MM are always selling in the late drop. In addition, the figure shows that PURE-HFT MM sell at the intermediate stage. These results provide further support for the notion that PURE-HFT MM tend to warehouse limited inventory, since their changes in trading imbalances are typically much smaller in magnitude and close to symmetric around zero.

To corroborate our analysis, we next dissect the trading imbalance into aggressive and passive trades (i.e., those trades where the trader group initiated (aggressive) or accepted (passive) the order). Thus, the imbalance of trades initiated ($\mathcal{D}$ = demand) and accepted ($\mathcal{S}$ = supply) by trader category $i$ over a period is given by:
\begin{equation} \label{equation:demand-supply-imbalance}
\mathcal{DI}^{i}_{ \textrm{period}} = \sum_{t \in { \textrm{period}}} s_t \cdot V_{t} \cdot I(t \textrm{ initiated by }i) \quad \text{and} \quad
\mathcal{SI}^{i}_{ \textrm{period}} = \sum_{t \in { \textrm{period}}} s_t \cdot V_{t} \cdot I(t \textrm{ accepted by }i),
\end{equation}
where $I( \cdot)$ is the indicator function.\footnote{Note that $\mathcal{TI}^{i}_{ \textrm{period}} = \mathcal{DI}^{i}_{ \textrm{period}} + \mathcal{SI}^{i}_{ \textrm{period}}$.}

Table \ref{table:var-a-p} provides selected coefficient estimates of the VAR model from equation \eqref{equation:var}, where we replace trading imbalance as the dependent variable with the aggressive (left-hand side of the table) and passive (right-hand side of the table) trading imbalances.\footnote{The full version of the table with all trader groups is reported in Tables \ref{table:var-full-a} -- \ref{table:var-full-p} in Appendix \ref{appendix:supplemental}.} Looking at the left-hand side with aggressive trading, IB-HFT OWN shows negative and significant coefficients for both the early and intermediate drops for both unsystematic and systematic drift burst EPMs. That this activity flows in the direction of the price change follows from the positive and significant coefficient on Return. In a systematic event, this is followed significantly in the late stage of the drop by IB-HFT MM. By contrast, for the passive trades, which are in the opposite direction of the return, the estimated coefficients show that only for unsystematic drift burst EPMs do IB-HFT MM offer immediacy during all stages. For a systematic event, DMMs (both PURE-HFT MM and IB-HFT MM) do offer immediacy, but only to the buy side, since the sign in front of the coefficients of their passive trades is mostly negative (and significant for the late drop). Overall, these results confirm the graphical assessment from Figure \ref{figure:trading-imbalance}. This therefore indicates that DMMs, even if receiving compensation for providing liquidity, actually consume it when the drift burst EPMs are systematic.

\begin{table}[p]
\setlength{ \tabcolsep}{0.03cm}
\begin{center}
\caption{Coefficient estimates from the VAR for aggressive and passive trading imbalance}
\label{table:var-a-p}
\begin{footnotesize}
\begin{tabular}{lccccccc}
\hline
\multicolumn{8}{l}{ \textit{Panel A: Unsystematic drift burst EPMs}} \\
 & \multicolumn{3}{c}{Aggressive} &  & \multicolumn{3}{c}{Passive} \\
\cline{2-4} \cline{6-8}
 & PURE-HFT MM & IB-HFT MM & IB-HFT OWN &  & PURE-HFT MM & IB-HFT MM & IB-HFT OWN \\
Pre-event& -0.006 & 0.003 & -0.018{*} &  & -0.006 & 0.033{*}{*}{*} & 0.004 \\
 & (-0.74) & (0.32) & (-1.95) &  & (-0.64) & (2.67) & (0.50) \\
Early drop & 0.021 & -0.037 & -0.167{*}{*}{*} &  & 0.098{*}{*} & 0.346{*}{*}{*} & 0.018 \\
 & (0.54) & (-1.11) & (-2.91) &  & (2.53) & (4.74) & (0.55) \\
Intermediate drop & 0.011 & 0.037 & -0.058{*} &  & -0.058{*} & 0.187{*}{*}{*} & 0.015 \\
 & (0.29) & (1.12) & (-1.67) &  & (-1.93) & (3.92) & (0.44) \\
Late drop & 0.055 & 0.013 & -0.065 &  & -0.063 & 0.238{*}{*}{*} & 0.094 \\
 & (0.99) & (0.26) & (-0.90) &  & (-1.23) & (3.85) & (1.61) \\
Recovery & -0.016 & -0.011 & -0.036{*}{*} &  & -0.021{*}{*} & -0.005 & 0.014 \\
 & (-1.46) & (-0.78) & (-2.44) &  & (-2.60) & (-0.32) & (0.62) \\
Bid-ask spread & 0.001 & 0.006 & -0.006 &  & 0.001 & 0.003 & -0.004 \\
 & (0.16) & (1.25) & (-1.03) &  & (0.11) & (0.58) & (-0.49) \\
Log(euro volume) & -0.020{*}{*}{*} & -0.013 & -0.015{*} &  & -0.002 & 0.036{*}{*}{*} & 0.024{*}{*} \\
 & (-2.85) & (-1.48) & (-1.69) &  & (-0.28) & (3.97) & (2.13) \\
Return & 0.239{*}{*}{*} & 0.142{*}{*}{*} & 0.199{*}{*}{*} &  & -0.300{*}{*}{*} & -0.312{*}{*}{*} & -0.208{*}{*}{*} \\
 & (22.13) & (12.89) & (22.29) &  & (-29.38) & (-22.60) & (-23.90) \\
\\
$\bar{R}^{2}$ & 0.054 & 0.029 & 0.050 &  & 0.082 & 0.106 & 0.058 \\
\\
\multicolumn{8}{l}{ \textit{Panel B: Systematic drift burst EPMs}} \\
 & \multicolumn{3}{c}{Aggressive} &  & \multicolumn{3}{c}{Passive} \\
\cline{2-4} \cline{6-8}
 & PURE-HFT MM & IB-HFT MM & IB-HFT OWN &  & PURE-HFT MM & IB-HFT MM & IB-HFT OWN \\
Pre-event & 0.047{*}{*} & 0.040{*} & 0.022 &  & -0.033{*}{*} & -0.051 & -0.055{*}{*} \\
 & (2.35) & (1.89) & (1.45) &  & (-2.21) & (-1.67) & (-2.18) \\
Early drop & -0.083{*} & 0.074 & -0.188{*}{*} &  & 0.017 & -0.053 & 0.000 \\
 & (-1.86) & (0.77) & (-2.07) &  & (0.39) & (-1.12) & (0.00) \\
Intermediate drop & -0.064 & -0.050 & -0.355{*}{*}{*} &  & -0.101{*}{*} & -0.145 & -0.036 \\
 & (-1.27) & (-0.41) & (-3.74) &  & (-2.10) & (-1.64) & (-0.79) \\
Late drop & 0.180{*}{*} & -0.309{*}{*} & -0.074 &  & -0.315{*}{*}{*} & -0.284{*}{*} & -0.017 \\
 & (2.68) & (-2.18) & (-1.01) &  & (-4.05) & (-2.20) & (-0.22) \\
Recovery & 0.010 & -0.078{*}{*}{*} & -0.031{*}{*} &  & 0.001 & -0.003 & 0.081{*}{*}{*} \\
 & (0.36) & (-2.93) & (-2.37) &  & (0.13) & (-0.14) & (2.92) \\
Bid-ask spread & 0.018{*}{*} & 0.010 & 0.011 &  & -0.012 & -0.015 & -0.020{*}{*}{*} \\
 & (2.14) & (1.08) & (1.69) &  & (-1.34) & (-1.47) & (-3.58) \\
Log(euro volume) & -0.010 & -0.108{*}{*}{*} & -0.064{*}{*}{*} &  & -0.018 & -0.089{*}{*}{*} & 0.048{*}{*} \\
 & (-0.74) & (-6.12) & (-3.15) &  & (-1.66) & (-6.45) & (2.71) \\
Return & 0.248{*}{*}{*} & 0.178{*}{*}{*} & 0.117{*}{*}{*} &  & -0.266{*}{*}{*} & -0.227{*}{*}{*} & -0.135{*}{*}{*} \\
 & (10.28) & (10.69) & (11.41) &  & (-14.22) & (-10.37) & (-12.29) \\
\\
$\bar{R}^{2}$ & 0.087 & 0.085 & 0.041 &  & 0.095 & 0.091 & 0.056 \\
\hline
\end{tabular}
\end{footnotesize}
\\
\medskip
\parbox{\textwidth}{\emph{Note.} We present the results of the VAR in equation \eqref{equation:var}. The model is estimated separately for unsystematic drift burst EPMs in Panel A and systematic ones in Panel B. The dependent variable is the 10-second trading imbalance, split into aggressive and passive orders, of the trader category indicated by the column label. The explanatory variables ``Pre-event'', ``Early drop'', ``Intermediate drop'', ``Late drop'', and ``Recovery'' refer to phase dummies of a drift burst EPM. ``Bid-ask spread'' is the quoted spread (in percent), ``Log(euro volume)'' is the log-euro volume, and ``Return'' is the contemporaneous stock return. We also include stock fixed effects. The non-dummy variables are standardized at the stock level. The coefficient estimates of lagged trading imbalances are not reported for brevity. The $t$-statistics (in parentheses) are based on heteroscedasticity-robust standard errors clustered at the stock and date level. *$P$-value$<$0.1; **$P$-value$<$0.05; ***$P$-value$<$0.01. nObs = 342,032 in Panel A and nObs = 79,404 in Panel B.}
\end{center}
\end{table}

Is the behavior of DMMs during drift burst EPMs affected by their previous inventory? Is it affected by the inventory of other traders? To shed light on this question, we conduct a cross-sectional analysis of trading imbalances by estimating the following regression:
\begin{equation} \label{equation:cross}
\mathcal{TI}_{ \textrm{Late}, k}^{i} = \alpha_{0}^{i} + \beta^{i} \mathcal{TI}_{ \textrm{Intermediate+Early},k}^{i} + \sum_{j} \gamma^{ij} \mathcal{TI}_{ \textrm{Intermediate+Early},k}^{j} + \alpha^{i} \mathrm{Controls}_{k} + \epsilon_{k}^{i},
\end{equation}
where $i$ and $j$ denote trader categories, $\mathcal{TI}_{ \textrm{period},k}^{ \bullet}$ is the trading imbalance of group $\bullet$ for the $k$-th drift burst EPM in a given period, $\textrm{Controls}_{k}$ is a vector of control variables and $\epsilon^{i}_{k}$ is the error term.

The dependent variable in equation \eqref{equation:cross} is the trading imbalance in the late stage of the drift burst EPM for DMMs (i.e., $i = $ PURE-HFT MM or IB-HFT MM). We add its lagged value to capture serial correlation. We further include the lagged trading imbalance from IB-HFT OWN and IB-CLIENT (i.e., $j = $ IB-HFT OWN and IB-CLIENT). As in equation \eqref{equation:var}, the control variables are the contemporaneous return, log-euro volume, and the average percentage bid-ask spread, measured from the start of the drift burst EPM to the beginning of the late stage. Thus, all explanatory variables are lagged in time relative to the dependent variable. We are mostly interested in the sign of the $\gamma$ coefficients. A positive sign means that DMMs start selling in the late stage of the drift burst EPM, when other traders were previously selling. This can be interpreted as DMMs adapting their trading activity in the lean-with-the-wind direction based on observed order flow.

Table \ref{table:cross} shows the coefficient estimates of equation \eqref{equation:cross}. The results support the role of PURE-HFT MM and IB-HFT MM as DMMs for unsystematic drift burst EPMs. Here, the estimated $\gamma$ coefficients are negative, so capital is flowing from the DMMs towards competing traders. The exception is the liquidity from PURE-HFT MM to IB-CLIENT, which has a positive coefficient estimate, but it is insignificant. As expected, we observe a strong persistence in trading imbalances. Moreover, the positive effect of the bid-ask spread suggests that DMMs hold positive (or less negative) trading imbalances in more illiquid stocks, which aligns with the function of a market maker during downward price pressure in such stocks. When the drift burst EPM is systematic, the outcome is mostly unchanged for PURE-HFT MM, which is consistent with the evolution of their trading imbalances in Figure \ref{figure:trading-imbalance}. However, it is a completely different story for IB-HFT MM, who prey on the previous order flow from IB-HFT OWN and IB-CLIENT. This in turn leads to negative serial correlation in own trading imbalances. Our cross-sectional analysis thus corroborates our previous findings.

\begin{table}[p]
\setlength{ \tabcolsep}{0.80cm}
\begin{center}
\caption{Cross-sectional analysis of DMMs trading imbalance}
\label{table:cross}
\begin{tabular}{lcccc}
\hline
\multicolumn{5}{l}{ \textit{Panel A: Unsystematic drift burst EPMs}} \\
 & & PURE-HFT MM & & IB-HFT MM \\
\cline{3-3} \cline{5-5}
Intercept & & $-0.17$ & & $-0.24$ \\
 & & $(-0.33)$ & & $(-0.61)$ \\
$\mathcal{TI}_{ \textrm{Intermediate+Early}}^{i}$ & & $0.49$ & & $0.35^{*}$ \\
 & & $(1.56)$ & & $(1.86)$ \\
$\mathcal{TI}_{ \textrm{Intermediate+Early}}^{ \text{IB-HFT OWN}}$ & & $-0.04$ & & $-0.06^{*}$ \\
 & & $(-0.79)$ & & $(-1.83)$ \\
$\mathcal{TI}_{ \textrm{Intermediate+Early}}^{ \text{IB-CLIENT}}$ & & $0.09$ & & $-0.08$ \\
 & & $(1.44)$ & & $(-1.52)$ \\
Log(euro volume) & & $0.01$ & & $0.01$ \\
 & & $(0.24)$ & & $(0.78)$ \\
Bid-ask spread & & $0.74$ & & $2.65^{*}$ \\
 & & $(0.54)$ & & $(1.72)$ \\
Return & & $-0.06$ & & $0.07^{**}$ \\
 & & $(-1.55)$ & & $(2.12)$ \\
\\
$\bar{R}^{2}$ & & $0.134$ & & $0.197$ \\
\\
\multicolumn{5}{l}{ \textit{Panel B: Systematic drift burst EPMs}} \\
 & & PURE-HFT MM & & IB-HFT MM \\
\cline{3-3} \cline{5-5}
Intercept & & $0.95$ & & $4.27^{***}$ \\
 & & $(0.80)$ & & $(2.76)$ \\
$\mathcal{TI}_{ \textrm{Intermediate+Early}}^{i}$ & & $0.75^{***}$ & & $-0.64^{**}$ \\
 & & $(2.67)$ & & $(-2.04)$ \\
$\mathcal{TI}_{ \textrm{Intermediate+Early}}^{ \text{IB-HFT OWN}}$ & & $-0.23^{*}$ & & $0.20^{**}$ \\
 & & $(-1.87)$ & & $(2.12)$ \\
$\mathcal{TI}_{ \textrm{Intermediate+Early}}^{ \text{IB-CLIENT}}$ & & $-0.25^{***}$ & & $0.25^{**}$ \\
 & & $(-2.59)$ & & $(2.13)$ \\
Log(euro volume) & & $-0.05$ & & $-0.22^{***}$ \\
 & & $(-1.00)$ & & $(-3.13)$ \\
Bid-ask spread & & $1.75$ & & $-0.47$ \\
 & & $(0.82)$ & & $(-0.14)$ \\
Return & & $-0.13$ & & $-0.27$ \\
 & & $(-0.84)$ & & $(-1.41)$ \\
\\
$\bar{R}^{2}$ & & $0.539$ & & $0.384$ \\
\hline
\end{tabular}
\\
\medskip
\parbox{ \textwidth}{\emph{Note.} This table presents coefficient estimates of the cross-sectional regression in equation \eqref{equation:cross} over unsystematic drift burst EPMs in Panel A and systematic ones in Panel B. The dependent variable is the trading imbalance of the DMM in the late stage of the event, $\mathcal{TI}_{ \textrm{Late}}^{i}$ (where either $i =$ PURE-HFT MM or IB-HFT MM). The explanatory variables are calculated over the combined early and intermediate stage and include the lagged value of the DMMs trading imbalance, $\mathcal{TI}_{ \textrm{Intermediate+Early}}^{i}$, the trading imbalance of IB-HFT OWN and IB-CLIENT, $\mathcal{TI}_{ \textrm{Intermediate+Early}}^{ \text{IB-HFT OWN}}$ and $\mathcal{TI}_{ \textrm{Intermediate+Early}}^{ \text{IB-CLIENT}}$, the log-euro volume, the average bid-ask spread, and the return. The $t$-statistics (in parentheses) are based on heteroscedasticity-robust standard errors. $\bar{R}^{2}$ is the adjusted coefficient of determination. *$P$-value$<$0.1; **$P$-value$<$0.05; ***$P$-value$<$0.01. nObs = 121 in Panel A and nObs = 27 in Panel B.}
\end{center}
\end{table}

We next examine whether DMMs change their trading behavior during drift burst EPMs compared to normal times. To perform this investigation, we borrow the framework of \citet{kirilenko-kyle-samadi-tuzun:17a}, who employ an error correction model to changes in inventory of different trader groups. They show that HFTs did not alter their behavior during the Flash Crash of May 6, 2010. We apply their econometric model to the 148 drift burst EPMs in our sample. The results of the full analysis are presented in Appendix \ref{appendix:kirilenko}, but the main message is in line with our previous findings, namely that DMMs modify their trading activity during systematic events, but not during unsystematic ones.

\begin{table}[p]
\setlength{\tabcolsep}{0.60cm}
\begin{center}
\caption{Monetary profit during a drift burst EPM}
\label{table:money_earned}
\begin{tabular}{lrrr}
\hline
& Systematic & & Unsystematic\\
\cline{2-2} \cline{4-4}
PURE CLIENT & $24.36$ & & $11.07$ \\
 & \footnotesize ($25.94$) & & \footnotesize ($124.21$) \\
PURE-HFT MM & $-377.78$ & & $-164.09$ \\
 & \footnotesize ($541.03$) & & \footnotesize ($444.22$) \\
PURE-HFT OWN & $-60.17$ & & $-27.52$ \\
 & \footnotesize ($108.69$) & & \footnotesize ($312.32$) \\
IB-CLIENT & $-1024.37$ & & $241.62$ \\
 & \footnotesize ($889.47$) & & \footnotesize ($774.40$) \\
IB-HFT MM & $-75.61$ & & $-2799.72^{***}$ \\
 & \footnotesize ($959.95$) & & \footnotesize ($706.58$) \\
IB-HFT OWN & $5396.13^{**}$ & & $2239.13^{*}$ \\
 & \footnotesize ($2713.46$) & & \footnotesize ($1204.43$) \\
IB-HFT PARENT & $-208.34$ & & $-423.07^{*}$ \\
 & \footnotesize ($484.79$) & & \footnotesize ($238.88$) \\
NON-HFT CLIENT & $-3164.53$ & & $-438.11$ \\
 & \footnotesize ($2250.18$) & & \footnotesize ($996.79$) \\
NON-HFT OWN & $-765.22$ & & $1375.07^{**}$ \\
 & \footnotesize ($1457.17$) & & \footnotesize ($690.90$) \\
\hline
\end{tabular}
\\
\medskip
\parbox{\textwidth}{\emph{Note.} This table presents the average monetary profit (in euros) during downward drift burst EPMs, divided into systematic and unsystematic events. Standard errors of the average are in parentheses.}
\end{center}
\end{table}

What motivates the change of strategy of DMMs and their switch to selling during a systematic event? If the sudden demand for immediacy is due to private information (something we can establish, on average, ex-post), then it is rational for DMMs to sell, as postulated by the back-running theory of \citet{yang-zhu:20a} or the predatory trading theory of \citet{brunnermeier-pedersen:05a}. Support to this hypothesis is provided by Table \ref{table:money_earned}, which reports the average profit (in euros) during a drift burst EPM for each trader category, together with an estimate of the standard error. The table shows that IB-HFT MM loses a significant amount of money, on average 2799.72 euros, by offering immediacy during an unsystematic drift burst EPM. They transfer this money, as expected, to IB-HFT OWN, who trade on information by selling at a higher price. It can be argued that DMMs still profit on average, since providing liquidity in tranquil times should be enough to cover these losses. However, if they have the option to fulfill the requirements of the SLP program without facing large losses, it is perfectly rational for them not to provide liquidity during distress. That is what they seem to be doing, since Table \ref{table:money_earned} shows that IB-HFT MM avoid losses during systematic events, effectively passing the ``hot potato'' to slow traders.

To provide additional insight on this important question, we calculate the profit and loss (P\&L) gross of fees and rebates in each 10-second interval as the monetary gain or loss on (1) new trades struck in that window and (2) existing positions from the previous 10-second interval, assuming a zero initial inventory at the beginning of each event. The mark-to-market is based on the most recent transaction price available at the end of each interval. We then estimate a VAR reminiscent to equation \eqref{equation:var}:
\begin{equation} \label{equation:PnL}
\textrm{PnL}^{i}_{j,t} = \alpha_{0}^{i} + \text{FE}_{ \mathrm{stock}} + \beta^{i} D_{j,t}+ \sum_{g=1}^{G} \sum_{l=1}^{5} \gamma_{g,l} \textrm{PnL}^{g}_{i,t-l} + \alpha^{i} \textrm{Controls}_{j,t} + \epsilon^{i}_{j,t},
\end{equation}
where $\textrm{PnL}^{i}_{j,t}$ is the P\&L for trader $i$ in the $t$-th 10-second interval of the $j$-th event, $\text{FE}_{ \mathrm{stock}}$ are stock fixed effects as before, and $D_{j,t}$ is a dummy, which is equal to one for the entirety of the drift burst EPM (spanning from the beginning of the pre-event until the end of the recovery), and is zero otherwise. The control variables and lags are as in equation \eqref{equation:var}. We primarily look at the $\beta^{i}$ coefficient. A negative value suggests a trader is losing money during a drift burst EPM, and vice versa.

Table \ref{table:PnL} reports the output of the estimation. In line with our prior expectations, the table shows that IB-HFT MM consistently lose money during unsystematic drift burst EPMs. On the flip side, PURE-HFT MM actually make a smaller amount, although the latter is not statistically significant. By contrast, during a systematic event, both IB-HFT MM and PURE-HFT MM reap substantial profits, which is mainly at the expense of NON-HFT. The other groups earning during a systematic event are IB-HFT OWN and, as shown in Appendix \ref{appendix:supplemental}, IB-HFT PARENT, which arguably crystallize their private information. Interestingly, IB-CLIENT suffers on their behalf, albeit to a much smaller extent.

\begin{table}[ht!]
\setlength{ \tabcolsep}{0.10cm}
\begin{center}
\caption{Profit and loss of trading activity}
\label{table:PnL}
\begin{tabular}{lcccccc}
\hline
\multicolumn{6}{l}{ \textit{Panel A: Unsystematic drift burst EPMs}} \\
 & PURE-HFT MM & IB-HFT MM & IB-CLIENT & IB-HFT OWN & NON-HFT \\
\cline{2-6}
Dummy & 0.124 & -0.349{*}{*}{*} & 0.432{*}{*}{*} & 0.895{*}{*} & 0.388{*} \\
 & (0.97) & (-2.96) & (2.96) & (2.54) & (1.79) \\
Bid-ask spread & 0.488{*}{*} & 0.007 & 0.236 & 0.730 & -0.097 \\
 & (2.59) & (0.07) & (1.21) & (1.10) & (-0.43) \\
Log(euro volume) & 0.580{*}{*}{*} & -0.072 & 0.131 & 0.360{*}{*} & -0.502{*}{*}{*} \\
 & (5.84) & (-1.21) & (1.29) & (2.39) & (-3.33) \\
Return & 0.064 & -0.098 & -0.988{*} & -0.428 & -0.310 \\
 & (0.22) & (-0.41) & (-1.89) & (-0.67) & (-0.66) \\
\\
$\bar{R}^{2}$ & 0.012 & 0.013 & 0.028 & 0.012 & 0.011 \\
\\
\multicolumn{6}{l}{ \textit{Panel B: Systematic drift burst EPMs}} \\
 & PURE-HFT MM & IB-HFT MM & IB-CLIENT & IB-HFT OWN & NON-HFT \\
\cline{2-6}
Dummy & 0.929{*}{*} & 0.739{*}{*} & -0.387 & 1.517{*}{*}{*} & -1.351{*} \\
 & (2.08) & (2.22) & (-1.63) & (2.93) & (-2.00) \\
Bid-ask spread & 0.315{*}{*} & -0.028 & 0.519{*}{*}{*} & 0.760{*}{*}{*} & -0.482{*} \\
 & (2.22) & (-0.16) & (3.01) & (2.94) & (-1.91) \\
Log(euro volume) & 1.047{*}{*}{*} & 0.047 & 0.083 & -0.056 & -1.198{*}{*}{*} \\
 & (3.58) & (0.38) & (0.61) & (-0.26) & (-4.13) \\
Return & -0.066 & -2.056{*}{*}{*} & 0.102 & -0.829 & 1.782{*}{*}{*} \\
 & (-0.19) & (-6.97) & (0.43) & (-1.62) & (4.42) \\
\\
$\bar{R}^{2}$ & 0.019 & 0.086 & 0.006 & 0.015 & 0.023 \\
\hline
\end{tabular}
\\
\medskip
\parbox{\textwidth}{\emph{Note.} This table presents coefficient estimates from the VAR in equation \eqref{equation:PnL}. The model is estimated separately for unsystematic drift burst EPMs in Panel A and systematic ones in Panel B. The dependent variable is the 10-second profit and loss (P\&L) of the trader category indicated by the column label. ``Dummy'' equals one for the entire duration of the drift burst EPM (from the beginning of the pre-event until the end of the recovery), zero otherwise, ``Bid-ask spread'' is the quoted spread (in percent), ``Log(euro volume)'' is the log-euro volume, and ``Return'' is the contemporaneous stock return. We also include stock fixed effects. The non-dummy variables are standardized at the stock level. The coefficient estimates of the lagged P\&L are not reported for brevity. The $t$-statistics (in parentheses) are based on heteroscedasticity-robust standard errors clustered at the stock and date level. *$P$-value$<$0.1; **$P$-value$<$0.05; ***$P$-value$<$0.01. nObs = 342,032 in Panel A and nObs = 79,404 in Panel B.}
\end{center}
\end{table}

Note that the compensation from the market making agreement does not offer much of an economic incentive for DMMs in and of itself. We estimate the average revenue earned by supplying liquidity during a drift burst EPM in our sample to be about 30 euro and 25 euro for PURE-HFT MM and IB-HFT MM.\footnote{We estimate these numbers by multiplying for each drift burst EPM the liquidity supplied by a DMM with the relative compensation, which is taken to be 0.20 bps per euro volume (in 2013, this changed from 0.20 to 0.22 and back again to 0.20), and then averaging across events.} Indeed, even if the DMMs had rested passively on the bid side in \textit{all} transactions during the detected drift burst EPMs, the average revenue is a meager 160 euro, which pales in comparison to the potential losses. This estimate does not include the standard gain from market making activity (i.e., the bid-ask spread), which is absorbed directly into the P\&L measure in Table \ref{table:PnL}.

Nevertheless, DMMs  still make profits, even if they suffer an occasional loss during a systematic drift burst EPM. We calculate the 2013 cumulative profit for DMMs and find that, not surprisingly, the DMM business is indeed quite valuable. Their gross profits are around 19 million euros at the end of the year, which is split into 13.5 million euros for PURE-HFT MM and 5.5 million euros for IB-HFT MM. In net terms (including fees), these yearly amounts reduce to 11.6 million euros and 4.9 million euros, respectively.\footnote{Since we do not observe the activity of individual DMMs, we have to rely on the aggregate profit proxy of all DMMs as a group under the assumption that every DMM fulfils their obligations for the calculation of fees (i.e., they all pay the lowest fee and receive the highest rebate). In the contractual agreements with SLP and DMM, the fees for DMMs are 0.30 bps for taking activity (aggressive trades) and a rebate (i.e., they receive a compensation) for liquidity provision (passive trades), which was 0.20 bps until May 2013, increased up to 0.22 bps in June 2013, and reverted back to 0.20 as of November 2013. We compute for all stocks of the CAC 40 Index the daily gross profit and loss, which is calculated as the total amount (in euro) bought minus the total amount sold. We assume that a trader starts the day with zero inventories, and end-of-day inventories are mark-to-market at closing price. Net daily profit and loss includes also the fees for the stocks traded at the auction, which is 0.6 bps for all market participants.}

In essence, the compensation from their liquidity provision during normal times appears to be large enough to offset such losses. However, if DMMs suspect the order flow is coming from informed trading and they can otherwise meet the requirements of their contractual agreement with the exchange without penalties, it is of course perfectly rational for them to withdraw from the market or even switch to a predatory lean-with-the-wind strategy, choosing to pass the obligation of market making (and the losses) onto slow traders. However, even if this behavior is rational, it is suboptimal, because it impairs market efficiency, since the combined selling pressure temporarily dislocates the market price below its fundamental value.

Summarizing, our results point out the limited willingness of DMMs to provide liquidity during a drift burst EPM depends on its cross-sectional extent. They provide liquidity during unsystematic events, but during systematic events PURE-HFT MM and, in particular, IB-HFT MM start selling in the late phase. Thus, their role as DMMs loses its effectiveness exactly when the market needs it the most. The role of liquidity providers is instead assumed by more traditional slow traders (i.e., NON-HFTs). In conclusion, from a policy perspective, our results trigger a question of the effectiveness of the compensation scheme offered by the exchange.

\subsection{An analysis of extreme selling pressure} \label{section:counterfactual}

In the previous subsection, we examined the behavior of DMMs during the occurrence of high selling pressure \textit{and} drift burst EPMs. In this subsection, we study the behavior of DMMs during high selling pressure. To this end, we identify in our sample non-drift burst time intervals with as much---or even more extreme---selling pressure than what we observe during a drift burst EPM. We contrast the behavior of DMMs with other trader categories during such episodes to examine whether the drift burst EPMs are more likely a result of the selling pressure itself \textit{or} a heterogeneous reaction of DMMs (and other traders) to it.

We compute an intuitive measure of selling pressure, which for each day, stock, and one-minute interval in our sample is defined as follows:
\begin{equation} \label{equation:sp}
\mathcal{SP}^{(day, stock)}_{m} = \gamma^{(stock)}_{m} \times \min_{u = 0,1, \dots, 30} \left( \sum_{t \in [m-u,m]} s_{t} \cdot V_{t} \right),
\end{equation}
where $m$ runs over a one-minute equispaced partition of the trading day, $\gamma^{(stock)}_{m}$ is an intraday periodicity factor, while $s_{t}$ and $V_{t}$ are as before. That is, for each time instant $m$, we compute the signed euro volume over the previous $u$ minutes of trading. Since large sell-side activity may sustain for different amounts of time, we do the calculation for intervals ranging from one to thirty minutes, matching the build-up from pre-event to trough of a typical drift burst EPM. To get a single number for each time point, we take the minimum value, which is usually negative and reflects the cumulative selling pressure. We correct $\mathcal{SP}^{(day,stock)}_{m}$ for intraday periodicity in trading activity by multiplying it by a factor $\gamma^{(stock)}_{m}$ that is computed as the reciprocal of the stock-specific average value of $\mathcal{SP}^{(day,stock)}_{m}$ for a fixed interval $m$, where the average is taken over all days in the sample for each stock. This ensures that the final number is comparable over time and across stocks.

\begin{figure}[t!]
\begin{center}
\caption{The distribution of selling pressure}
\label{figure:selling-pressure}
\begin{tabular}{c}
\includegraphics[height=10cm,width=0.9\textwidth]{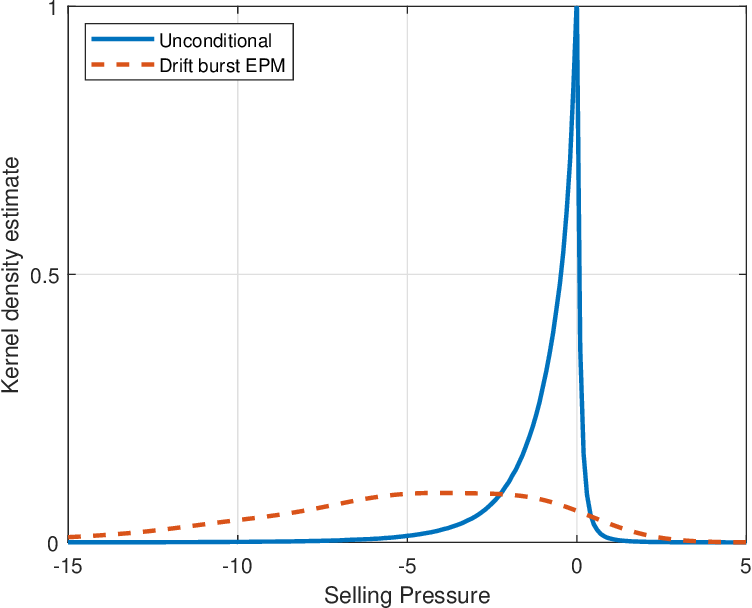}
\end{tabular}
\smallskip
\parbox{\textwidth}{\emph{Note.} We show the unconditional distribution for the measure of selling pressure, $\mathcal{SP}^{(day,stock)}_{m}$, pooled over each day, stock, and one-minute interval in our sample. We also plot the distribution of selling pressure during a drift burst EPM.}
\end{center}
\end{figure}

Figure \ref{figure:selling-pressure} shows the unconditional distribution of $\mathcal{SP}^{(day,stock)}_{m}$ over the entire sample. As expected, we observe that selling pressure is strongly left-skewed. Moreover, the distribution of $\mathcal{SP}^{(day,stock)}_{m}$ conditional on a drift burst EPM is centered at the tail of the unconditional distribution. This shows that the most extreme values of $\mathcal{SP}^{(day,stock)}_{m}$ correspond to candidate times with potential drift burst EPMs.

\begin{figure}[t!]
\begin{center}
\caption{Change in trading imbalance during extreme selling pressure}
\label{figure:sp}
\begin{center}
\end{center}
\smallskip
\begin{tabular}{ccc}
\small{Panel A: None} & \small{Panel B: Unsystematic} & \small{Panel C: Systematic} \\
\includegraphics[height=4cm,width=0.32\textwidth]{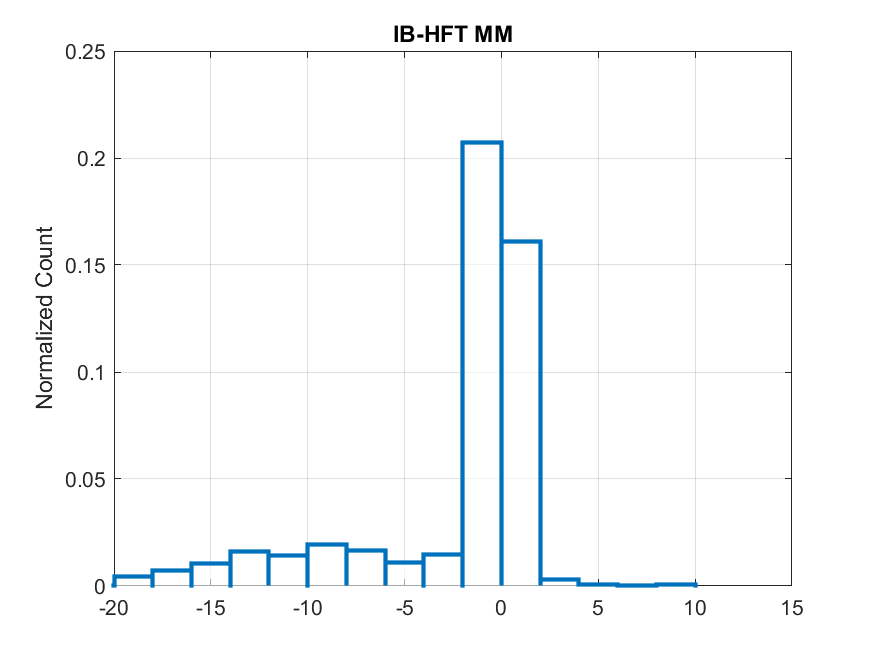} &
\includegraphics[height=4cm,width=0.32\textwidth]{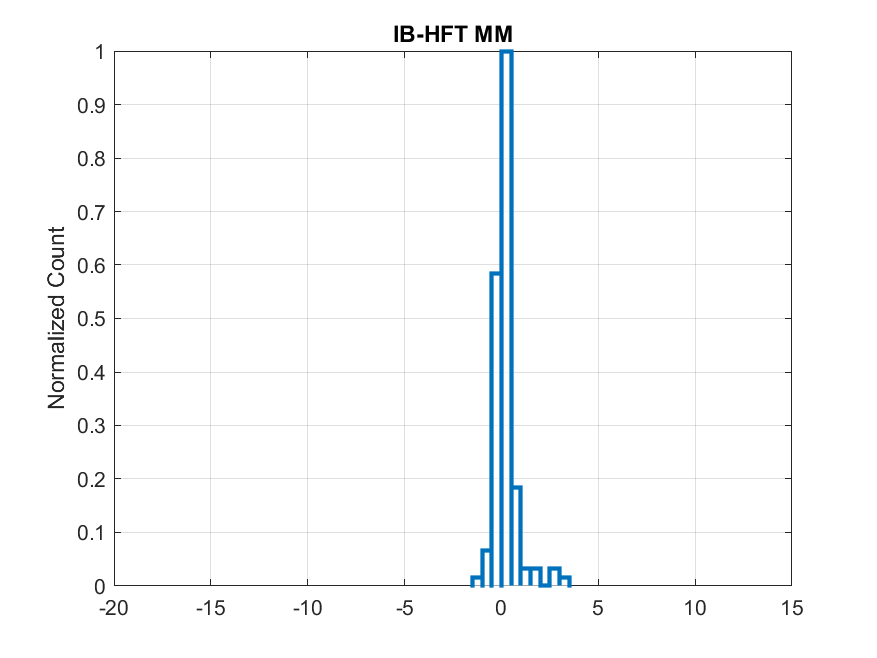} &
\includegraphics[height=4cm,width=0.32\textwidth]{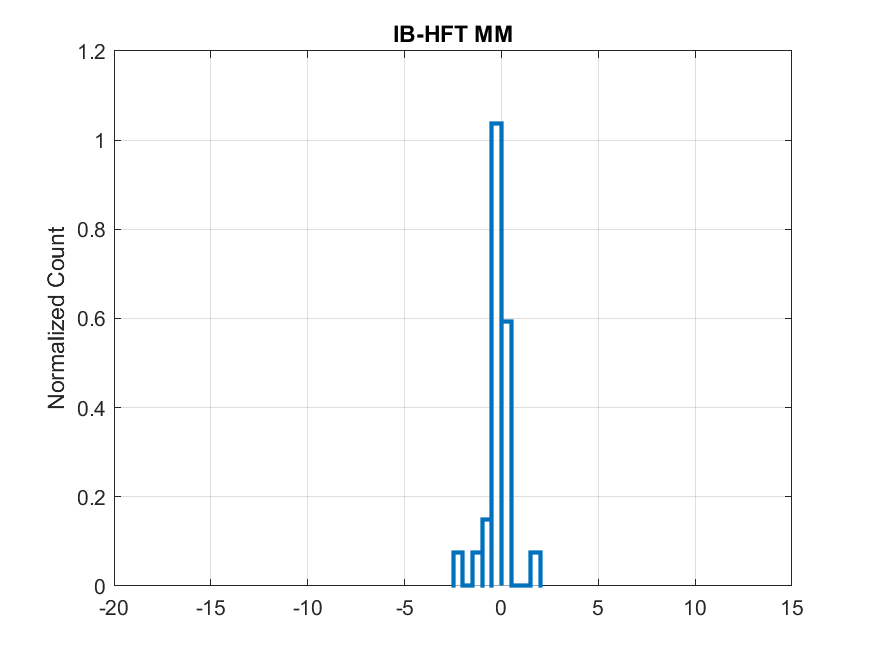} \\
\includegraphics[height=4cm,width=0.32\textwidth]{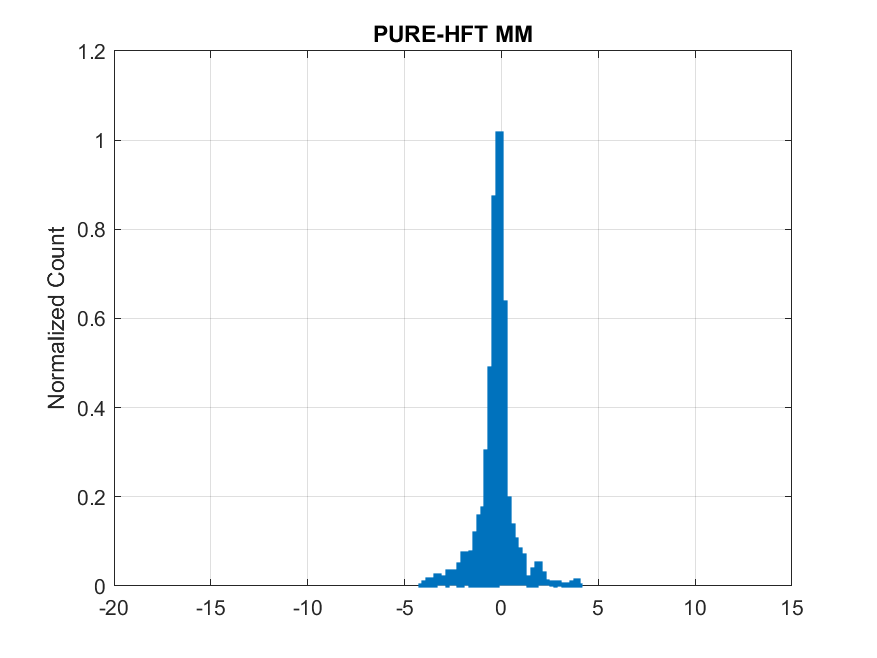} &
\includegraphics[height=4cm,width=0.32\textwidth]{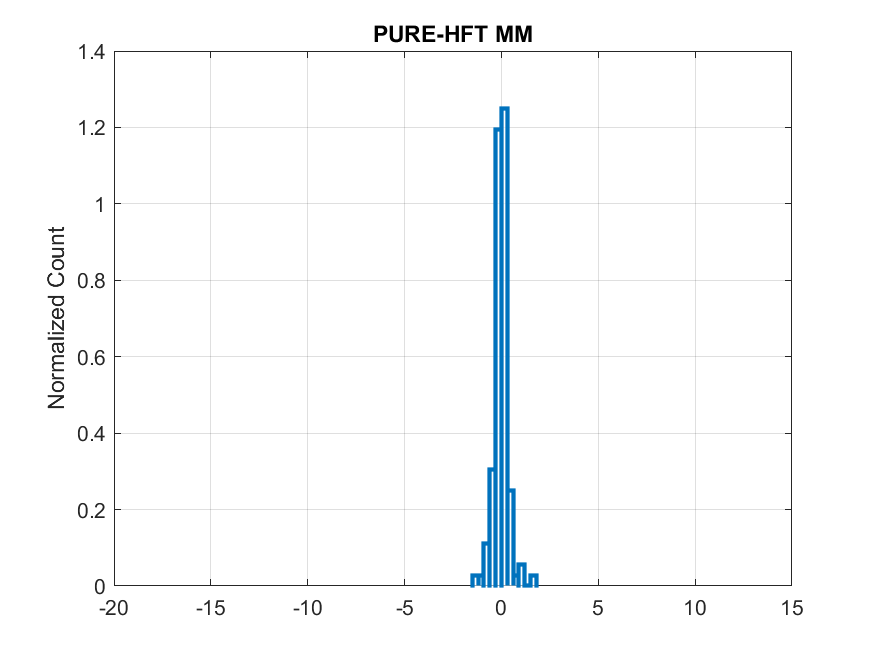} &
\includegraphics[height=4cm,width=0.32\textwidth]{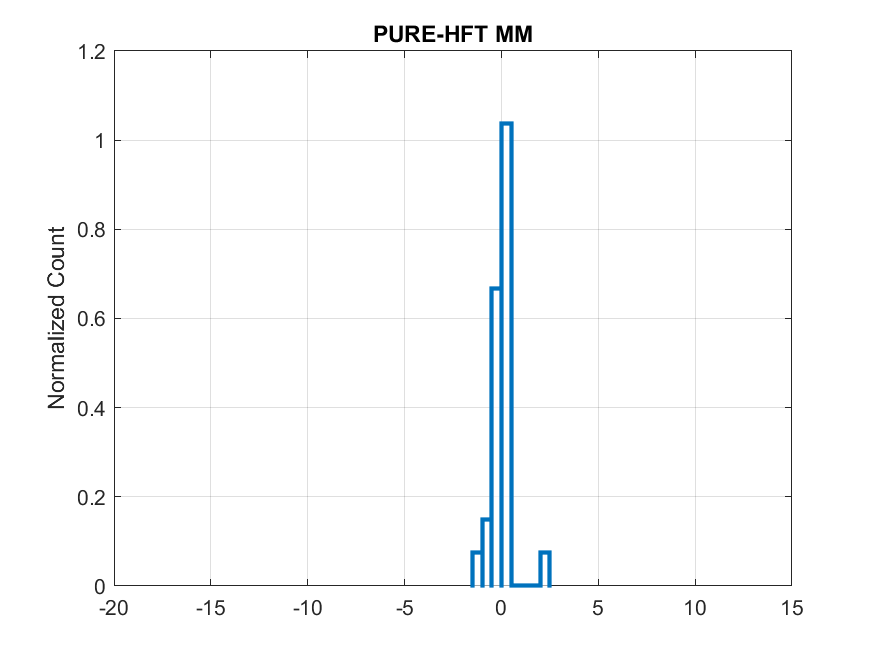} \\
\includegraphics[height=4cm,width=0.32\textwidth]{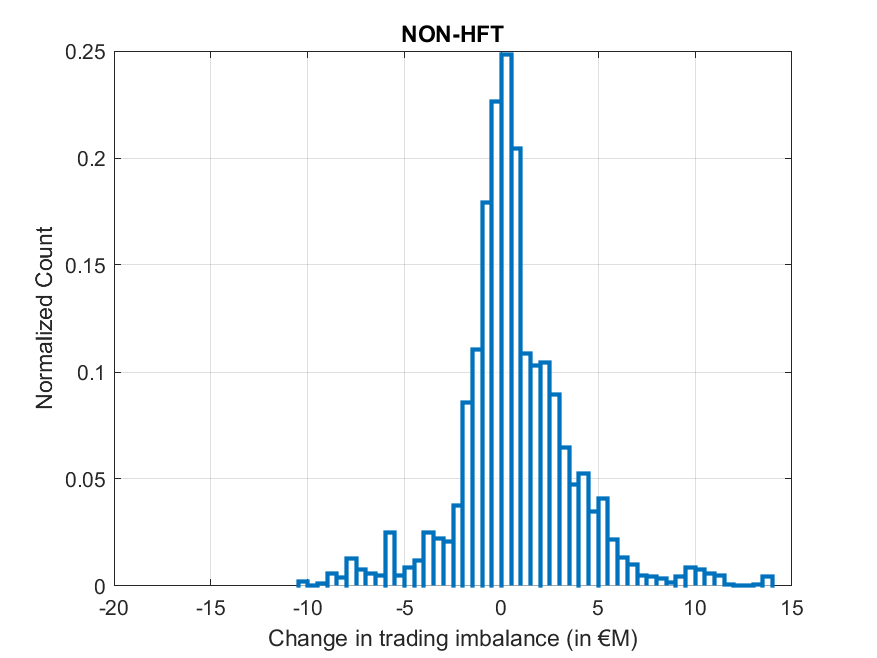} &
\includegraphics[height=4cm,width=0.32\textwidth]{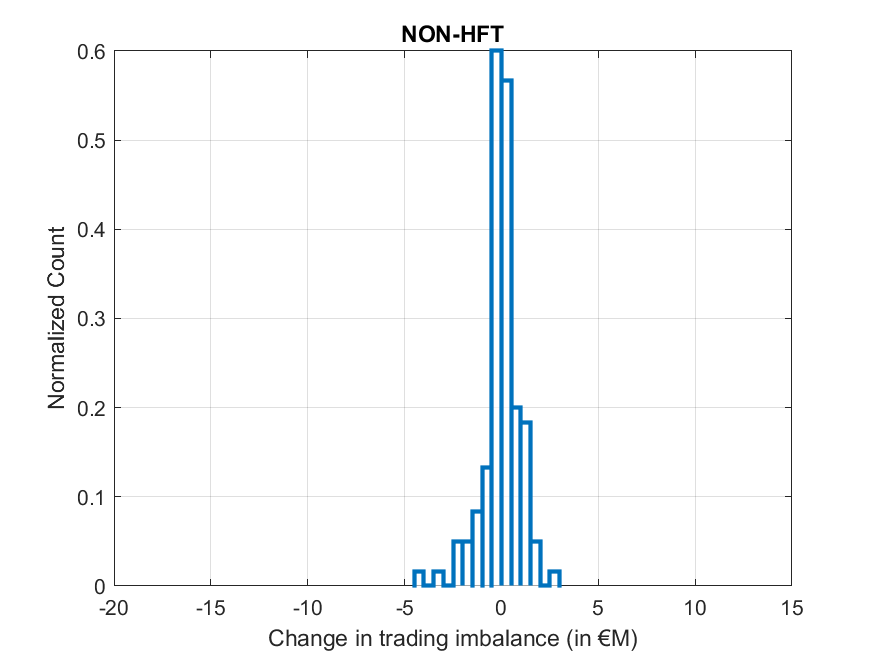} &
\includegraphics[height=4cm,width=0.32\textwidth]{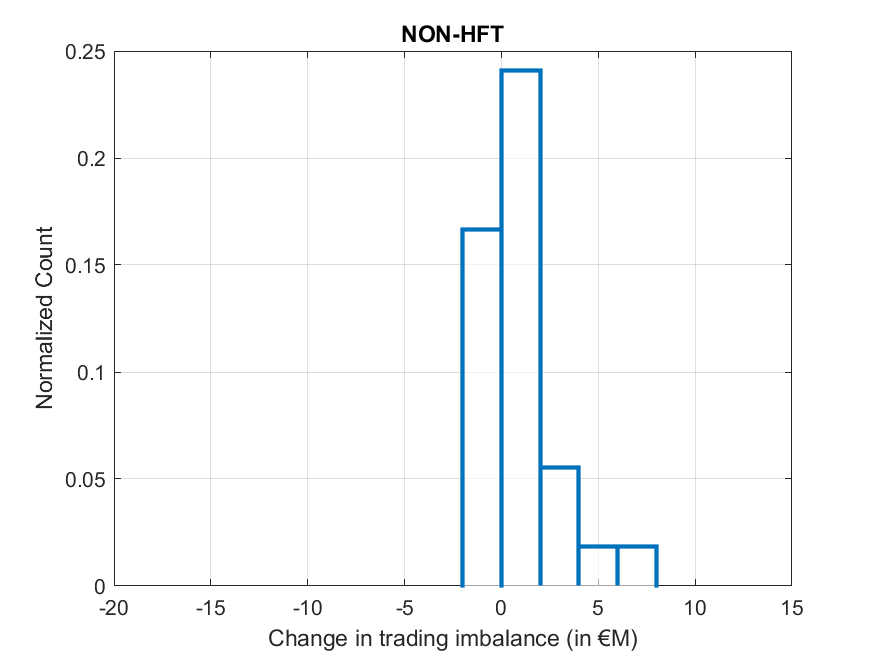} \\
\end{tabular}
\smallskip
\parbox{\textwidth}{\emph{Note.} We plot the distribution of the change in trading imbalances of IB-HFT MM, PURE-HFT MM, and NON-HFT during extreme selling pressure. We extract the $0.1\%$ of time points with the most selling pressure from the unconditional distribution of the $\mathcal{SP}^{(day,stock)}_{m}$ measure defined in \eqref{equation:sp}. We split these observations into three mutually exclusive sets, depending on whether they are associated with no drift burst EPM (Panel A), an unsystematic one (Panel B), or a systematic one (Panel C).}
\end{center}
\end{figure}

We extract the $0.1\%$ of time points with the most extreme selling pressure from the unconditional distribution of $\mathcal{SP}^{(day,stock)}_{m}$. In each of these time points, we compute the change in the trading imbalance for each trader group over the time window $[m-u^{*},m]$, where $u^{*}$ corresponds to the argmax of the minimum value selected in equation \eqref{equation:sp}. In Panel A of Figure \ref{figure:sp}, we show a histogram of the trading imbalance for IB-HFT MM, PURE-HFT MM, and NON-HFT during extreme selling pressure not identified as drift burst EPMs. As a comparison, we show the corresponding results for unsystematic and systematic drift burst EPMs in Panel B and Panel C, respectively. This figure can be compared with the one-minute changes in the trading imbalance of IB-HFT MM, PURE-HFT MM, and IB-HFT OWN during different stages of the drop and recovery for systematic drift burst EPMs in Figure \ref{figure:delta-imbalance-s}.

It is evident that a crucial difference between the occurrence or non-occurrence of a drift burst EPM lies in the actions of NON-HFT. As shown in Panel A, NON-HFTs support the market with a lot more capital during extreme selling pressure that does not overlap with a drift burst EPM. Furthermore, even here part of the sell-side activity emanates from DMMs (the distribution of changes in trading imbalances is heavily skewed to the left for IB-HFT MM and---to a lesser extent---for PURE-HFT MM). So again, we observe that DMMs do not always offer immediacy. The instances in which a drift burst EPM occurs (Panel B and Panel C), neither NON-HFTs nor DMMs bring enough money to the table, although NON-HFTs remain the primary source of liquidity. These findings indicate that slow traders are preventing a larger price drop when selling pressure is extreme. However, in their absence (either when they leave the order book or their quoted depth is small), a drift burst EPM occurs. This result reinforces the view that DMMs are not incentivized enough from the compensation scheme offered by the exchange to fully support the market with liquidity when it is demanded the most.

\section{Conclusion} \label{section:conclusion}

This paper shows that DMMs do not always fulfil their role as liquidity providers, despite the fact that this is their mandate. They even consume liquidity when it is needed the most. We apply the novel econometric technique of \citet{christensen-oomen-reno:22a}---the drift burst hypothesis---to detect 148 downward drift burst EPMs, in which market prices are being revised sharply lower over short time horizons, in the year 2013 of blue-chip companies, whose shares are traded in the French equity market. A unique feature of our database is that it allows us to separate the trading activity coming from different trader groups.

The results show that DMMs offer immediacy during isolated drift burst EPMs, as expected, but they reverse course and start devouring the limit order book at the late stage of events that affect multiple stocks, thus shifting to a lean-with-the-wind strategy. This behavior is different from what DMMs tend do in normal times with a balanced order flow, but we argue it is rational. Indeed, when a drift burst EPM affects several stocks, DMMs transition from  buyer to seller as the event is unfolding, probably to avoid big losses in fear that trading is informed, which it is on average. We thus document DMMs respond in disagreement with the intention of their contractual market making obligation, which impairs the efficient functioning of the market in times of need. Indeed, the collective impact of trading activity leads to a nontrivial, but possibly avoidable, overshooting effect. In place of DMMs, the main trader category that helps to stop the price from dropping in the wake of the significant selling pressure, and also aiding with the recovery, is NON-HFT, who arguably invests from a fundamental valuation perspective and looking to buy at a discount.

Electronic traders are widely known to offer immediacy with limited net inventory hovering around zero. Hence, the risk-bearing capacity of DMMs, which are strictly electronic in the Paris branch of the NYSE Euronext market, may also be exhausted too rapidly during systematic market downturns, which can further help to explain their sudden change of behavior. In view of this, our empirical results also bear an important policy implication in this regard and can be informative for a redesign of the market architecture, or an entirely different structure \citep[see, e.g.,][]{budish-cramton-smith:15a}, such that DMMs are presented with the right and sufficient incentives to maintain their presence in the market, despite potential mark-to-market losses, during widespread stress.

We recommend an evaluation of the recent MiFID II regulation, which recognizes algorithmic liquidity provision as a pivotal instrument to the functioning of financial markets. It endorses automatic liquidity provision by electronic market makers, imposing binding agreements between the exchange and the trading firms. We show that this rule, already in place at the NYSE Euronext Paris stock exchange, is not sufficient to prevent drift burst EPMs. This suggests that the rule should be revised in light of its objective. A possible solution to this problem is to change the compensation scheme. Another option is to change the mechanics of trading halts, which do not work as intended during drift burst EPMs, since the latter never triggered a pause in trading during our sample and is thus not associated with ``excess volatility.''

\pagebreak

\bibliographystyle{agsm}
\bibliography{userref}

\pagebreak

\processdelayedfloats
\csname efloat@restorefloats\endcsname

\appendix

\renewcommand{\theequation}{\thesection.\arabic{equation}}
\setcounter{equation}{0}

\section*{Appendices}

\renewcommand\thefigure{\thesection.\arabic{figure}}
\renewcommand\thetable{\thesection.\arabic{figure}}

\setcounter{page}{1}
\pagenumbering{roman}
\numberwithin{table}{section}

\section{A comparison with volatility-based EPMs} \label{appendix:epm}

\setcounter{figure}{0}
\setcounter{table}{0}

\citet{brogaard-carrion-moyaert-riordan-shkilko-sokolov:18a} identify EPMs from a sample of returns calculated over 10-second intervals. First, they extract the subset of such returns belonging to the 99.9th percentile of the 10-second absolute return distribution. Second, the calculation is repeated on residuals from a return autoregression that controls for ten lagged 10-second returns, to capture serial correlation induced by microstructure noise, and also incorporates the ten lagged returns for the market index, to correct for systematic risk. Loosely speaking, these techniques are designed to extract high-volatility returns. Third, they employ the jump test statistic of \citet{lee-mykland:08a}.

In this appendix, we implement these approaches and compare the outcome with the drift burst EPMs. We denote the first and second approach as ``return EPMs'' and ``residual EPMs.'' We exclude the jump test, because---by construction---the drift burst statistic is resistant to the price jump component.

In Table \ref{appendix:EPMs}, we compare the negative half of identified return- and residual-based EPMs with the downward drift burst EPMs. On average, for each stock, both volatility- and drift burst-based detection label 0.05\% of the 10-second intervals as EPMs, which is a mechanical effect of the confidence level. However, there is only a modest overlap between the identified EPMs. The return EPMs intersect with 26.35\% of drift burst EPMs, while the residual EPMs detect a lesser 18.92\%. We conclude that the overlap, while not negligible, is nevertheless limited. Hence, our sample of drift burst EPMs is to a large extent different from \citet{brogaard-carrion-moyaert-riordan-shkilko-sokolov:18a}.

\begin{table}[ht!]
\setlength{ \tabcolsep}{0.3cm}
\begin{center}
\caption{Comparison of volatility- and drift burst-based EPMs} \label{appendix:EPMs}
\medskip
\begin{tabular}{lrrrrr}
\hline
 & \multicolumn{2}{c}{EPMs per stock} && \multicolumn{2}{c}{Overlap to drift burst EPMs} \\
\hline
& Absolute (\#) & Fraction (\%) && Absolute (\#) & Fraction (\%)\\
\cline{2-3} \cline{5-6}
Return EPMs & 286.48 & 0.05 && 39 & 26.35 \\
Residual EPMs & 282.19 & 0.05 && 28 & 18.92 \\
\hline
\end{tabular}
\\
\medskip
\parbox{\textwidth}{\emph{Note.} The table reports the average number of EPMs detected for a single stock in absolute value and as a fraction (in per cent) of the total number of $10$-second returns in our sample, which is composed of 37 stocks traded on NYSE Euronext Paris that belong to the CAC 40 Index for the year 2013. It also shows the number of drift burst EPMs that are also detected by the return- and residual-based EPM calculation of \citet{brogaard-carrion-moyaert-riordan-shkilko-sokolov:18a}.}
\end{center}
\end{table}

\begin{figure}[ht!]
\begin{center}
\caption{Average cumulative return during volatility- and drift burst-based EPM} \label{figure:volatility-EPM}
\includegraphics[height=10cm,width=0.9\textwidth]{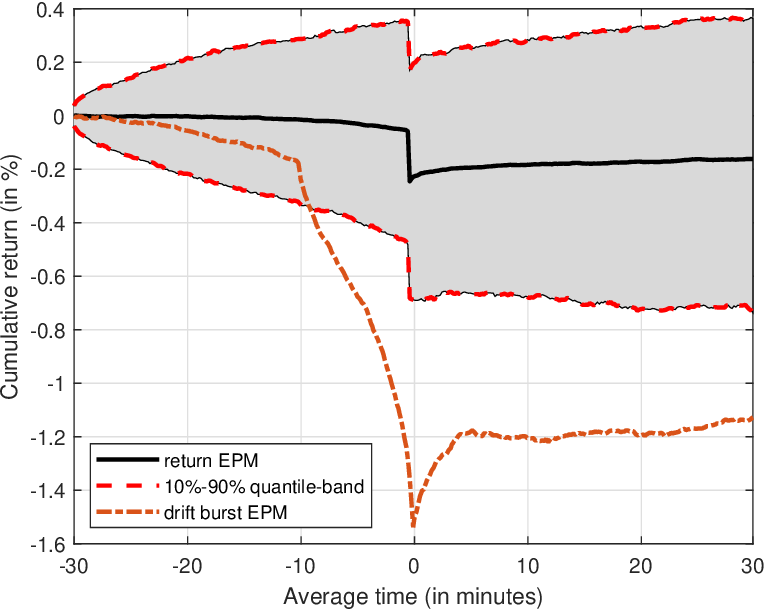}
\parbox{\textwidth}{\emph{Note.} This figure reports the average price evolution in a one-hour window centered around a volatility-based EPM. We follow \citet{brogaard-carrion-moyaert-riordan-shkilko-sokolov:18a} and label each $10$-second return at or above the 99.9th percentile of distribution of absolute 10-second returns as an EPM. The figure is based on the negative half of that sample. We superimpose a 10\%-90\% quantile-band for each 10-second interval over the interval. As a comparison, we also show the average cumulative return of a drift burst-based EPM, separately for unsystematic and systematic events.}
\end{center}
\end{figure}

Figure \ref{figure:volatility-EPM} replicates Figure \ref{figure:drift-burst-EPM} by showing the price evolution in a one-hour interval centered around the identified negative return EPMs. To get a better impression of the dynamic, we restrict attention to isolated observations that are not surrounded by any other return EPMs inside the window. As a comparison, we show the return dynamic for an average drift burst EPM. Evidently, return EPMs tend to extract volatility-based returns that---while being sampled from the left tail of the 10-second return distribution---do not otherwise lead to persistent changes in the fundamental value of the stock before or after the initial decline. By contrast, Figure \ref{figure:volatility-EPM} strongly supports the conjecture that drift burst EPMs are a different sample that allows to extract a sequence of consecutive returns around which the market was highly directional.

\clearpage

\section{Supplementary material} \label{appendix:supplemental}

\setcounter{figure}{0}
\setcounter{table}{0}

This appendix presents information about the sample of drift burst EPMs. Moreover, it contains supplemental material relative to the exposition in the main text.

\begin{figure}[ht!]
\begin{center}
\caption{Individual stock price changes during systematic drift burst EPMs}
\label{figure:systematic}
\begin{tabular}{cc}
\small{Panel A: April 17, 2013.} & \small{Panel B: September 3, 2013.} \\
\includegraphics[height=8cm,width=0.48\textwidth]{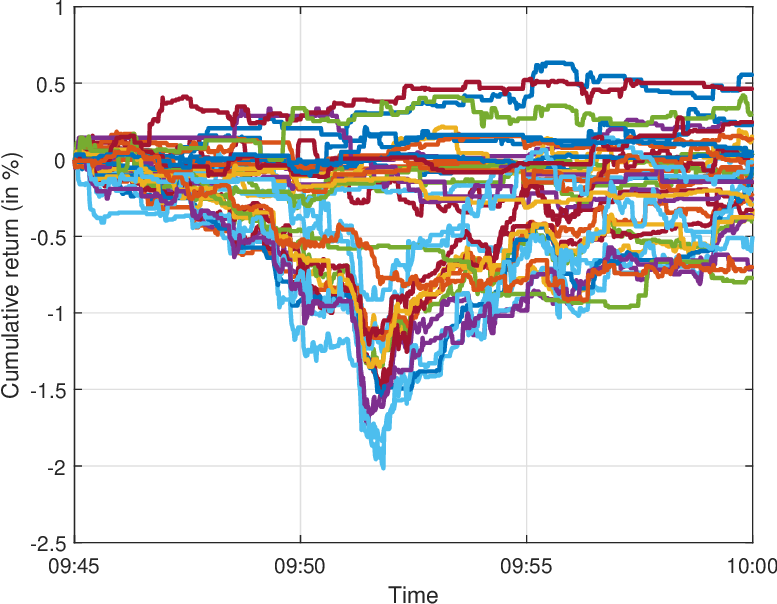} &
\includegraphics[height=8cm,width=0.48\textwidth]{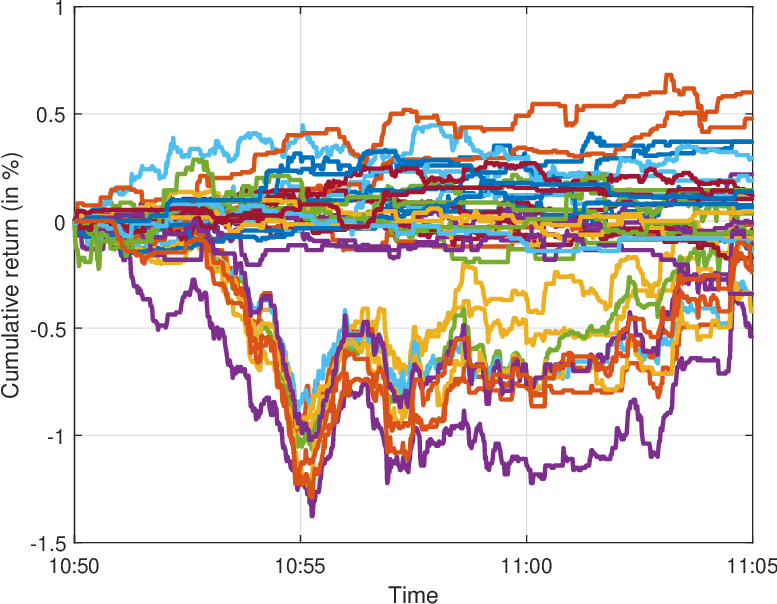} \\
\end{tabular}
\parbox{\textwidth}{\emph{Note.} This figure shows the price evolution of the equities included in our analysis during the systematic drift bursts EPMs detected on April 17, 2013 (in Panel A) and September 3, 2013 (in Panel B). We include a 15-minute window spanning the trough.}
\end{center}
\end{figure}

\begin{figure}[ht!]
\begin{center}
\caption{Changes in trading imbalance for an unsystematic drift burst EPM}
\label{figure:delta-imbalance-u}
\begin{tabular}{ccc}
\small{Panel A: IB-HFT OWN} & \small{Panel B: IB-HFT MM} & \small{Panel C: PURE-HFT MM} \\
\includegraphics[height=4.5cm,width=0.31\textwidth]{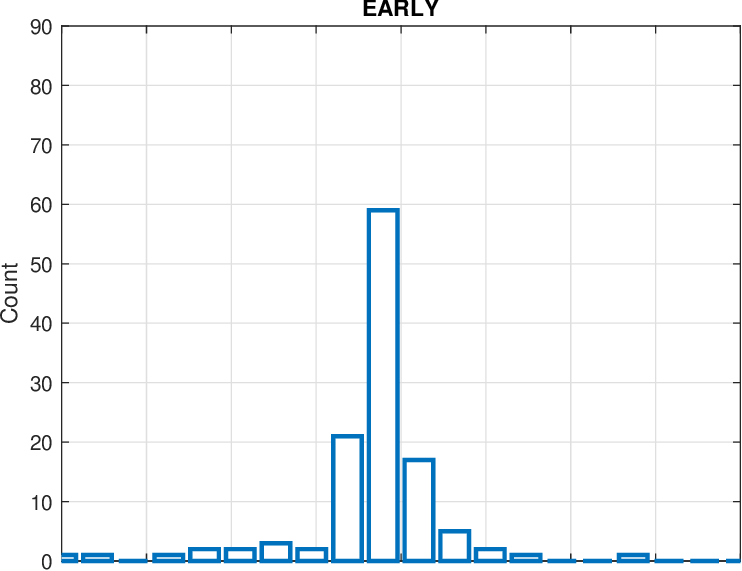} &
\includegraphics[height=4.5cm,width=0.31\textwidth]{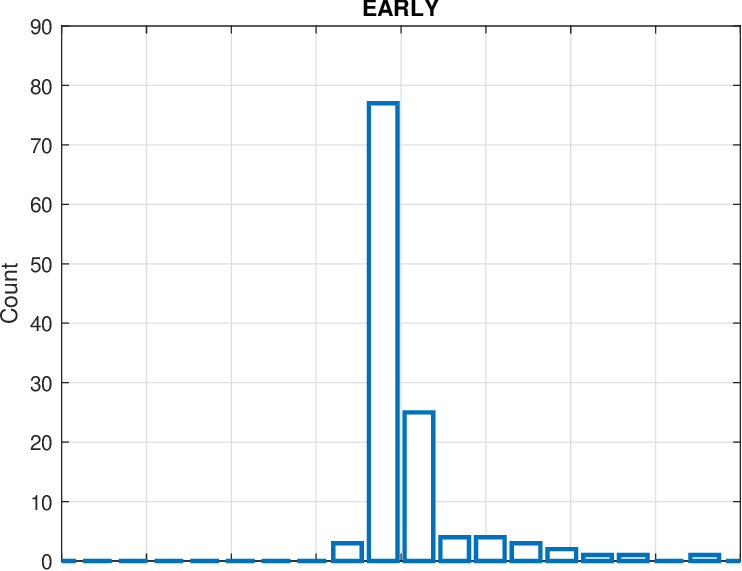} &
\includegraphics[height=4.5cm,width=0.31\textwidth]{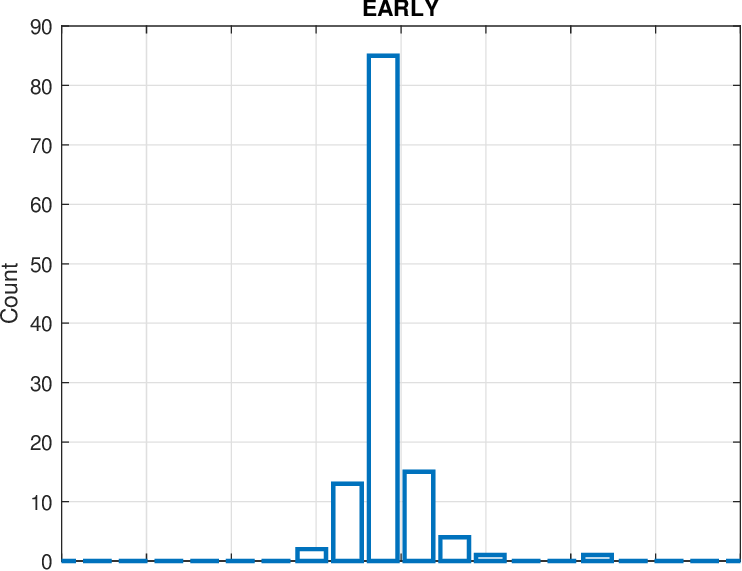} \\
\includegraphics[height=4.5cm,width=0.31\textwidth]{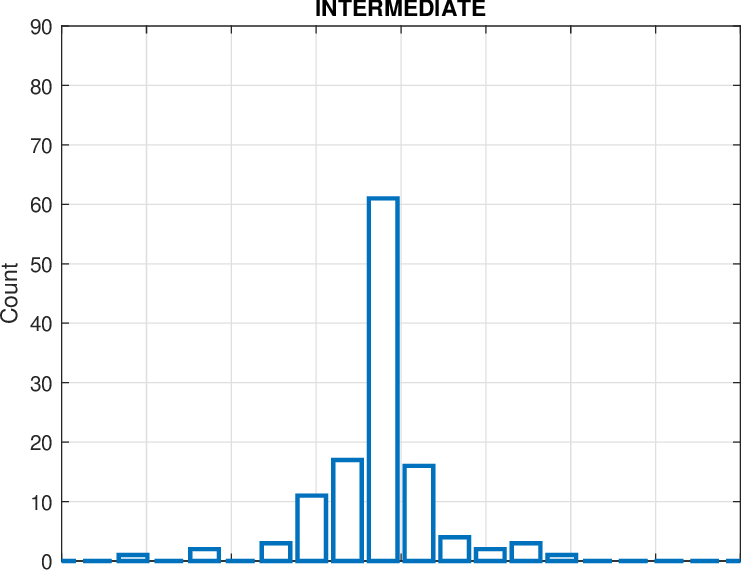} &
\includegraphics[height=4.5cm,width=0.31\textwidth]{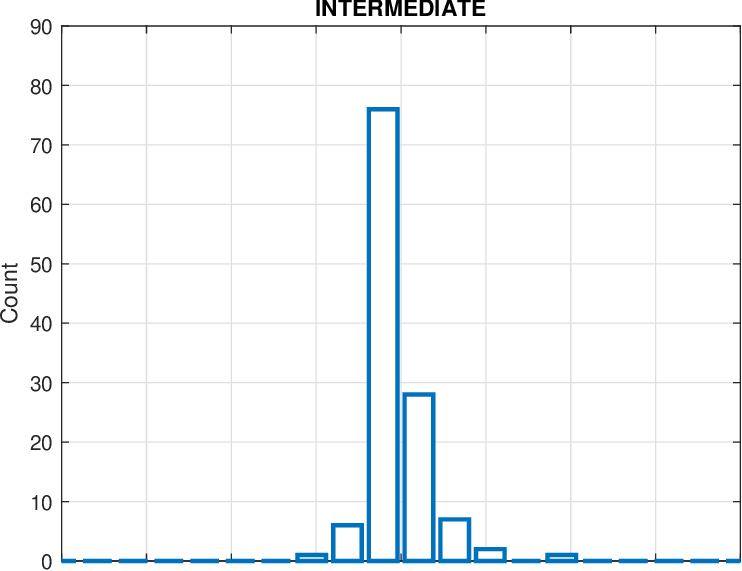} &
\includegraphics[height=4.5cm,width=0.31\textwidth]{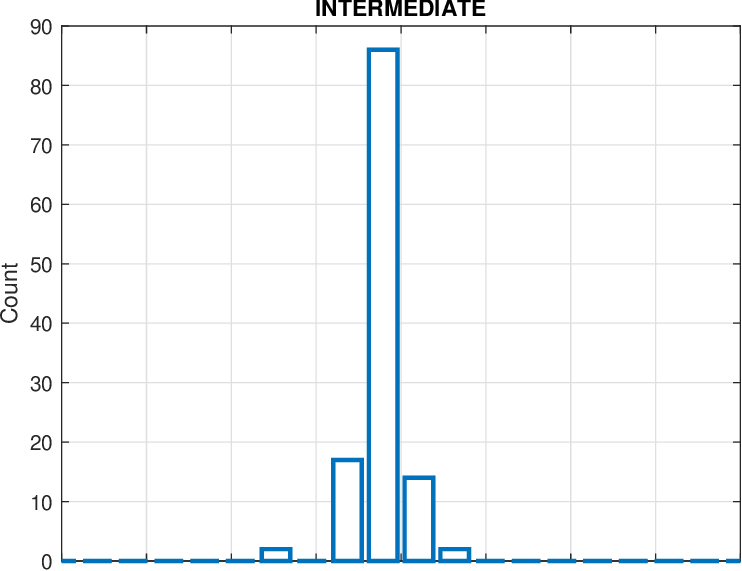} \\
\includegraphics[height=4.5cm,width=0.31\textwidth]{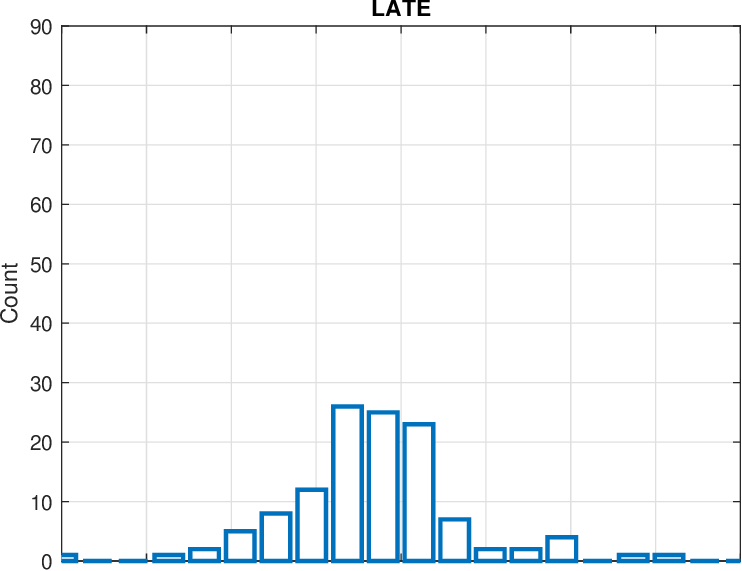} &
\includegraphics[height=4.5cm,width=0.31\textwidth]{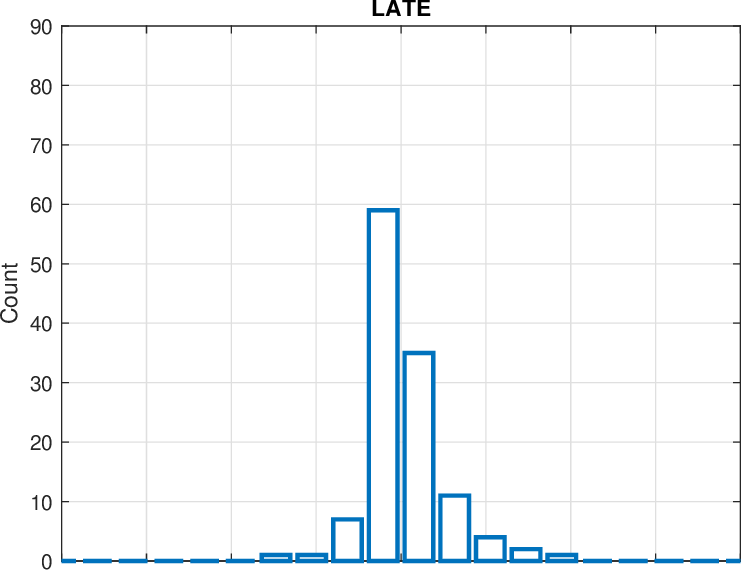} &
\includegraphics[height=4.5cm,width=0.31\textwidth]{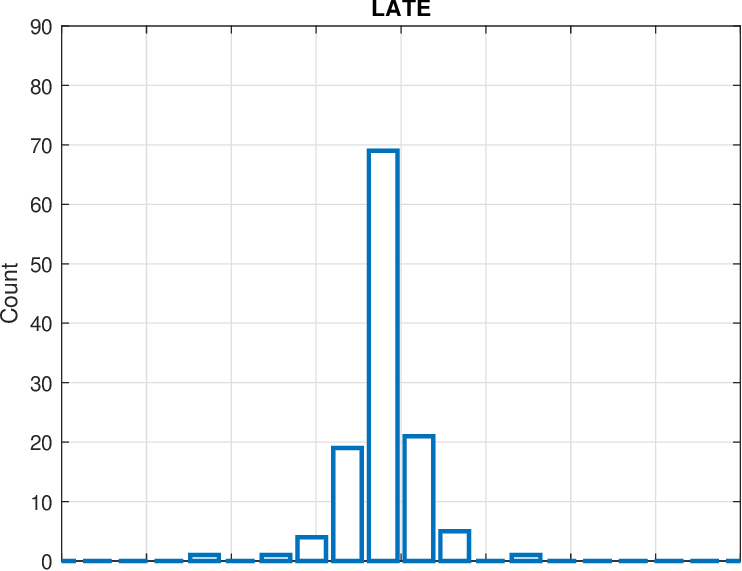} \\
\includegraphics[height=4.5cm,width=0.31\textwidth]{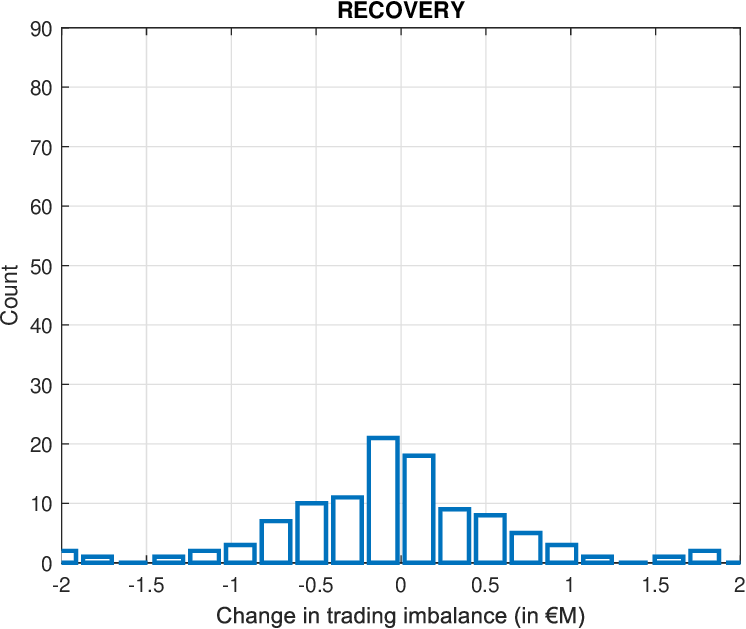} &
\includegraphics[height=4.5cm,width=0.31\textwidth]{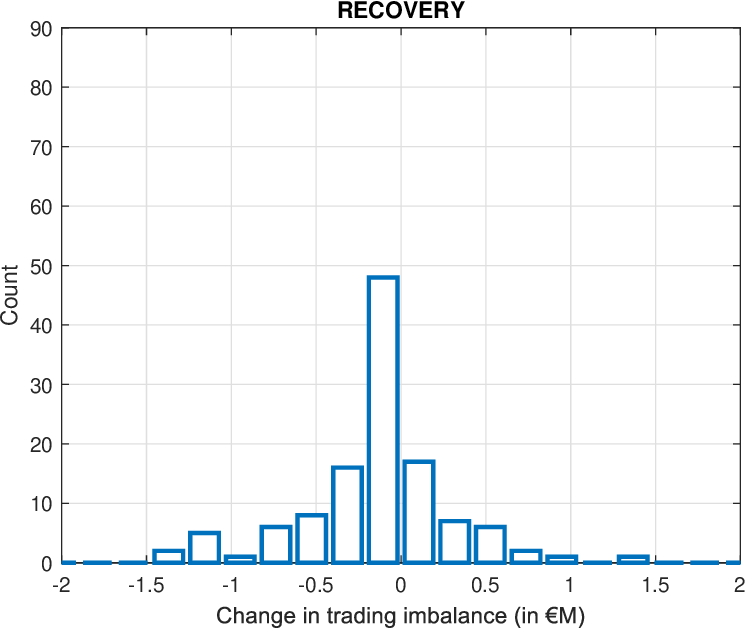} &
\includegraphics[height=4.5cm,width=0.31\textwidth]{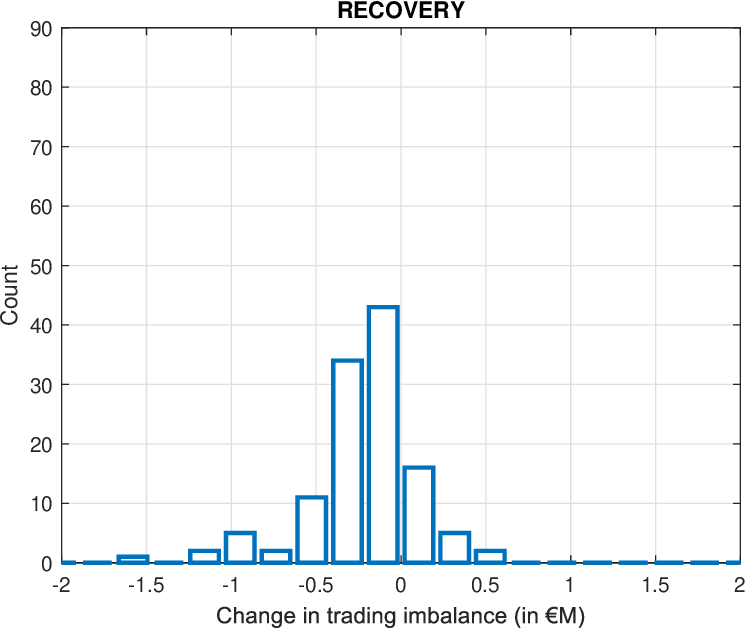} \\
\end{tabular}
\parbox{\textwidth}{\emph{Note.} The figure depicts a histogram of the empirical distribution of one-minute changes in trading imbalance for IB-HFT OWN (Panel A), IB-HFT MM (Panel B), and PURE-HFT MM (Panel C) at different stages of an unsystematic drift burst EPM. A negative number means that the trader category is collectively selling either via aggressive or passive trades, and vice versa.}
\end{center}
\end{figure}

\begin{itemize}
\item Table \ref{table:drift-burst-epm-individual} shows a detailed list of the individual drift burst EPMs in our sample, while in Table \ref{table:drift-burst-epm-aggregated} these EPMs are grouped by stock.

\item Table \ref{table:var-full} shows estimates of the VAR model in equation \eqref{equation:var} and is the expanded version of Table \ref{table:var} that includes all trader groups.

\item In Table \ref{table:var-full-a} -- \ref{table:var-full-p}, we break the dependent variable from equation \eqref{equation:var} into aggressive and passive trading imbalances. Otherwise, the estimated equation and the table formatting are identical to Table \ref{table:var-full}.

\item In Figure \ref{figure:systematic}, we show the time series of transaction prices for each stock during the systematic drift burst EPMs detected on April 17, 2013 and September 3, 2013.

\item Figure \ref{figure:delta-imbalance-u} replicates Figure \ref{figure:delta-imbalance-s} on trading imbalances for unsystematic drift burst EPMs.
\end{itemize}

The full versions of the tables that present coefficient estimates from the entire list of covariates included in the VAR model are rather long and are available upon request.

\clearpage

\input{tables/drift-burst-epm-individual.tex}

\clearpage

\input{tables/drift-burst-epm-aggregated.tex}

\clearpage

\begin{sidewaystable}[ht!]
\setlength{ \tabcolsep}{0.01cm}
\begin{center}
\caption{Coefficient estimates from the VAR for trading imbalance (full version)}
\label{table:var-full}
\begin{scriptsize}
\begin{tabular}{lccccccccc}
\hline
\multicolumn{10}{l}{ \textit{Panel A: Unsystematic drift burst EPMs}} \\
 & PURE-CLIENT & PURE-HFT OWN & PURE-HFT MM & IB-HFT MM & IB-CLIENT & IB-HFT OWN & IB-HFT PARENT & NON-HFT CLIENT & NON-HFT OWN \\
\cline{2-10}
Pre-event & 0.020 & 0.006 & -0.007 & 0.031{*}{*} & -0.019 & -0.012 & 0.013 & 0.027{*} & -0.010 \\
 & (1.14) & (0.42) & (-0.77) & (2.46) & (-1.47) & (-1.07) & (1.40) & (1.75) & (-0.96) \\
Early drop & -0.002 & 0.073{*}{*}{*} & 0.078{*} & 0.290{*}{*}{*} & -0.020 & -0.144{*}{*} & 0.055 & 0.038 & -0.112{*}{*} \\
 & (-0.04) & (3.03) & (1.69) & (4.10) & (-0.50) & (-2.27) & (1.64) & (1.17) & (-2.01) \\
Intermediate drop & 0.193{*}{*} & 0.020 & -0.025 & 0.176{*}{*}{*} & -0.085{*}{*} & -0.055 & 0.102{*}{*}{*} & 0.013 & -0.050 \\
 & (2.44) & (0.55) & (-0.74) & (3.86) & (-2.03) & (-1.47) & (3.52) & (0.36) & (-1.10) \\
Late drop & 0.113 & 0.025 & 0.017 & 0.227{*}{*}{*} & -0.267{*}{*}{*} & -0.006 & 0.206{*}{*}{*} & 0.103{*} & -0.150{*}{*} \\
 & (1.17) & (0.58) & (0.30) & (3.36) & (-3.49) & (-0.08) & (5.15) & (1.78) & (-2.17) \\
Recovery & 0.006 & 0.014 & -0.024{*}{*} & -0.010 & -0.016 & -0.018 & 0.006 & 0.054{*}{*}{*} & -0.009 \\
 & (0.34) & (1.26) & (-2.46) & (-0.67) & (-0.91) & (-0.89) & (0.36) & (3.41) & (-0.54) \\
Bid-ask spread & 0.005 & 0.012{*} & 0.002 & 0.006 & 0.001 & -0.008 & 0.005 & -0.003 & -0.001 \\
 & (0.86) & (1.93) & (0.50) & (1.20) & (0.16) & (-1.45) & (1.10) & (-0.46) & (-0.10) \\
Log(euro volume) & 0.008 & 0.004 & -0.018{*}{*} & 0.025{*}{*} & -0.026{*} & 0.008 & -0.019{*}{*}{*} & 0.016 & -0.005 \\
 & (1.32) & (0.64) & (-2.22) & (2.33) & (-1.83) & (0.63) & (-2.82) & (0.96) & (-0.46) \\
Return & 0.012 & 0.012 & 0.021{*} & -0.189{*}{*}{*} & 0.031{*}{*}{*} & 0.011 & 0.023{*}{*}{*} & 0.002 & 0.024{*}{*}{*} \\
 & (1.55) & (1.26) & (1.78) & (-12.05) & (3.66) & (1.23) & (3.01) & (0.23) & (2.99) \\
\\
$\bar{R}^{2}$ & 0.010 & 0.006 & 0.003 & 0.057 & 0.039 & 0.024 & 0.016 & 0.020 & 0.024 \\
\\
\multicolumn{10}{l}{ \textit{Panel B: Systematic drift burst EPMs}} \\
 & PURE-CLIENT & PURE-HFT OWN & PURE-HFT MM & IB-HFT MM & IB-CLIENT & IB-HFT OWN & IB-HFT PARENT & NON-HFT CLIENT & NON-HFT OWN \\
\cline{2-10}
Pre-event & -0.007 & -0.017 & 0.015 & -0.010 & 0.018 & -0.024 & -0.011 & 0.039{*} & -0.017 \\
 & (-0.85) & (-0.92) & (0.79) & (-0.39) & (1.00) & (-1.28) & (-0.66) & (1.84) & (-1.01) \\
Early drop & -0.012 & 0.008 & -0.059 & 0.003 & 0.040 & -0.152{*}{*} & 0.009 & 0.305{*}{*}{*} & -0.129 \\
 & (-0.52) & (0.51) & (-1.05) & (0.04) & (0.46) & (-2.07) & (0.12) & (3.30) & (-1.03) \\
Intermediate drop & 0.038 & 0.130{*} & -0.136{*}{*} & -0.164 & 0.131 & -0.349{*}{*}{*} & -0.244{*}{*} & 0.455{*}{*}{*} & 0.154 \\
 & (1.66) & (1.90) & (-2.32) & (-1.29) & (1.05) & (-3.51) & (-2.08) & (3.08) & (0.92) \\
Late drop & 0.082 & 0.146 & -0.088 & -0.374{*}{*} & -0.002 & -0.087 & -0.708{*}{*}{*} & 0.490{*}{*}{*} & 0.227{*}{*}{*} \\
 & (1.14) & (1.26) & (-1.27) & (-2.27) & (-0.02) & (-0.97) & (-4.59) & (3.62) & (2.82) \\
Recovery & 0.012 & -0.010 & 0.012 & -0.041 & -0.015 & 0.028 & -0.044{*}{*} & 0.016 & 0.007 \\
 & (0.96) & (-0.82) & (0.49) & (-1.38) & (-0.58) & (1.45) & (-2.49) & (1.00) & (0.38) \\
Bid-ask spread & -0.005 & 0.005 & 0.006 & -0.007 & 0.003 & -0.003 & 0.003 & -0.006 & 0.013{*} \\
 & (-0.92) & (0.79) & (0.63) & (-0.69) & (0.26) & (-0.53) & (0.44) & (-0.90) & (1.83) \\
Log(euro volume) & 0.010{*}{*} & 0.006 & -0.023 & -0.143{*}{*}{*} & 0.047{*}{*} & -0.022 & -0.102{*}{*}{*} & 0.060{*}{*} & 0.074{*}{*} \\
 & (2.35) & (1.19) & (-1.29) & (-7.61) & (2.33) & (-1.13) & (-7.34) & (2.75) & (2.68) \\
Return & -0.000 & 0.018{*}{*}{*} & 0.031 & -0.076{*}{*}{*} & 0.030{*}{*} & 0.006 & 0.054{*}{*}{*} & -0.028{*}{*} & -0.020{*}{*} \\
 & (-0.03) & (3.51) & (1.09) & (-3.64) & (2.14) & (0.61) & (5.11) & (-2.58) & (-2.10) \\
\\
$\bar{R}^{2}$ & 0.041 & 0.004 & 0.009 & 0.070 & 0.031 & 0.022 & 0.029 & 0.031 & 0.032 \\
\hline
\end{tabular}
\end{scriptsize}
\\
\medskip
\parbox{\textwidth}{\emph{Note.} We present the results of the VAR in equation \eqref{equation:var}. The model is estimated separately for unsystematic drift burst EPMs in Panel A and systematic ones in Panel B. The dependent variable is the 10-second trading imbalance of the trader category indicated by the column label. The explanatory variables ``Pre-event'', ``Early drop'', ``Intermediate drop'', ``Late drop'', and ``Recovery'' refer to phase dummies of a drift burst EPM. ``Bid-ask spread'' is the quoted spread (in percent), ``Log(euro volume)'' is the log-euro volume, and ``Return'' is the contemporaneous stock return. We also include stock fixed effects. The non-dummy variables are standardized at the stock level. The coefficient estimates of lagged trading imbalances are not reported for brevity. The $t$-statistics (in parentheses) are based on heteroscedasticity-robust standard errors clustered at the stock and date level. *$P$-value$<$0.1; **$P$-value$<$0.05; ***$P$-value$<$0.01. nObs = 342,032 in Panel A and nObs = 79,404 in Panel B.}
\end{center}
\end{sidewaystable}

\clearpage

\begin{sidewaystable}[ht!]
\setlength{ \tabcolsep}{0.01cm}
\begin{center}
\caption{Coefficient estimates from the VAR for aggressive trading imbalance (full version)}
\label{table:var-full-a}
\begin{scriptsize}
\begin{tabular}{lccccccccc}
\hline
\multicolumn{10}{l}{ \textit{Panel A: Unsystematic drift burst EPMs}} \\
 & PURE-CLIENT & PURE-HFT OWN & PURE-HFT MM & IB-HFT MM & IB-CLIENT & IB-HFT OWN & IB-HFT PARENT & NON-HFT CLIENT & NON-HFT OWN \\
\cline{2-10}
Pre-event & -0.004 & -0.005 & -0.006 & 0.003 & -0.024{*}{*} & -0.018{*} & 0.004 & -0.008 & -0.006 \\
 & (-0.30) & (-0.44) & (-0.75) & (0.31) & (-1.99) & (-1.95) & (0.37) & (-1.01) & (-0.87) \\
Early drop & -0.011 & 0.034 & 0.021 & -0.037 & -0.082{*} & -0.167{*}{*}{*} & 0.048 & -0.067{*}{*} & -0.110{*}{*} \\
 & (-0.28) & (1.27) & (0.54) & (-1.13) & (-1.82) & (-2.92) & (1.49) & (-2.42) & (-2.62) \\
Intermediate drop & 0.058{*} & 0.003 & 0.011 & 0.037 & -0.116{*}{*}{*} & -0.058{*} & 0.083{*}{*}{*} & -0.054{*} & -0.054 \\
 & (1.78) & (0.12) & (0.29) & (1.10) & (-2.69) & (-1.67) & (3.09) & (-1.98) & (-1.25) \\
Late drop & 0.019 & 0.046 & 0.055 & 0.012 & -0.341{*}{*}{*} & -0.065 & 0.189{*}{*}{*} & -0.188{*}{*}{*} & -0.166{*}{*}{*} \\
 & (0.41) & (1.17) & (0.99) & (0.24) & (-4.54) & (-0.91) & (4.69) & (-2.73) & (-2.79) \\
Recovery & 0.008 & 0.010 & -0.016 & -0.011 & -0.023 & -0.036{*}{*} & 0.001 & 0.036{*} & -0.002 \\
 & (0.45) & (0.71) & (-1.47) & (-0.80) & (-1.33) & (-2.45) & (0.07) & (1.92) & (-0.20) \\
Bid-ask spread & 0.005 & 0.010 & 0.001 & 0.006 & 0.008 & -0.006 & 0.010{*} & -0.007 & 0.004 \\
 & (1.14) & (1.60) & (0.15) & (1.26) & (1.14) & (-1.02) & (1.72) & (-1.03) & (0.51) \\
Log(euro volume) & -0.007 & -0.009{*} & -0.020{*}{*}{*} & -0.013 & -0.033{*}{*}{*} & -0.015{*} & -0.020{*}{*}{*} & -0.015 & -0.014 \\
 & (-1.26) & (-1.68) & (-2.85) & (-1.48) & (-2.77) & (-1.69) & (-3.17) & (-1.50) & (-1.52) \\
Return & 0.086{*}{*}{*} & 0.091{*}{*}{*} & 0.239{*}{*}{*} & 0.142{*}{*}{*} & 0.161{*}{*}{*} & 0.199{*}{*}{*} & 0.113{*}{*}{*} & 0.173{*}{*}{*} & 0.117{*}{*}{*} \\
 & (14.89) & (9.12) & (22.11) & (12.90) & (18.90) & (22.27) & (16.10) & (21.73) & (13.17) \\
\\
$\bar{R}^{2}$ & 0.008 & 0.010 & 0.054 & 0.029 & 0.049 & 0.050 & 0.019 & 0.035 & 0.028 \\
\\
\multicolumn{10}{l}{ \textit{Panel B: Systematic drift burst EPMs}} \\
 & PURE-CLIENT & PURE-HFT OWN & PURE-HFT MM & IB-HFT MM & IB-CLIENT & IB-HFT OWN & IB-HFT PARENT & NON-HFT CLIENT & NON-HFT OWN \\
\cline{2-10}
Pre-event & 0.017 & -0.002 & 0.047{*}{*} & 0.040{*} & 0.031 & 0.022 & -0.007 & 0.036{*}{*} & 0.011 \\
 & (1.56) & (-0.12) & (2.33) & (1.90) & (1.40) & (1.44) & (-0.39) & (2.19) & (0.67) \\
Early drop & -0.003 & 0.004 & -0.086{*} & 0.074 & -0.071 & -0.188{*} & 0.005 & 0.109{*}{*}{*} & -0.148 \\
 & (-0.13) & (0.12) & (-1.86) & (0.76) & (-1.03) & (-2.03) & (0.08) & (3.05) & (-1.00) \\
Intermediate drop & 0.069 & 0.102{*} & -0.066 & -0.049 & -0.021 & -0.354{*}{*}{*} & -0.278{*}{*} & 0.074 & -0.027 \\
 & (1.32) & (1.92) & (-1.29) & (-0.40) & (-0.52) & (-3.76) & (-2.30) & (1.22) & (-0.35) \\
Late drop & 0.117{*}{*}{*} & 0.203{*}{*} & 0.182{*}{*} & -0.309{*}{*} & -0.109 & -0.074 & -0.509{*}{*}{*} & 0.068 & 0.209{*}{*}{*} \\
 & (4.15) & (2.49) & (2.72) & (-2.19) & (-1.31) & (-1.01) & (-3.32) & (0.87) & (3.30) \\
Recovery & -0.022 & -0.023{*} & 0.010 & -0.078{*}{*}{*} & -0.040{*}{*} & -0.031{*}{*} & -0.045{*}{*} & -0.010 & -0.034 \\
 & (-1.41) & (-1.87) & (0.35) & (-2.92) & (-2.53) & (-2.34) & (-2.62) & (-0.58) & (-1.62) \\
Bid-ask spread & 0.001 & 0.003 & 0.018{*}{*} & 0.010 & 0.007 & 0.011 & 0.009 & -0.010 & 0.021{*}{*} \\
 & (0.27) & (0.37) & (2.13) & (1.09) & (0.64) & (1.68) & (1.43) & (-1.36) & (2.51) \\
Log(euro volume) & 0.005 & -0.000 & -0.010 & -0.108{*}{*}{*} & -0.003 & -0.064{*}{*}{*} & -0.086{*}{*}{*} & 0.005 & 0.043{*}{*}{*} \\
 & (1.17) & (-0.04) & (-0.74) & (-6.12) & (-0.21) & (-3.14) & (-9.13) & (0.32) & (3.29) \\
Return & 0.073{*}{*}{*} & 0.074{*}{*}{*} & 0.248{*}{*}{*} & 0.178{*}{*}{*} & 0.125{*}{*}{*} & 0.117{*}{*}{*} & 0.115{*}{*}{*} & 0.088{*}{*}{*} & 0.065{*}{*}{*} \\
 & (9.45) & (10.20) & (10.29) & (10.70) & (6.22) & (11.38) & (13.38) & (9.49) & (7.40) \\
\\
$\bar{R}^{2}$ & 0.010 & 0.013 & 0.087 & 0.085 & 0.042 & 0.041 & 0.038 & 0.019 & 0.019 \\
\hline
\end{tabular}
\end{scriptsize}
\\
\medskip
\parbox{\textwidth}{\emph{Note.} We present the results of the VAR in equation \eqref{equation:var}. The model is estimated separately for unsystematic drift burst EPMs in Panel A and systematic ones in Panel B. The dependent variable is the 10-second aggressive trading imbalance of the trader category indicated by the column label. The explanatory variables ``Pre-event'', ``Early drop'', ``Intermediate drop'', ``Late drop'', and ``Recovery'' refer to phase dummies of a drift burst EPM. ``Bid-ask spread'' is the quoted spread (in percent), ``Log(euro volume)'' is the log-euro volume, and ``Return'' is the contemporaneous stock return. We also include stock fixed effects. The non-dummy variables are standardized at the stock level. The coefficient estimates of lagged trading imbalances are not reported for brevity. The $t$-statistics (in parentheses) are based on heteroscedasticity-robust standard errors clustered at the stock and date level. *$P$-value$<$0.1; **$P$-value$<$0.05; ***$P$-value$<$0.01. nObs = 342,032 in Panel A and nObs = 79,404 in Panel B.}
\end{center}
\end{sidewaystable}

\clearpage

\begin{sidewaystable}[ht!]
\setlength{ \tabcolsep}{0.01cm}
\begin{center}
\caption{Coefficient estimates from the VAR for passive trading imbalance (full version)}
\label{table:var-full-p}
\begin{scriptsize}
\begin{tabular}{lccccccccc}
\hline
\multicolumn{10}{l}{ \textit{Panel A: Unsystematic drift burst EPMs}} \\
 & PURE-CLIENT & PURE-HFT OWN & PURE-HFT MM & IB-HFT MM & IB-CLIENT & IB-HFT OWN & IB-HFT PARENT & NON-HFT CLIENT & NON-HFT OWN \\
\cline{2-10}
Pre-event & 0.021 & 0.004 & -0.006 & 0.033{*}{*}{*} & -0.007 & 0.005 & 0.018{*}{*}{*} & 0.038{*}{*} & -0.010 \\
 & (1.41) & (0.27) & (-0.59) & (2.68) & (-0.65) & (0.51) & (2.66) & (2.30) & (-0.77) \\
Early drop & 0.009 & 0.031 & 0.099{*}{*} & 0.346{*}{*}{*} & 0.075 & 0.018 & 0.038 & 0.093{*}{*}{*} & -0.034 \\
 & (0.27) & (1.42) & (2.54) & (4.75) & (1.59) & (0.56) & (1.07) & (2.67) & (-0.80) \\
Intermediate drop & 0.146{*} & 0.008 & -0.057{*} & 0.187{*}{*}{*} & -0.020 & 0.016 & 0.065{*}{*} & 0.055 & -0.008 \\
 & (1.82) & (0.30) & (-1.89) & (3.93) & (-0.65) & (0.45) & (2.00) & (1.32) & (-0.22) \\
Late drop & 0.053 & -0.015 & -0.060 & 0.239{*}{*}{*} & -0.003 & 0.095 & 0.074{*}{*}{*} & 0.254{*}{*}{*} & -0.026 \\
 & (0.74) & (-0.43) & (-1.18) & (3.86) & (-0.04) & (1.63) & (2.75) & (3.97) & (-0.39) \\
Recovery & -0.005 & 0.006 & -0.021{*}{*} & -0.005 & -0.007 & 0.014 & 0.011 & 0.039{*}{*}{*} & -0.005 \\
 & (-0.63) & (0.45) & (-2.58) & (-0.32) & (-0.43) & (0.63) & (0.98) & (3.00) & (-0.27) \\
Bid-ask spread & -0.004 & 0.004 & 0.001 & 0.003 & -0.006 & -0.004 & -0.003 & 0.003 & -0.007 \\
 & (-0.86) & (0.89) & (0.11) & (0.58) & (-0.82) & (-0.50) & (-0.57) & (0.52) & (-1.24) \\
Log(euro volume) & 0.014{*}{*}{*} & 0.013{*}{*} & -0.002 & 0.036{*}{*}{*} & -0.000 & 0.024{*}{*} & -0.008 & 0.034{*}{*} & 0.008 \\
 & (2.66) & (2.15) & (-0.29) & (3.98) & (-0.00) & (2.13) & (-1.01) & (2.29) & (0.90) \\
Return & -0.082{*}{*}{*} & -0.086{*}{*}{*} & -0.300{*}{*}{*} & -0.312{*}{*}{*} & -0.111{*}{*}{*} & -0.208{*}{*}{*} & -0.116{*}{*}{*} & -0.129{*}{*}{*} & -0.077{*}{*}{*} \\
 & (-14.40) & (-8.43) & (-29.38) & (-22.59) & (-11.40) & (-23.90) & (-14.13) & (-13.38) & (-10.29) \\
\\
$\bar{R}^{2}$ & 0.009 & 0.009 & 0.082 & 0.106 & 0.035 & 0.060 & 0.023 & 0.041 & 0.022 \\
\\
\multicolumn{10}{l}{ \textit{Panel B: Systematic drift burst EPMs}} \\
 & PURE-CLIENT & PURE-HFT OWN & PURE-HFT MM & IB-HFT MM & IB-CLIENT & IB-HFT OWN & IB-HFT PARENT & NON-HFT CLIENT & NON-HFT OWN \\
\cline{2-10}
Pre-event & -0.031{*}{*} & -0.035{*} & -0.033{*}{*} & -0.051 & -0.009 & -0.055{*}{*} & -0.012 & 0.024 & -0.033{*} \\
 & (-2.30) & (-1.97) & (-2.21) & (-1.67) & (-0.59) & (-2.17) & (-0.80) & (1.05) & (-1.74) \\
Early drop & -0.018 & 0.004 & 0.019 & -0.053 & 0.075 & 0.001 & 0.038 & 0.298{*}{*} & -0.016 \\
 & (-0.52) & (0.11) & (0.41) & (-1.15) & (0.84) & (0.02) & (0.81) & (2.70) & (-0.44) \\
Intermediate drop & -0.044 & 0.060 & -0.100{*}{*} & -0.145 & 0.165 & -0.036 & -0.018 & 0.503{*}{*}{*} & 0.255 \\
 & (-0.85) & (0.61) & (-2.09) & (-1.66) & (1.09) & (-0.77) & (-0.36) & (3.21) & (1.38) \\
Late drop & -0.071 & -0.138{*}{*}{*} & -0.314{*}{*}{*} & -0.284{*}{*} & 0.051 & -0.018 & -0.345{*}{*}{*} & 0.551{*}{*}{*} & 0.143 \\
 & (-1.22) & (-3.31) & (-4.10) & (-2.20) & (0.55) & (-0.23) & (-3.85) & (3.24) & (1.48) \\
Recovery & 0.025 & 0.011 & 0.001 & -0.003 & 0.008 & 0.080{*}{*}{*} & -0.012 & 0.033 & 0.040{*}{*} \\
 & (1.54) & (0.62) & (0.14) & (-0.14) & (0.30) & (2.92) & (-0.63) & (1.52) & (2.14) \\
Bid-ask spread & -0.001 & -0.000 & -0.012 & -0.014 & -0.003 & -0.020{*}{*}{*} & -0.009 & 0.002 & -0.000 \\
 & (-0.13) & (-0.00) & (-1.33) & (-1.44) & (-0.70) & (-3.54) & (-1.60) & (0.24) & (-0.06) \\
Log(euro volume) & 0.001 & 0.010 & -0.018 & -0.089{*}{*}{*} & 0.066{*}{*}{*} & 0.048{*}{*} & -0.042{*}{*} & 0.068{*}{*}{*} & 0.061{*}{*} \\
 & (0.09) & (1.21) & (-1.66) & (-6.45) & (3.79) & (2.72) & (-2.65) & (3.37) & (2.13) \\
Return & -0.084{*}{*}{*} & -0.072{*}{*}{*} & -0.266{*}{*}{*} & -0.227{*}{*}{*} & -0.077{*}{*}{*} & -0.135{*}{*}{*} & -0.077{*}{*}{*} & -0.099{*}{*}{*} & -0.074{*}{*}{*} \\
 & (-9.81) & (-8.64) & (-14.22) & (-10.36) & (-9.62) & (-12.24) & (-8.04) & (-7.32) & (-6.60) \\
\\
$\bar{R}^{2}$ & 0.014 & 0.013 & 0.095 & 0.091 & 0.041 & 0.056 & 0.015 & 0.053 & 0.036 \\
\hline
\end{tabular}
\end{scriptsize}
\\
\medskip
\parbox{\textwidth}{\emph{Note.} We present the results of the VAR in equation \eqref{equation:var}. The model is estimated separately for unsystematic drift burst EPMs in Panel A and systematic ones in Panel B. The dependent variable is the 10-second passive trading imbalance of the trader category indicated by the column label. The explanatory variables ``Pre-event'', ``Early drop'', ``Intermediate drop'', ``Late drop'', and ``Recovery'' refer to phase dummies of a drift burst EPM. ``Bid-ask spread'' is the quoted spread (in percent), ``Log(euro volume)'' is the log-euro volume, and ``Return'' is the contemporaneous stock return. We also include stock fixed effects. The non-dummy variables are standardized at the stock level. The coefficient estimates of lagged trading imbalances are not reported for brevity. The $t$-statistics (in parentheses) are based on heteroscedasticity-robust standard errors clustered at the stock and date level. *$P$-value$<$0.1; **$P$-value$<$0.05; ***$P$-value$<$0.01. nObs = 342,032 in Panel A and nObs = 79,404 in Panel B.}
\end{center}
\end{sidewaystable}

\clearpage

\begin{sidewaystable}[ht!]
\setlength{ \tabcolsep}{0.01cm}
\begin{center}
\caption{Profit and loss of trading activity (full version)}
\label{table:var-full-profits}
\begin{scriptsize}
\begin{tabular}{lccccccccc}
\hline
\multicolumn{10}{l}{ \textit{Panel A: Unsystematic drift burst EPMs}} \\
 & PURE-CLIENT & PURE-HFT OWN & PURE-HFT MM & IB-HFT MM & IB-CLIENT & IB-HFT OWN & IB-HFT PARENT & NON-HFT CLIENT & NON-HFT OWN \\
\cline{2-10}
Dummy & -0.013 & 0.060 & 0.108 & -0.347{*}{*}{*} & 0.419{*}{*}{*} & 0.842{*}{*} & -0.145{*}{*} & -0.162 & 0.539{*}{*}{*}\tabularnewline
 & (-0.21) & (0.76) & (0.84) & (-2.99) & (2.85) & (2.35) & (-2.54) & (-0.99) & (2.69) \\
Bid-ask spread & -0.030{*} & 0.092 & 0.483{*}{*} & 0.005 & 0.230 & 0.710 & 0.052 & 0.022 & -0.122 \\
 & (-1.76) & (1.28) & (2.55) & (0.05) & (1.19) & (1.08) & (0.83) & (0.10) & (-0.74) \\
Log(euro Volume) & -0.006 & 0.068 & 0.581{*}{*}{*} & -0.073 & 0.133 & 0.356{*}{*} & 0.084{*}{*} & -0.609{*}{*}{*} & 0.110 \\
 & (-0.48) & (1.30) & (5.82) & (-1.23) & (1.31) & (2.37) & (2.56) & (-4.04) & (1.15) \\
Return & -0.005 & -0.036 & 0.066 & -0.101 & -0.986{*} & -0.436 & -0.003 & 0.148 & -0.451 \\
 & (-0.08) & (-0.24) & (0.22) & (-0.42) & (-1.88) & (-0.69) & (-0.02) & (0.32) & (-1.59) \\
\\
$\bar{R}^{2}$ & 0.008 & 0.011 & 0.013 & 0.013 & 0.029 & 0.014 & 0.008 & 0.018 & 0.008 \\
\\
\multicolumn{10}{l}{ \textit{Panel B: Systematic drift burst EPMs}} \\
 & PURE-CLIENT & PURE-HFT OWN & PURE-HFT MM & IB-HFT MM & IB-CLIENT & IB-HFT OWN & IB-HFT PARENT & NON-HFT CLIENT & NON-HFT OWN \\
\cline{2-10}
Dummy & 0.039{*}{*} & 0.013 & 0.894{*} & 0.729{*}{*} & -0.398 & 1.514{*}{*}{*} & 0.803{*}{*}{*} & -0.964{*}{*} & -0.391 \\
 & (2.35) & (0.17) & (2.05) & (2.22) & (-1.64) & (2.93) & (2.86) & (-2.31) & (-1.06) \\
Bid-ask spread & 0.038{*} & -0.057 & 0.308{*}{*} & -0.030 & 0.518{*}{*}{*} & 0.755{*}{*}{*} & 0.117 & -0.264{*}{*} & -0.217 \\
 & (2.02) & (-1.53) & (2.25) & (-0.18) & (2.99) & (2.97) & (1.52) & (-2.20) & (-1.41) \\
Log(euro Volume) & -0.005 & -0.016 & 1.035{*}{*}{*} & 0.040 & 0.081 & -0.068 & 0.361{*}{*}{*} & -0.743{*}{*}{*} & -0.457{*}{*} \\
 & (-0.44) & (-0.44) & (3.58) & (0.33) & (0.60) & (-0.31) & (4.89) & (-3.61) & (-2.44) \\
Return & 0.006 & -0.016 & -0.054 & -2.049{*}{*}{*} & 0.104 & -0.816 & -1.226{*}{*}{*} & 0.839{*}{*}{*} & 0.936{*}{*}{*} \\
 & (0.17) & (-0.38) & (-0.16) & (-7.02) & (0.44) & (-1.60) & (-5.78) & (3.17) & (4.00) \\
\\
$\bar{R}^{2}$ &  0.058 & 0.008 & 0.022 & 0.087 & 0.006 & 0.016 & 0.053 & 0.022 & 0.012 \\
\hline
\end{tabular}
\end{scriptsize}
\\
\medskip
\parbox{\textwidth}{\emph{Note.} This table presents coefficient estimates from the VAR in equation \eqref{equation:PnL}. The model is estimated separately for unsystematic drift burst EPMs in Panel A and systematic ones in Panel B. The dependent variable is the 10-second profit and loss (P\&L) of the trader category indicated by the column label. ``Dummy'' equals one for the entire duration of the drift burst EPM (from the beginning of the pre-event until the end of the recovery), zero otherwise, ``Bid-ask spread'' is the quoted spread (in percent), ``Log(euro volume)'' is the log-euro volume, and ``Return'' is the contemporaneous stock return. We also include stock fixed effects. The non-dummy variables are standardized at the stock level. The coefficient estimates of the lagged P\&L are not reported for brevity. The $t$-statistics (in parentheses) are based on heteroscedasticity-robust standard errors clustered at the stock and date level. *$P$-value$<$0.1; **$P$-value$<$0.05; ***$P$-value$<$0.01. nObs = 342,032 in Panel A and nObs = 79,404 in Panel B.}
\end{center}
\end{sidewaystable}

\clearpage

\clearpage

\clearpage

\section{Alternative analysis of trading activity} \label{appendix:kirilenko}

\setcounter{figure}{0}
\setcounter{table}{0}

In this appendix, we look at the question of whether DMMs change their trading activity during drift burst EPMs. We rely on \citet{kirilenko-kyle-samadi-tuzun:17a}, who employ an error correction model to changes in inventory of different trader groups. They show that HFTs did not alter their behavior during the Flash Crash of May 6, 2010. We arrive at the opposite conclusion regarding DMMs in our sample of drift burst EPMs, which is possibly due to the fact that we analyze 148 events instead of a single one.

The econometric model reads:
\begin{align} \label{equation:kirilenko}
\begin{split}
\Delta y_{t} &= \alpha + \phi \Delta y_{t-1} + \delta y_{t-1} + \sum_{i=0}^{3} \beta_{i} \Delta p_{t-i} \\
&+ D^{D} \left( \alpha^{D} + \phi^{D} \Delta y_{t-1} + \delta^{D} y_{t-1} + \sum_{i=0}^{3} \beta_{i}^{D} \Delta p_{t-i} \right) \\
&+ D^{U} \left( \alpha^{U} + \phi^{U} \Delta y_{t-1} + \delta^{U} y_{t-1} + \sum_{i=0}^{3} \beta_{i}^{U} \Delta p_{t-i} \right) + \varepsilon_{t},
\end{split}
\end{align}
where $D^{D}$ is a dummy activated during the drift burst EPM (from $t_{ \mathrm{start}}$ to $t_{ \mathrm{trough}}$ according to the definitions in the main text), while $D^{U}$ is a dummy activated during the recovery (from $t_{ \mathrm{trough}}$ to $t_{ \mathrm{end}}$), $y_{t}$ is the net inventory of each trader category at time $t$ (measured in million euros and assumed to start from zero at the beginning of each day), $\Delta y_{t}$ is the first difference of $y_{t}$, $p_{t}$ is the log-price at time $t$ (measured as the midpoint of the prevailing bid and ask quote), and $\Delta p_{t}$ is the log-return from $t-1$ to $t$.\footnote{The model is slightly modified compared to \citet{kirilenko-kyle-samadi-tuzun:17a}. In particular, we use three 10-second lags (as opposed to 20 one-second lags in their paper), and we regress on log-returns (as opposed to price changes in their paper) to ensure a greater uniformity across stocks and days.}

The model allows the slopes on the explanatory variables to shift during the day. We estimate equation \eqref{equation:kirilenko} for each day with a drift burst EPM and for each trader category separately. The data are recorded on a 10-second grid. A change in trading attitude during a particular phase should thus emerge via significant coefficient estimates in front of the drift burst EPM or recovery explanatory variables.

\begin{table}[ht!]
\setlength{\tabcolsep}{0.35cm}
\begin{center}
\caption{Inventory changes during unsystematic drift burst EPMs}
\label{table:kirilenko-u}
\begin{scriptsize}
\begin{tabular}{lccccccc}
\hline
 & PURE-HFT & PURE-HFT & IB-HFT & IB-HFT & IB     & NON-HFT & NON-HFT\\
 & MM       & OWN      & MM     & OWN    & CLIENT & CLIENT  & OWN \\
\hline
\multicolumn{3}{l}{ \textit{Panel A: Outside drift burst EPM}} \\
constant & $-0.000^{***}$ & $0.000$ & $0.000^{*}$ & $0.000$ & $-0.000^{***}$ & $0.000$ & $0.000$ \\
 & ($-4.62$) & ($0.73$) & ($1.74$) & ($1.09$) & ($-4.85$) & ($1.22$) & ($1.34$) \\
$\Delta y_{t-1}$ & $0.004$ & $0.029^{***}$ & $0.065^{***}$ & $0.091^{***}$ & $0.099^{***}$ & $0.081^{***}$ & $0.080^{***}$ \\
 & ($1.15$) & ($3.73$) & ($15.01$) & ($19.59$) & ($13.81$) & ($14.70$) & ($8.35$) \\
$y_{t-1}$ & $-0.002^{***}$ & $-0.001$ & $0.000$ & $0.000$ & $-0.000$ & $-0.001^{***}$ & $-0.000$ \\
 & ($-13.82$) & ($-1.37$) & ($0.91$) & ($1.12$) & ($-0.77$) & ($-7.44$) & ($-1.55$) \\
$\Delta p_{t}$ & $0.181^{***}$ & $0.002$ & $-0.227^{***}$ & $-0.064^{**}$ & $0.050$ & $-0.107^{***}$ & $0.061^{***}$ \\
 & ($11.34$) & ($0.13$) & ($-14.12$) & ($-2.37$) & ($1.60$) & ($-3.99$) & ($2.61$) \\
$\Delta p_{t-1}$ & $-0.086^{***}$ & $-0.020^{**}$ & $0.030^{***}$ & $0.097^{***}$ & $0.018$ & $-0.054^{***}$ & $0.020$ \\
 & ($-9.13$) & ($-2.22$) & ($3.82$) & ($5.24$) & ($1.39$) & ($-2.88$) & ($1.20$) \\
$\Delta p_{t-2}$ & $-0.041^{***}$ & $-0.008$ & $-0.011^{*}$ & $0.003$ & $0.039^{***}$ & $-0.004$ & $0.004$ \\
 & ($-4.79$) & ($-1.06$) & ($-1.75$) & ($0.13$) & ($3.81$) & ($-0.20$) & ($0.28$) \\
$\Delta p_{t-3}$ & $-0.050^{***}$ & $-0.003$ & $-0.011^{*}$ & $0.045^{***}$ & $0.025^{**}$ & $-0.012$ & $-0.006$ \\
 & ($-5.84$) & ($-0.49$) & ($-1.74$) & ($2.93$) & ($2.34$) & ($-0.73$) & ($-0.30$) \\
\hline
\multicolumn{3}{l}{ \textit{Panel B: During drift burst EPM ($D^{D}$)}} \\
constant & $-0.000$ & $0.000$ & $0.001^{***}$ & $0.001$ & $0.003$ & $0.006^{***}$ & $-0.000$ \\
 & ($-0.26$) & ($0.27$) & ($4.47$) & ($0.81$) & ($0.86$) & ($2.90$) & ($-0.21$) \\
$\Delta y_{t-1}$ & $-0.029^{*}$ & $-0.063^{**}$ & $-0.076^{***}$ & $-0.051^{***}$ & $-0.040^{*}$ & $-0.061^{***}$ & $-0.032$ \\
 & ($-1.87$) & ($-2.08$) & ($-4.37$) & ($-3.04$) & ($-1.85$) & ($-3.85$) & ($-1.37$) \\
$y_{t-1}$ & $-0.092^{***}$ & $-0.047^{***}$ & $-0.040^{***}$ & $-0.061^{***}$ & $-0.035^{***}$ & $-0.027^{***}$ & $-0.036^{***}$ \\
 & ($-9.86$) & ($-3.22$) & ($-7.38$) & ($-7.68$) & ($-4.35$) & ($-3.73$) & ($-4.88$) \\
$\Delta p_{t}$ & $0.045$ & $0.012$ & $-0.676^{***}$ & $1.034^{***}$ & $0.076$ & $-0.834^{***}$ & $0.205^{*}$ \\
 & ($0.47$) & ($0.14$) & ($-7.47$) & ($7.04$) & ($0.63$) & ($-4.45$) & ($1.91$) \\
$\Delta p_{t-1}$ & $-0.063$ & $0.068$ & $-0.088$ & $-0.282^{**}$ & $0.148$ & $0.001$ & $0.307^{***}$ \\
 & ($-0.91$) & ($1.40$) & ($-1.40$) & ($-2.36$) & ($1.48$) & ($0.00$) & ($2.70$) \\
$\Delta p_{t-2}$ & $0.014$ & $0.059$ & $0.040$ & $0.085$ & $0.095$ & $0.030$ & $-0.084$ \\
 & ($0.25$) & ($1.49$) & ($0.79$) & ($0.80$) & ($1.06$) & ($0.28$) & ($-1.07$) \\
$\Delta p_{t-3}$ & $0.054$ & $0.025$ & $-0.113^{**}$ & $-0.164^{*}$ & $0.046$ & $-0.045$ & $0.127^{*}$ \\
 & ($0.91$) & ($0.60$) & ($-2.20$) & ($-1.74$) & ($0.59$) & ($-0.48$) & ($1.73$) \\
\hline
\multicolumn{3}{l}{ \textit{Panel C: Recovery phase ($D^{U}$)}} \\
constant & $-0.001^{*}$ & $-0.001^{*}$ & $0.003^{***}$ & $0.001$ & $-0.003^{**}$ & $0.003^{***}$ & $-0.001^{***}$ \\
 & ($-1.78$) & ($-1.94$) & ($6.70$) & ($0.82$) & ($-2.35$) & ($2.96$) & ($-3.68$) \\
$\Delta y_{t-1}$ & $-0.019$ & $0.042^{*}$ & $-0.043^{***}$ & $-0.037^{***}$ & $-0.013$ & $-0.022$ & $-0.029$ \\
 & ($-1.60$) & ($1.66$) & ($-3.05$) & ($-2.83$) & ($-0.82$) & ($-1.59$) & ($-1.27$) \\
$y_{t-1}$ & $-0.052^{***}$ & $-0.071^{***}$ & $-0.054^{***}$ & $-0.047^{***}$ & $-0.033^{***}$ & $-0.036^{***}$ & $-0.037^{***}$ \\
 & ($-12.37$) & ($-5.59$) & ($-9.18$) & ($-11.21$) & ($-9.89$) & ($-8.67$) & ($-6.37$) \\
$\Delta p_{t}$ & $-0.303^{***}$ & $0.226^{**}$ & $-0.222^{***}$ & $0.167^{*}$ & $-0.107$ & $0.424^{***}$ & $-0.101$ \\
 & ($-4.99$) & ($2.10$) & ($-4.30$) & ($1.84$) & ($-1.40$) & ($3.67$) & ($-1.45$) \\
$\Delta p_{t-1}$ & $0.090^{**}$ & $0.023$ & $-0.058^{*}$ & $-0.129^{*}$ & $-0.017$ & $-0.024$ & $0.006$ \\
 & ($2.31$) & ($0.24$) & ($-1.74$) & ($-1.85$) & ($-0.26$) & ($-0.37$) & ($0.12$) \\
$\Delta p_{t-2}$ & $0.010$ & $0.205^{*}$ & $0.023$ & $-0.061$ & $0.123^{**}$ & $-0.052$ & $-0.071$ \\
 & ($0.31$) & ($1.89$) & ($0.81$) & ($-1.10$) & ($2.14$) & ($-0.89$) & ($-1.45$) \\
$\Delta p_{t-3}$ & $0.086^{**}$ & $0.167$ & $-0.017$ & $0.015$ & $-0.069$ & $-0.004$ & $0.027$ \\
 & ($2.02$) & ($1.29$) & ($-0.52$) & ($0.25$) & ($-1.32$) & ($-0.05$) & ($0.50$) \\
\hline
\end{tabular}
\end{scriptsize}
\\
\medskip
\parbox{\textwidth}{\emph{Note.} The significance of the regression coefficient estimate is evaluated with a standard $t$-test assuming independence but adjusted for heteroskedasticity (in parentheses). *$P$-value$<$0.1; **$P$-value$<$0.05; ***$P$-value$<$0.01.}
\end{center}
\end{table}

\begin{table}[ht!]
\setlength{\tabcolsep}{0.35cm}
\begin{center}
\caption{Inventory changes during systematic drift burst EPMs}
\label{table:kirilenko-s}
\begin{scriptsize}
\begin{tabular}{lccccccc}
\hline
 & PURE-HFT & PURE-HFT & IB-HFT & IB-HFT & IB     & NON-HFT & NON-HFT\\
 & MM       & OWN      & MM     & OWN    & CLIENT & CLIENT  & OWN \\
\hline
\multicolumn{3}{l}{ \textit{Panel A: Outside drift burst EPM}} \\
constant & $-0.000^{***}$ & $-0.000$ & $0.000$ & $0.000^{**}$ & $0.000$ & $0.000$ & $-0.000$ \\
 & ($-4.11$) & ($-1.11$) & ($0.46$) & ($2.17$) & ($1.51$) & ($1.18$) & ($-0.13$) \\
$\Delta y_{t-1}$ & $-0.006$ & $0.011$ & $0.060^{***}$ & $0.083^{***}$ & $0.037^{*}$ & $0.058^{***}$ & $0.097^{***}$ \\
 & ($-0.73$) & ($1.21$) & ($5.60$) & ($6.23$) & ($1.93$) & ($5.98$) & ($4.29$) \\
$y_{t-1}$ & $-0.005^{***}$ & $0.000$ & $0.000^{***}$ & $0.000$ & $0.000$ & $-0.000$ & $0.000$ \\
 & ($-8.43$) & ($0.20$) & ($5.16$) & ($0.08$) & ($1.56$) & ($-1.06$) & ($1.34$) \\
$\Delta p_{t}$ & $-0.036$ & $0.006$ & $-0.038^{*}$ & $0.021$ & $0.081^{***}$ & $0.025$ & $-0.072^{**}$ \\
 & ($-1.19$) & ($0.69$) & ($-1.87$) & ($0.43$) & ($3.78$) & ($0.59$) & ($-2.35$) \\
$\Delta p_{t-1}$ & $-0.067^{***}$ & $-0.004$ & $0.040^{***}$ & $0.050^{**}$ & $0.010$ & $-0.005$ & $-0.030^{*}$ \\
 & ($-4.35$) & ($-0.97$) & ($3.95$) & ($1.97$) & ($0.52$) & ($-0.46$) & ($-1.92$) \\
$\Delta p_{t-2}$ & $-0.048^{***}$ & $-0.003$ & $0.012$ & $0.007$ & $0.029^{***}$ & $0.019$ & $0.003$ \\
 & ($-3.73$) & ($-1.17$) & ($1.29$) & ($0.34$) & ($2.64$) & ($1.64$) & ($0.14$) \\
$\Delta p_{t-3}$ & $-0.038^{***}$ & $0.001$ & $0.014$ & $-0.006$ & $0.022^{*}$ & $0.009$ & $0.026$ \\
 & ($-2.82$) & ($0.23$) & ($1.47$) & ($-0.27$) & ($1.70$) & ($0.66$) & ($1.24$) \\
\hline
\multicolumn{3}{l}{ \textit{Panel B: During drift burst EPM ($D^{D}$)}} \\
constant & $0.000$ & $-0.000$ & $0.000$ & $0.000$ & $-0.000^{*}$ & $0.001^{***}$ & $0.000$ \\
 & ($1.02$) & ($-0.99$) & ($0.19$) & ($0.06$) & ($-1.89$) & ($2.71$) & ($1.53$) \\
$\Delta y_{t-1}$ & $-0.039$ & $-0.043$ & $-0.075^{*}$ & $-0.020$ & $0.045$ & $-0.048$ & $-0.010$ \\
 & ($-1.06$) & ($-1.19$) & ($-1.77$) & ($-0.64$) & ($1.17$) & ($-1.48$) & ($-0.19$) \\
$y_{t-1}$ & $-0.089^{***}$ & $-0.055^{***}$ & $0.006$ & $-0.060^{***}$ & $-0.032^{***}$ & $-0.025^{***}$ & $-0.017$ \\
 & ($-5.14$) & ($-2.71$) & ($0.48$) & ($-3.25$) & ($-2.71$) & ($-2.67$) & ($-1.51$) \\
$\Delta p_{t}$ & $0.341^{***}$ & $0.036$ & $0.119^{**}$ & $-0.019$ & $0.029$ & $-0.652^{***}$ & $-0.397^{*}$ \\
 & ($3.50$) & ($0.90$) & ($2.05$) & ($-0.11$) & ($0.25$) & ($-4.45$) & ($-1.65$) \\
$\Delta p_{t-1}$ & $-0.095$ & $0.016^{*}$ & $0.079^{*}$ & $0.128$ & $-0.080$ & $-0.138^{*}$ & $0.032$ \\
 & ($-1.27$) & ($1.76$) & ($1.88$) & ($1.07$) & ($-1.10$) & ($-1.88$) & ($0.24$) \\
$\Delta p_{t-2}$ & $-0.052$ & $0.020$ & $-0.008$ & $0.016$ & $0.050$ & $-0.118$ & $0.198$ \\
 & ($-0.79$) & ($1.22$) & ($-0.17$) & ($0.09$) & ($0.61$) & ($-1.24$) & ($1.28$) \\
$\Delta p_{t-3}$ & $-0.038$ & $0.010$ & $-0.061$ & $0.211^{*}$ & $-0.033$ & $-0.039$ & $-0.037$ \\
 & ($-0.74$) & ($0.98$) & ($-1.44$) & ($1.85$) & ($-0.59$) & ($-0.56$) & ($-0.53$) \\
\hline
\multicolumn{3}{l}{ \textit{Panel C: Recovery phase ($D^{U}$)}} \\
constant & $-0.001^{***}$ & $-0.000^{**}$ & $-0.001^{***}$ & $0.001$ & $0.000$ & $0.001$ & $0.005^{***}$ \\
 & ($-4.67$) & ($-2.45$) & ($-4.46$) & ($1.12$) & ($0.52$) & ($0.77$) & ($6.50$) \\
$\Delta y_{t-1}$ & $-0.001$ & $-0.013$ & $-0.068^{***}$ & $-0.021$ & $0.002$ & $-0.037$ & $-0.037$ \\
 & ($-0.03$) & ($-0.52$) & ($-2.59$) & ($-0.81$) & ($0.05$) & ($-1.20$) & ($-1.14$) \\
$y_{t-1}$ & $-0.085^{***}$ & $-0.036^{***}$ & $-0.023^{***}$ & $-0.050^{***}$ & $-0.059^{***}$ & $-0.016^{**}$ & $-0.119^{***}$ \\
 & ($-6.42$) & ($-3.91$) & ($-4.55$) & ($-4.61$) & ($-4.59$) & ($-2.09$) & ($-10.76$) \\
$\Delta p_{t}$ & $0.153^{***}$ & $-0.006$ & $-0.052$ & $-0.072$ & $-0.089$ & $-0.062$ & $-0.010$ \\
 & ($2.70$) & ($-0.60$) & ($-1.27$) & ($-0.68$) & ($-1.47$) & ($-0.82$) & ($-0.14$) \\
$\Delta p_{t-1}$ & $-0.000$ & $0.010^{*}$ & $0.078^{***}$ & $-0.055$ & $0.007$ & $0.082$ & $-0.101$ \\
 & ($-0.00$) & ($1.90$) & ($2.64$) & ($-0.67$) & ($0.13$) & ($1.35$) & ($-1.18$) \\
$\Delta p_{t-2}$ & $0.073^{**}$ & $0.011^{**}$ & $0.016$ & $0.012$ & $0.007$ & $0.014$ & $-0.056$ \\
 & ($2.00$) & ($2.22$) & ($0.66$) & ($0.15$) & ($0.19$) & ($0.26$) & ($-1.18$) \\
$\Delta p_{t-3}$ & $0.089^{**}$ & $0.002$ & $0.074^{***}$ & $-0.077$ & $0.033$ & $-0.004$ & $-0.147^{**}$ \\
 & ($2.38$) & ($0.38$) & ($2.62$) & ($-0.73$) & ($0.87$) & ($-0.08$) & ($-2.10$) \\
\hline
\end{tabular}
\end{scriptsize}
\\
\medskip
\parbox{\textwidth}{\emph{Note.} The significance of the regression coefficient estimate is evaluated with a standard $t$-test assuming independence but adjusted for heteroskedasticity (in parentheses). *$P$-value$<$0.1; **$P$-value$<$0.05; ***$P$-value$<$0.01.}
\end{center}
\end{table}

The average coefficient estimates, together with their associated $t$-statistics, are reported in Tables \ref{table:kirilenko-u} -- \ref{table:kirilenko-s} for unsystematic and systematic events. To compute the $t$-statistic of the average regression coefficient, we assume the individual estimates are independent, but we allow for heteroskedasticity with a White correction. Apart from the fact that the regression yields significant coefficients in a lot of instances, below are our key findings.

\citet{kirilenko-kyle-samadi-tuzun:17a} report a significant negative coefficient on the lagged inventory level in their pre-flash crash period (which is interpreted as an indication of mean-reversion), but no significant changes during the flash crash or in its aftermath. Instead, we find that mean-reversion is much stronger for all trader categories during the drift burst EPM and recovery compared to the pre-event and post-recovery period. There is one exception to this regularity, which is the positive but insignificant coefficient of IB-HFT MM during the drift burst EPM (and also in normal times, where it is significant) for systematic events, denoting a lack of mean reversion. This observation is consistent with the reported trading activity in Panel A of Figure \ref{figure:trading-imbalance}.

\citet{kirilenko-kyle-samadi-tuzun:17a} find that HFTs inventory changes are positively related to contemporaneous and lagged price changes for the first few seconds of the flash crash, before turning negatively correlated. Furthermore, inventory changes of market makers are negatively related to contemporaneous price change and positively correlated with lagged ones. We broadly confirm this with appropriate interpretation of various trader groups. For example, in \citet{kirilenko-kyle-samadi-tuzun:17a} IB-HFT MM and IB-HFT OWN correspond to market makers, while PURE-HFT MM correspond to HFTs. The main difference is the significant changes for the price/inventory coefficients during the drift burst EPM and recovery. While IB-HFT MM inventories are more significantly negatively correlated to contemporaneous price changes during unsystematic drift burst EPMs, signalling a larger inventory absorption, the opposite holds for systematic events, since they are net sellers (see Figure \ref{figure:trading-imbalance}). The slow traders (i.e., NON-HFT CLIENT and NON-HFT OWN) exhibit negative correlations of their inventory changes with contemporaneous price changes during systematic events, consistent with the fact that they act as liquidity providers.

Summarizing, the analysis of the regression model in equation \eqref{equation:kirilenko} yields the following conclusion. First, it provides further statistical evidence in favor of our results about trading imbalances of DMMs, since it also underlines significant changes in inventory management of those traders during drift burst EPMs at an alternative frequency and with a different statistical model. Second, the significant change in inventory is robust to interaction with prices. That is, when price changes are part of the model, the behavior of DMMs still changes during the drift burst EPMs compared to normal times, in line with our findings in the main text. In particular, the estimated model parameters confirm that DMMs consume liquidity during systematic events, and that the role of liquidity providers is implicitly overtaken by fundamental players, i.e., NON-HFTs.

\end{document}

%% file: tables/drift-burst-epm-individual.tex
\begin{spacing}{1.2}
\begin{landscape}
\setlength{ \tabcolsep}{0.18cm}
\begin{scriptsize}
\begin{longtable}{lllrrrclllrrr}
\caption{The sample of drift burst EPMs -- Individual level}
\label{table:drift-burst-epm-individual} \\
\hline
Date & ISIN & Name & Begin & Trough & Duration & & Date & ISIN & Company & Begin & Trough & Duration \\ 
\hline
01/04 & FR0000131708 & Technip & 10:19:27 & 10:29:29 & 10:01 & & 07/23 & FR0010220475 & Alstom & 11:18:39 & 11:22:57 & 04:18 \\
01/08 & FR0010220475 & Alstom & 14:47:54 & 15:03:40 & 15:46 & & 07/25 & FR0000131708 & Technip & 11:52:35 & 12:05:16 & 12:41 \\
01/09 & FR0000120172 & Carrefour & 10:07:42 & 10:25:18 & 17:36 & & 07/31 & FR0000120404 & Accor & 13:59:14 & 14:03:35 & 04:21 \\
01/11 & FR0000130809 & Soci\'{e}t\'{e} G\'{e}n\'{e}rale & 10:53:06 & 10:57:35 & 04:30 & & 07/31 & NL0000235190 & EADS & 16:54:15 & 17:01:15 & 07:00 \\
01/11 & FR0000131708 & Technip & 10:39:28 & 10:54:05 & 14:37 & & 08/16 & FR0000120578 & Sanofi Synth\'{e}labo & 15:30:00 & 15:35:45 & 05:45 \\
01/14 & FR0000125338 & Cap G\'{e}mini & 16:45:52 & 16:52:20 & 06:28 & & 08/20 & FR0000120537 & Lafarge & 9:54:11 & 10:07:20 & 13:09 \\
01/14 & FR0000125007 & Saint-Gobain & 15:21:35 & 15:27:45 & 06:10 & & 08/23 & FR0000124141 & Veolia Environnement & 14:48:18 & 14:48:39 & 00:21 \\
01/14 & FR0000120354 & Vallourec & 15:48:07 & 15:52:15 & 04:08 & & 08/27 & FR0000120578 & Sanofi Synth\'{e}labo & 9:39:30 & 9:51:30 & 12:00 \\
01/15 & FR0010307819 & Legrand & 12:04:25 & 12:08:24 & 03:58 & & 08/27 & FR0000125007 & Saint-Gobain & 9:44:07 & 9:51:35 & 07:28 \\
01/22 & FR0000127771 & Vivendi Universal & 13:46:51 & 13:50:10 & 03:19 & & 08/27 & FR0000120537 & Lafarge & 9:44:47 & 9:51:56 & 07:09 \\
01/22 & FR0000121261 & Michelin & 9:54:07 & 10:06:05 & 11:58 & & 08/27 & FR0000121667 & Essilor International & 9:44:59 & 9:51:29 & 06:30 \\
01/23 & FR0010220475 & Alstom & 15:59:27 & 15:59:30 & 00:04 & & 08/29 & FR0000120693 & Pernod Ricard & 12:25:04 & 12:59:15 & 34:12 \\
01/29 & FR0000120537 & Lafarge & 9:52:51 & 10:05:06 & 12:15 & & 08/29 & FR0000121667 & Essilor International & 15:01:36 & 15:04:20 & 02:44 \\
01/29 & FR0000045072 & Credit Agricole & 10:02:42 & 10:15:25 & 12:43 & & 09/02 & FR0000131906 & Renault & 16:44:04 & 16:44:31 & 00:26 \\
01/31 & FR0000120644 & Danone & 11:54:52 & 12:12:21 & 17:28 & & 09/03 & FR0000121485 & Kering & 10:52:46 & 10:54:56 & 02:10 \\
02/04 & FR0000131104 & BNP & 14:08:05 & 14:15:11 & 07:06 & & 09/03 & FR0000121014 & Lvmh Moet & 10:52:45 & 10:54:54 & 02:09 \\
02/11 & FR0000120578 & Sanofi Synth\'{e}labo & 15:31:45 & 15:36:10 & 04:25 & & 09/03 & FR0000131104 & BNP & 10:53:21 & 10:55:15 & 01:55 \\
02/20 & FR0000133308 & Orange & 9:32:28 & 9:49:00 & 16:33 & & 09/03 & FR0000130809 & Soci\'{e}t\'{e} G\'{e}n\'{e}rale & 10:53:18 & 10:55:10 & 01:52 \\
02/22 & FR0000131906 & Renault & 14:59:51 & 15:15:05 & 15:14 & & 09/03 & FR0000120628 & Axa & 10:53:09 & 10:55:10 & 02:01 \\
02/26 & NL0000235190 & EADS & 16:29:03 & 16:38:35 & 09:32 & & 09/03 & FR0000125486 & Vinci & 10:52:56 & 10:55:14 & 02:18 \\
03/01 & FR0000045072 & Credit Agricole & 13:17:21 & 13:20:15 & 02:54 & & 09/03 & FR0000120073 & Air Liquide & 10:51:15 & 10:54:55 & 03:40 \\
03/06 & FR0000133308 & Orange & 15:22:29 & 15:24:25 & 01:56 & & 09/03 & FR0000120578 & Sanofi Synth\'{e}labo & 10:52:54 & 10:54:55 & 02:01 \\
03/06 & FR0000120404 & Accor & 15:23:22 & 15:34:03 & 10:41 & & 09/03 & FR0000120271 & Total & 10:53:00 & 10:54:55 & 01:55 \\
03/07 & FR0000130809 & Soci\'{e}t\'{e} G\'{e}n\'{e}rale & 10:42:26 & 10:49:45 & 07:19 & & 09/03 & FR0000125007 & Saint-Gobain & 10:51:03 & 10:55:04 & 04:01 \\
03/08 & FR0000125338 & Cap G\'{e}mini & 13:59:34 & 14:10:35 & 11:01 & & 09/03 & NL0000235190 & EADS & 10:52:56 & 10:54:55 & 01:59 \\
03/12 & NL0000226223 & STMicroelectronics  & 17:18:25 & 17:24:55 & 06:30 & & 09/03 & FR0000120537 & Lafarge & 10:52:40 & 10:54:59 & 02:19 \\
03/12 & FR0000121261 & Michelin & 16:22:32 & 16:28:04 & 05:33 & & 09/03 & FR0000121972 & Schneider & 10:52:38 & 10:55:07 & 02:29 \\
03/15 & FR0000130809 & Soci\'{e}t\'{e} G\'{e}n\'{e}rale & 11:51:19 & 11:55:10 & 03:52 & & 09/06 & FR0000121014 & Lvmh Moet & 15:39:26 & 15:47:35 & 08:09 \\
03/20 & FR0000073272 & Safran & 15:17:00 & 15:45:28 & 28:28 & & 09/06 & FR0000124141 & Veolia Environnement & 15:24:16 & 15:43:00 & 18:44 \\
03/21 & FR0000125338 & Cap G\'{e}mini & 15:15:37 & 15:33:26 & 17:49 & & 09/09 & FR0000125338 & Cap G\'{e}mini & 16:45:34 & 17:12:40 & 27:07 \\
03/25 & FR0010220475 & Alstom & 16:00:55 & 16:20:45 & 19:49 & & 09/10 & FR0000120354 & Vallourec & 14:47:32 & 14:50:04 & 02:32 \\
03/25 & NL0000235190 & EADS & 16:14:00 & 16:23:06 & 09:05 & & 09/16 & FR0000120644 & Danone & 11:23:54 & 11:30:20 & 06:26 \\
03/26 & FR0000073272 & Safran & 12:50:52 & 13:06:22 & 15:30 & & 09/17 & FR0000120354 & Vallourec & 16:37:51 & 16:40:15 & 02:24 \\
04/04 & NL0000235190 & EADS & 16:19:32 & 16:29:26 & 09:54 & & 09/20 & FR0010208488 & ENGIE & 10:37:10 & 10:40:05 & 02:55 \\
04/05 & FR0000131104 & BNP & 11:49:35 & 11:53:15 & 03:39 & & 09/24 & FR0000121485 & Kering & 15:37:05 & 15:44:04 & 06:58 \\
04/05 & FR0010242511 & EDF & 14:53:00 & 14:54:55 & 01:55 & & 09/24 & FR0000121261 & Michelin & 15:25:05 & 15:30:00 & 04:55 \\
04/05 & FR0000120073 & Air Liquide & 11:47:43 & 11:53:20 & 05:37 & & 09/24 & FR0000120693 & Pernod Ricard & 16:18:15 & 16:25:20 & 07:05 \\
04/10 & FR0000125338 & Cap G\'{e}mini & 13:11:55 & 13:13:52 & 01:57 & & 10/02 & FR0000121261 & Michelin & 14:34:11 & 14:45:30 & 11:19 \\
04/15 & FR0000130577 & Publicis Groupe SA & 14:46:08 & 14:56:04 & 09:56 & & 10/02 & FR0000120693 & Pernod Ricard & 17:21:58 & 17:25:05 & 03:08 \\
04/16 & FR0010208488 & ENGIE & 12:36:04 & 12:47:55 & 11:51 & & 10/02 & FR0010220475 & Alstom & 15:39:45 & 15:43:00 & 03:15 \\
04/17 & FR0000131906 & Renault & 9:30:08 & 9:51:45 & 21:37 & & 10/02 & FR0000121667 & Essilor International & 15:21:32 & 15:28:50 & 07:18 \\
04/17 & FR0010220475 & Alstom & 9:30:01 & 9:51:26 & 21:24 & & 10/03 & FR0010220475 & Alstom & 9:29:40 & 9:48:00 & 18:20 \\
04/17 & FR0010208488 & ENGIE & 9:44:54 & 9:51:34 & 06:41 & & 10/09 & FR0000120172 & Carrefour & 14:01:46 & 14:05:45 & 04:00 \\
04/17 & FR0000073272 & Safran & 9:29:55 & 9:51:56 & 22:00 & & 10/09 & FR0000120537 & Lafarge & 12:30:21 & 12:38:28 & 08:07 \\
04/17 & FR0000127771 & Vivendi Universal & 9:42:54 & 9:51:24 & 08:30 & & 10/18 & FR0000120172 & Carrefour & 11:23:12 & 11:34:48 & 11:35 \\
04/17 & FR0000125486 & Vinci & 9:30:09 & 9:51:45 & 21:36 & & 10/23 & NL0000226223 & STMicroelectronics  & 16:16:06 & 16:24:16 & 08:10 \\
04/17 & FR0000120628 & Axa & 9:42:50 & 9:51:45 & 08:56 & & 10/24 & FR0000073272 & Safran & 14:29:39 & 14:42:41 & 13:02 \\
04/17 & FR0000120354 & Vallourec & 9:30:02 & 9:51:25 & 21:23 & & 10/25 & FR0000120628 & Axa & 15:31:14 & 15:47:16 & 16:02 \\
04/17 & FR0000120578 & Sanofi Synth\'{e}labo & 9:42:55 & 9:51:30 & 08:35 & & 10/28 & FR0000120628 & Axa & 13:00:01 & 13:11:11 & 11:11 \\
04/17 & FR0000120172 & Carrefour & 9:42:34 & 9:51:29 & 08:55 & & 10/28 & FR0000125486 & Vinci & 17:25:06 & 17:29:35 & 04:29 \\
04/17 & FR0000120073 & Air Liquide & 9:41:34 & 9:51:30 & 09:56 & & 10/28 & FR0000125007 & Saint-Gobain & 13:02:45 & 13:15:25 & 12:39 \\
04/17 & FR0000121261 & Michelin & 9:42:50 & 9:51:40 & 08:50 & & 10/29 & FR0000124141 & Veolia Environnement & 9:29:59 & 9:40:05 & 10:06 \\
04/17 & FR0000120271 & Total & 9:30:09 & 9:51:34 & 21:26 & & 10/30 & FR0000131708 & Technip & 12:29:41 & 12:31:48 & 02:07 \\
04/17 & FR0000121972 & Schneider & 9:30:00 & 9:51:29 & 21:29 & & 11/05 & FR0000120693 & Pernod Ricard & 14:55:14 & 15:10:07 & 14:53 \\
04/23 & FR0000120404 & Accor & 16:50:24 & 17:14:07 & 23:43 & & 11/05 & FR0000130809 & Soci\'{e}t\'{e} G\'{e}n\'{e}rale & 14:46:45 & 14:50:15 & 03:30 \\
04/26 & FR0000120693 & Pernod Ricard & 10:18:56 & 10:27:19 & 08:23 & & 11/05 & FR0000120628 & Axa & 14:40:25 & 14:48:38 & 08:13 \\
04/29 & FR0000127771 & Vivendi Universal & 13:55:11 & 14:02:29 & 07:19 & & 11/07 & FR0000130809 & Soci\'{e}t\'{e} G\'{e}n\'{e}rale & 16:32:20 & 16:36:35 & 04:15 \\
04/30 & FR0000127771 & Vivendi Universal & 15:48:25 & 15:57:40 & 09:15 & & 11/20 & FR0000131708 & Technip & 15:37:23 & 16:09:40 & 32:17 \\
05/07 & NL0000235190 & EADS & 15:46:36 & 16:01:39 & 15:03 & & 11/21 & NL0000226223 & STMicroelectronics  & 13:56:25 & 13:58:30 & 02:06 \\
05/07 & FR0000121972 & Schneider & 12:17:56 & 12:27:00 & 09:04 & & 11/22 & FR0000133308 & Orange & 12:17:19 & 12:27:59 & 10:41 \\
05/10 & FR0010220475 & Alstom & 10:01:03 & 10:12:20 & 11:17 & & 11/22 & FR0000127771 & Vivendi Universal & 12:16:49 & 12:22:49 & 06:00 \\
05/22 & FR0000125338 & Cap G\'{e}mini & 15:44:54 & 15:47:05 & 02:11 & & 11/25 & FR0000125338 & Cap G\'{e}mini & 10:08:52 & 10:13:24 & 04:32 \\
05/23 & FR0000120172 & Carrefour & 15:46:17 & 15:48:51 & 02:34 & & 11/26 & FR0000120404 & Accor & 11:49:31 & 11:59:05 & 09:34 \\
05/29 & FR0000120271 & Total & 16:13:10 & 16:17:20 & 04:10 & & 11/26 & FR0000121667 & Essilor International & 16:37:43 & 16:41:02 & 03:18 \\
06/05 & FR0000120172 & Carrefour & 11:23:47 & 11:26:35 & 02:48 & & 11/27 & FR0000120404 & Accor & 13:27:19 & 13:47:55 & 20:36 \\
06/07 & FR0000121485 & Kering & 12:02:29 & 12:09:09 & 06:41 & & 12/03 & FR0000133308 & Orange & 9:58:45 & 10:03:55 & 05:10 \\
06/14 & FR0010220475 & Alstom & 15:01:31 & 15:02:23 & 00:52 & & 12/03 & FR0000131906 & Renault & 10:09:54 & 10:13:25 & 03:30 \\
06/14 & FR0000127771 & Vivendi Universal & 15:39:05 & 15:51:09 & 12:05 & & 12/03 & FR0000131708 & Technip & 9:53:52 & 10:03:48 & 09:56 \\
06/19 & FR0010220475 & Alstom & 13:08:27 & 13:10:54 & 02:27 & & 12/05 & FR0000130577 & Publicis Groupe SA & 15:31:10 & 15:43:04 & 11:55 \\
06/24 & FR0000130809 & Soci\'{e}t\'{e} G\'{e}n\'{e}rale & 10:15:26 & 10:33:25 & 17:59 & & 12/17 & FR0000120404 & Accor & 11:05:41 & 11:16:16 & 10:35 \\
06/24 & FR0000125007 & Saint-Gobain & 10:15:02 & 10:33:25 & 18:23 & & 12/18 & FR0000120354 & Vallourec & 9:30:00 & 9:46:41 & 16:41 \\
07/01 & FR0000120537 & Lafarge & 16:39:15 & 16:41:45 & 02:30 & & 12/19 & FR0000045072 & Credit Agricole & 14:48:57 & 14:53:20 & 04:23 \\
07/11 & FR0010242511 & EDF & 11:13:35 & 11:18:55 & 05:20 & & 12/19 & FR0000131906 & Renault & 16:54:09 & 17:07:10 & 13:01 \\
07/18 & FR0000130577 & Publicis Groupe SA & 13:51:35 & 13:51:56 & 00:21 & & 12/23 & FR0000133308 & Orange & 10:46:27 & 11:02:27 & 16:00 \\
\hline
\end{longtable}
\end{scriptsize}
\begin{spacing}{0.5}
\noindent \emph{Note.}
The table reports detailed information about the 148 drift burst EPMs in our sample.
In particular, we report the date of occurrence, and the name and ISIN code of the affected stock.
Moreover, we show the time of the beginning and trough of each event, and its duration (in mm:ss).
\end{spacing}
\end{landscape}\end{spacing}

%% file: tables/drift-burst-epm-aggregated.tex
\begin{sidewaystable}[ht!]
\setlength{ \tabcolsep}{0.15cm}
\begin{center}
\caption{The sample of drift burst EPMs -- Aggregated by stock}
\label{table:drift-burst-epm-aggregated}
\begin{footnotesize}
\begin{tabular}{lllrrrrrrrcrrr}
\hline
&&&&&&& \multicolumn{3}{c}{Return (in \%)} & & \multicolumn{3}{c}{Duration (mm:ss)} \\
\cline{8-10} \cline{12-14}
ISIN & Name & Ticker & Market Cap & Trades & Volume & \# EPMs & Mean & Median & Std. && Mean & Median & Std. \\
\hline
FR0000045072 & Credit Agricole & ACA & 23,221 & 12,774 & 88 & 3 & -1.13 & -0.89 & 0.47 & & 06:40 & 04:23 & 05:17 \\
FR0000073272 & Safran & SAF & 21,064 & 8,569 & 60 & 4 & -1.11 & -0.98 & 0.53 & & 19:45 & 18:45 & 06:56 \\
FR0000120073 & Air Liquide & AI & 32,047 & 12,821 & 128 & 3 & -0.99 & -0.99 & 0.30 & & 06:24 & 05:37 & 03:12 \\
FR0000120172 & Carrefour & CA & 20,858 & 13,152 & 116 & 6 & -1.53 & -1.57 & 0.74 & & 07:55 & 06:27 & 05:58 \\
FR0000120271 & Total & FP & 145,995 & 26,025 & 348 & 3 & -1.23 & -1.01 & 0.49 & & 09:10 & 04:10 & 10:40 \\
FR0000120354 & Vallourec & VK & 5,035 & 9,902 & 51 & 5 & -1.18 & -0.78 & 0.60 & & 09:26 & 04:08 & 08:57 \\
FR0000120404 & Accor & AC & 7,822 & 7,872 & 49 & 6 & -1.10 & -1.07 & 0.48 & & 13:15 & 10:38 & 07:21 \\
FR0000120537 & Lafarge & LG & 15,652 & 11,512 & 76 & 6 & -1.12 & -0.97 & 0.40 & & 07:35 & 07:38 & 04:37 \\
FR0000120578 & Sanofi Synth\'{e}labo & SAN & 101,851 & 27,238 & 400 & 5 & -0.94 & -0.87 & 0.57 & & 06:33 & 05:45 & 03:52 \\
FR0000120628 & Axa & CS & 48,784 & 19,042 & 200 & 5 & -1.24 & -1.02 & 0.71 & & 09:16 & 08:56 & 05:05 \\
FR0000120644 & Danone & BN & 30,688 & 14,526 & 176 & 2 & -1.40 & -1.40 & 0.02 & & 11:57 & 11:57 & 07:49 \\
FR0000120693 & Pernod Ricard & RI & 21,799 & 10,385 & 96 & 5 & -0.88 & -0.69 & 0.59 & & 13:32 & 08:23 & 12:18 \\
FR0000121014 & Lvmh Moet & MC & 66,353 & 13,133 & 200 & 2 & -1.27 & -1.27 & 0.33 & & 05:09 & 05:09 & 04:14 \\
FR0000121261 & Michelin & ML & 14,350 & 12,618 & 99 & 5 & -1.50 & -1.77 & 0.72 & & 08:31 & 08:50 & 03:13 \\
FR0000121485 & Kering  & KER & 19,395 & 6,899 & 83 & 3 & -0.89 & -0.84 & 0.29 & & 05:16 & 06:41 & 02:42 \\
FR0000121667 & Essilor International & EI & 16,592 & 10,950 & 86 & 4 & -0.98 & -0.93 & 0.15 & & 04:58 & 04:54 & 02:17 \\
FR0000121972 & Schneider  & SU & 35,628 & 16,767 & 164 & 3 & -1.79 & -1.43 & 1.15 & & 11:01 & 09:04 & 09:39 \\
FR0000124141 & Veolia Environnement & VIE & 6,338 & 10,989 & 68 & 3 & -2.46 & -2.79 & 1.21 & & 09:44 & 10:06 & 09:12 \\
FR0000125007 & Saint-Gobain & SGO & 22,193 & 13,639 & 116 & 5 & -1.36 & -1.34 & 0.33 & & 09:44 & 07:28 & 05:47 \\
FR0000125338 & Cap G\'{e}mini & CAP & 7,876 & 9,677 & 61 & 7 & -0.87 & -0.87 & 0.28 & & 10:09 & 06:28 & 09:20 \\
FR0000125486 & Vinci & DG & 28,713 & 14,361 & 124 & 3 & -1.25 & -1.18 & 0.79 & & 09:28 & 04:29 & 10:34 \\
FR0000127771 & Vivendi Universal & VIV & 25,660 & 13,320 & 143 & 6 & -1.39 & -1.34 & 0.66 & & 07:45 & 07:54 & 02:59 \\
FR0000130577 & Publicis Groupe SA & PUB & 13,740 & 9,466 & 75 & 3 & -1.52 & -1.26 & 0.54 & & 07:24 & 09:56 & 06:11 \\
FR0000130809 & Soci\'{e}t\'{e} G\'{e}n\'{e}rale & GLE & 33,722 & 32,204 & 317 & 7 & -1.62 & -1.52 & 0.83 & & 06:11 & 04:15 & 05:27 \\
FR0000131104 & BNP & BNP & 70,354 & 33,015 & 364 & 3 & -1.14 & -1.03 & 0.28 & & 04:13 & 03:39 & 02:38 \\
FR0000131708 & Technip & TEC & 7,942 & 10,665 & 75 & 6 & -1.17 & -1.08 & 0.48 & & 13:36 & 11:21 & 10:06 \\
FR0000131906 & Renault & RNO & 17,064 & 14,722 & 118 & 5 & -1.57 & -1.33 & 0.88 & & 10:46 & 13:01 & 08:41 \\
FR0000133308 & Orange & ORA & 23,630 & 21,114 & 167 & 5 & -1.47 & -1.25 & 0.82 & & 10:04 & 10:41 & 06:28 \\
FR0010208488 & ENGIE & ENGI & 40,349 & 13,831 & 148 & 3 & -2.02 & -1.68 & 0.81 & & 07:09 & 06:41 & 04:29 \\
FR0010220475 & Alstom & ALO & 8,126 & 11,838 & 81 & 10 & -1.45 & -1.47 & 0.67 & & 09:45 & 07:47 & 08:29 \\
FR0010242511 & EDF & EDF & 47,729 & 9,368 & 64 & 2 & -1.40 & -1.40 & 0.67 & & 03:37 & 03:37 & 02:25 \\
FR0010307819 & Legrand & LR & 10,633 & 6,387 & 45 & 1 & -1.08 & -1.08 & & & 03:58 & 03:58 &  \\
NL0000226223 & STMicroelectronics  & STM & 7,098 & 8,668 & 44 & 3 & -2.60 & -1.41 & 2.16 & & 05:35 & 06:30 & 03:08 \\
NL0000235190 & EADS & AIR & 43,550 & 20,886 & 212 & 6 & -0.88 & -0.96 & 0.25 & & 08:46 & 09:19 & 04:15 \\
\hline
\end{tabular}
\end{footnotesize}
\\
\medskip
\parbox{ \textwidth}{ \emph{Note.}
We study tick-by-tick order-level data in a sample composed of 37 stocks that were traded on NYSE Euronext Paris in 2013 and are members of the CAC 40 index
(we exclude Arcelor Mittal, Gemalto, and Solvay, because their main trading venue is not the Paris branch of NYSE Euronext).
The table reports the average daily market capitalization (in million euros, from Bloomberg), number of transactions, and trading volume (in million euros) of each stock.
We apply the test statistic of \citet{christensen-oomen-reno:22a} to screen for downward drift burst EPMs, which shows 148 drift burst EPMs affecting 34 of the 37 included equities.
\# EPMs is the number of detected events in each stock.
We report the mean, median, and standard deviation of the rate of return (from beginning to trough) and duration (in mm:ss) of stock-level events.
}
\end{center}
\end{sidewaystable}

%% file: userref.bib
@MISC{amf:17a,
 AUTHOR = {{Autorit\'{e} des March\'{e}s Financiers}},
 YEAR = {2017},
 TITLE = {{Study of the behaviour of high-frequency traders on Euronext Paris}},
 NOTE = {Available at \url{https://www.amf-france.org/en/news-publications/publications/reports-research-and-analysis/study-behaviour-high-frequency-traders-euronext-paris}}
}

@ARTICLE{anand-venkataraman:16a,
 AUTHOR = {A. Anand and K. Venkataraman},
 YEAR = {2016},
 TITLE = {{Market conditions, fragility, and the economics of market making}},
 JOURNAL = {Journal of Financial Economics},
 VOLUME = {121},
 NUMBER = {2},
 PAGES = {327--349}
}

@ARTICLE{barclay-warner:93a,
 AUTHOR = {M. J. Barclay and J. B. Warner},
 YEAR = {1993},
 TITLE = {{Stealth trading and volatility: Which trades move prices?}},
 JOURNAL = {Journal of Financial Economics},
 VOLUME = {34},
 NUMBER = {3},
 PAGES = {281--305}
}

@ARTICLE{bellia-pelizzon-subrahmanyam-yuferova:25a,
 AUTHOR = {M. Bellia and L. Pelizzon and M. G. Subrahmanyam and D. Yuferova},
 YEAR = {2025},
 TITLE = {{Market liquidity and competition among designated market makers}},
 JOURNAL = {Management Science},
 VOLUME = {71},
 NUMBER = {1},
 PAGES = {184--201}
}

@MISC{benos-sagade:12a,
 AUTHOR = {E. Benos and S. Sagade},
 YEAR = {2012},
 TITLE = {{High-frequency trading behaviour and its impact on market quality: Evidence from the UK equity market}},
 NOTE = {Working paper}
}

@ARTICLE{bessembinder-hao-zheng:20a,
 AUTHOR = {H. Bessembinder and J. Hao and K. Zheng},
 YEAR = {2020},
 TITLE = {{Liquidity provision contracts and market quality: Evidence from the New York Stock Exchange}},
 JOURNAL = {Review of Financial Studies},
 VOLUME = {33},
 NUMBER = {1},
 PAGES = {44--74}
}

@MISC{brogaard:10a,
 AUTHOR = {J. A. Brogaard},
 YEAR = {2010},
 TITLE = {{High frequency trading and its impact on market quality}},
 NOTE = {Working paper}
}

@ARTICLE{brogaard-carrion-moyaert-riordan-shkilko-sokolov:18a,
 AUTHOR = {J. A. Brogaard and A. Carrion and T. Moyaert and R. Riordan and A. Shkilko and K. Sokolov},
 YEAR = {2018},
 TITLE = {{High frequency trading and extreme price movements}},
 JOURNAL = {Journal of Financial Economics},
 VOLUME = {128},
 NUMBER = {2},
 PAGES = {253--265}
}

@ARTICLE{brunnermeier-pedersen:05a,
 AUTHOR = {M. K. Brunnermeier and L. H. Pedersen},
 YEAR = {2005},
 TITLE = {{Predatory trading}},
 JOURNAL = {Journal of Finance},
 VOLUME = {60},
 NUMBER = {4},
 PAGES = {1825--1863}
}

@ARTICLE{budish-cramton-smith:15a,
 AUTHOR = {E. Budish and P. Cramton and J. Smith},
 YEAR = {2015},
 TITLE = {{The high-frequency trading arms race: Frequent batch auctions as a market design response}},
 JOURNAL = {Quarterly Journal of Economics},
 VOLUME = {130},
 NUMBER = {4},
 PAGES = {1547--1621}
}

@ARTICLE{cespa-foucault:14a,
 AUTHOR = {G. Cespa and T. Foucault},
 YEAR = {2014},
 TITLE = {{Illiquidity contagion and liquidity crashes}},
 JOURNAL = {Review of Financial Studies},
 VOLUME = {27},
 NUMBER = {6},
 PAGES = {1615--1660}
}

@MISC{cespa-vives:25a,
 AUTHOR = {G. Cespa and X. Vives},
 YEAR = {2025},
 TITLE = {{Market opacity and fragility: Why liquidity evaporates when it is most needed}},
 NOTE = {Working paper}
}

@ARTICLE{chaboud-chiquoine-hjalmarsson-vega:14a,
 AUTHOR = {A. P. Chaboud and B. Chiquoine and E. Hjalmarsson and C. Vega},
 YEAR = {2014},
 TITLE = {{Rise of the machines: Algorithmic trading in the foreign exchange market}},
 JOURNAL = {Journal of Finance},
 VOLUME = {69},
 NUMBER = {5},
 PAGES = {2045--2084}
}

@ARTICLE{gao-mizrach:16a,
 AUTHOR = {C. Gao and B. Mizrach},
 YEAR = {2016},
 TITLE = {{Market quality breakdowns in equities}},
 JOURNAL = {Journal of Financial Markets},
 VOLUME = {28},
 NUMBER = {1},
 PAGES = {1--23}
}

@ARTICLE{christensen-oomen-reno:22a,
 AUTHOR = {K. Christensen and R. C. A. Oomen and R. Ren\`{o}},
 YEAR = {2022},
 TITLE = {{The drift burst hypothesis}},
 JOURNAL = {Journal of Econometrics},
 VOLUME = {227},
 NUMBER = {2},
 PAGES = {461--497}
}

@ARTICLE{clark-joseph-ye-zi:17a,
 AUTHOR = {A. D. Clark-Joseph and M. Ye and C. Zi},
 YEAR = {2017},
 TITLE = {{Designated market makers still matter: Evidence from two natural experiments}},
 JOURNAL = {Journal of Financial Economics},
 VOLUME = {126},
 NUMBER = {3},
 PAGES = {652--667}
}

@MISC{euronext:12a,
 AUTHOR = {NYSE--Euronext},
 YEAR = {2012},
 TITLE = {{Euronext Cash Market, Info Flash of 26 January 2012}},
 NOTE = {Available at \url{https://euronext.com/}}
}

@MISC{euronext:13a,
 AUTHOR = {NYSE--Euronext},
 YEAR = {2013},
 TITLE = {{Euronext Cash Market, Info Flash of 9 May 2013}},
 NOTE = {Available at \url{https://euronext.com/}}
}

@MISC{euronext:25a,
 AUTHOR = {Euronext},
 YEAR = {2025},
 TITLE = {{Euronext Rule Book -- Book I: Harmonised Rules}},
 NOTE = {The latest version is available at \url{https://www.euronext.com/en/regulation/euronext-regulated-market}}
}

@ARTICLE{glosten:87a,
 AUTHOR = {L. R. Glosten},
 YEAR = {1987},
 TITLE = {{Components of the bid-ask spread and the statistical properties of transaction prices}},
 JOURNAL = {Journal of Finance},
 VOLUME = {42},
 NUMBER = {5},
 PAGES = {1293--1307}
}

@ARTICLE{goldstein-kavajecz:04a,
 AUTHOR = {M. A. Goldstein and K. A. Kavajecz},
 YEAR = {2004},
 TITLE = {{Trading strategies during circuit breakers and extreme market movements}},
 JOURNAL = {Journal of Financial Markets},
 VOLUME = {7},
 NUMBER = {3},
 PAGES = {301--333}
}

@ARTICLE{hendershott-jones-menkveld:11a,
 AUTHOR = {T. Hendershott and C. M. Jones and A. J. Menkveld},
 YEAR = {2011},
 TITLE = {{Does algorithmic trading improve liquidity?}},
 JOURNAL = {Journal of Finance},
 VOLUME = {66},
 NUMBER = {1},
 PAGES = {1--33}
}

@ARTICLE{huang-wang:09a,
 AUTHOR = {J. Huang and J. Wang},
 YEAR = {2009},
 TITLE = {{Liquidity and market crashes}},
 JOURNAL = {Review of Financial Studies},
 VOLUME = {22},
 NUMBER = {7},
 PAGES = {2607--2643}
}

@ARTICLE{jacod-li-mykland-podolskij-vetter:09a,
 AUTHOR = {J. Jacod and Y. Li and P. A. Mykland and M. Podolskij and M. Vetter},
 YEAR = {2009},
 TITLE = {{Microstructure noise in the continuous case: The pre-averaging approach}},
 JOURNAL = {Stochastic Processes and their Applications},
 VOLUME = {119},
 NUMBER = {7},
 PAGES = {2249--2276}
}

@MISC{jones:13a,
 AUTHOR = {C. M. Jones},
 YEAR = {2013},
 TITLE = {{What do we know about high-frequency trading?}},
 NOTE = {Working paper}
}

@ARTICLE{kervel-menkveld:19a,
 AUTHOR = {V. van Kervel and A. J. Menkveld},
 YEAR = {2019},
 TITLE = {{High-frequency trading around large institutional orders}},
 JOURNAL = {Journal of Finance},
 VOLUME = {74},
 NUMBER = {3},
 PAGES = {1091--1137}
}

@ARTICLE{kirilenko-kyle-samadi-tuzun:17a,
 AUTHOR = {A. Kirilenko and A. S. Kyle and M. Samadi and T. Tuzun},
 YEAR = {2017},
 TITLE = {{The Flash Crash: High frequency trading in an electronic market}},
 JOURNAL = {Journal of Finance},
 VOLUME = {72},
 NUMBER = {3},
 PAGES = {967--998}
}

@ARTICLE{korajczyk-murphy:19a,
 AUTHOR = {R. A. Korajczyk and D. Murphy},
 YEAR = {2019},
 TITLE = {{High-frequency market making to large institutional trades}},
 JOURNAL = {Review of Financial Studies},
 VOLUME = {32},
 NUMBER = {3},
 PAGES = {1034--1067}
}

@ARTICLE{lee-mykland:08a,
 AUTHOR = {S. S. Lee and P. A. Mykland},
 YEAR = {2008},
 TITLE = {{Jumps in financial markets: A new nonparametric test and jump dynamics}},
 JOURNAL = {Review of Financial Studies},
 VOLUME = {21},
 NUMBER = {6},
 PAGES = {2535--2563}
}

@ARTICLE{menkveld:14a,
 AUTHOR = {A. J. Menkveld},
 YEAR = {2014},
 TITLE = {{High-frequency traders and market structure}},
 JOURNAL = {Financial Review},
 VOLUME = {49},
 NUMBER = {2},
 PAGES = {333--344}
}

@MISC{mifid:21a,
 AUTHOR = {{ESMA}},
 YEAR = {2021},
 TITLE = {{MiFID II review report on Algorithmic Trading}},
 NOTE = {Available at \url{https://www.esma.europa.eu/sites/default/files/library/esma70-156-4572_mifid_ii_final_report_on_algorithmic_trading.pdf}}
}

@MISC{sec:14a,
 AUTHOR = {{U.S. Securities and Exchange Commission}},
 YEAR = {2014},
 TITLE = {{Equity market structure literature review -- Part II: High frequency trading}},
 NOTE = {Report by the staff of the Division of Trading and Markets, available at \url{https://www.sec.gov/marketstructure/research/hft_lit_review_march_2014.pdf}}
}

@ARTICLE{yang-zhu:20a,
 AUTHOR = {L. Yang and H. Zhu},
 YEAR = {2020},
 TITLE = {{Back-running: Seeking and hiding fundamental information in order flows}},
 JOURNAL = {Review of Financial Studies},
 VOLUME = {33},
 NUMBER = {4},
 PAGES = {1484--1533}
}
